\documentclass[12pt]{article}
\def\be{\begin{equation}}
\def\bea{\begin{eqnarray}}
\def\ee{\end{equation}}
\def\eea{\end{eqnarray}}

\def\d{\partial}
\input epsf
\begin{document}

\begin{flushright}
OHSTPY-HEP-T-99-018 \\
% MPI H-V \\
hep-th/9910222
\end{flushright}
\vspace{20mm}

\begin{center}
{\LARGE SDLCQ\\ Supersymmetric Discrete Light Cone Quantization }
\\
\vspace{20mm}
{\bf O. Lunin, S. Pinsky \\}
\vspace{4mm}
Department of Physics,\\ The Ohio State University,\\ Columbus, OH 43210, USA\\
\vspace{4mm}
\end{center}
\vspace{10mm}

\begin{abstract}
In these lectures we discuss the application of discrete light cone
quantization (DLCQ) to supersymmetric field theories. We will see
that it is possible to formulate DLCQ so that supersymmetry is
exactly preserved in the discrete approximation. We call this formulation
of DLCQ,
SDLCQ and it combines the power of DLCQ with all of the
beauty of supersymmetry. In these lecture we will review the application of
SDLCQ
to several interesting supersymmetric theories. We will discuss two dimensional
theories with (1,1), (2,2) and (8,8) supersymmetry, zero modes, vacuum
degeneracy,
massless states, mass gaps, theories in higher dimensions, and the Maldacena
conjecture among other subjects.
\end{abstract}

\vspace{10mm}
\begin{center}
To be published in:\\
{\it New Directions in Quantum Chromodynamics,} C.R.~Ji, Ed.,\\
American Institute of Physics, New York, 1999.\\
(Proceedings of a Summer School in Seoul, Korea, 26 May-- 18 June 1999).
\end{center}
\newpage

\subsection*{Introduction.}

In the last decade there have been significant improvements in our
understanding of
gauge theories and important breakthroughs in the
nonperturbative description of supersymmetric gauge theories
\cite{seiberg,seibergwitten}. In the last few years various relations
between string
theory, brane theory and gauge fields \cite{givkut,agmoo} have also emerged.
While these developments give us some insight into strongly coupled gauge
theories
\cite{seibergwitten}, they do not offer a direct method for non-perturbative
calculations. In these lectures we discuss some recent developments in
light cone
quantization approaches to non-perturbative problems. We will see that
these methods
have the potential to expand our understanding of strongly coupled gauge
theories
in directions not previously available.

The original idea was formulated half of a century ago \cite{Dirac}, but
apart from several technical clarifications \cite{MaYam} it remained mostly
undeveloped. The first change came in the mid 80--th when the Discrete
Light Cone
Quantization (DLCQ) was suggested as a practical way for calculating the
masses and wavefunctions of hadrons \cite{BP85}. Although the direct
application of the method to realistic problems meets some difficulties (for
review see \cite{BPP}), DLCQ has been successful in studying various two
dimensional models. Given the importance of supersymmetric theories, it is not
surprising that light cone quantization was ultimately applied to such models
\cite{kut93,bdk93,kub94}. Even in this early work the mass spectrum was
shown to be supersymmetric in the continuum and a great deal of information
about the properties of bound states in supersymmetric theories was extracted.
However the straightforward application of DLCQ to the supersymmetric systems
had one disadvantage: the supersymmetry was lost in the discrete formulation.
The way to solve this problem was suggested in \cite{sakai}, where an
alternative formulation of DLCQ was introduced. Namely it was noted that since
the supercharge is the "square root" of Hamiltonian one can define a new DLCQ
procedure based on the supercharge. We will study this formulation (called
SDLCQ) in these lectures.

These lectures have the following organization. In section \ref{ChLCGauge} we
introduce the basic concepts of DLCQ and SDLCQ. We also define the systems to
be studied in the remaining lectures. We will concentrate our
attention on  two dimensional models with adjoint matter, several examples
of such systems can be constructed from supersymmetric Yang--Mills theories
in higher dimensions using  dimensional reduction. In section \ref{ChVac}
we address the problem of the DLCQ vacuum. In the continuum theory the
light cone
vacuum is very simple: it coincides with the usual Fock vacuum. This property
is related to the decoupling between positive and negative frequency modes
on the light cone and does not occur for equal time quantization. In DLCQ
however one encounters the problem of zero modes which complicates the
structure of vacuum and allows us to reproduce the
correct vacuum degeneracy in certain theories. We continue to analyze zero
modes in
the section
\ref{ChExSUSY} and they are shown to play an important role in  explaining
the difference between DLCQ and SDLCQ regularization procedures.

Section \ref{ChMsls} is devoted to the study of massless states. Our
numerical analysis \cite{sakai,alp2,appt,alppt} shows an important
property of the mass spectrum for some supersymmetric theories. We find
that unlike the QCD--like models \cite{thooft}, such systems appear to have a
lot of massless bound states and in fact the supersymmetric $SU(\infty)$ gauge
theory seems to have an infinite number of such states in the continuum
limit. Since the states with zero mass dominate the partition function
for low enough temperatures they deserve to be studied very carefully and in
section \ref{ChMsls} we analyze the structure of such states.

As we already mentioned in the beginning of this introduction, the relation
between string theory and gauge fields has attracted a lot of attention in
recent years. In particular it was conjectured \cite{adscft} that one can
extract some information about strongly coupled gauge theory from supergravity
calculations. The problem however is that in the relevant regime  usual
field theoretic methods do not work, so it is hard to really test the
conjecture. For two dimensional systems, however  DLCQ gives  solutions of the
bound state problem which are valid beyond perturbation theory, so the results
can be used to test the conjecture. We report the results of this first test
in section \ref{ChMaldac}. For realistic systems with eight supersymmetries we
still
don't have enough computer power to compare the results with the supergravity
predictions. The general techniques described in this section can also be
used to
calculate other correlation functions in the nonperturbative regime.

Finally in section \ref{Ch3D} we make a first attempt to move beyond two
dimensions.  We present the general ideas for formulating SDLCQ in more
than two
dimensions. As an example we present the numerical results of SYM for the
simplest
case when only one transverse momentum mode is introduced.

\section{ \bf Supersymmetric Yang--Mills Theory in the Light--Cone Gauge.}
\label{ChLCGauge}
\renewcommand{\theequation}{1.\arabic{equation}}
\setcounter{equation}{0}

\subsection{DLCQ and Its Supersymmetric Version.}

In this work we will study the bound state problem for various supersymmetric
matrix models in two dimensions. The examples of such models may be
constructed by dimensional reduction of supersymmetric Yang--Mills theory in
higher dimensions. In this subsection we will consider such reduction for three
dimensional SYM.
Before we begin the detailed analysis of the bound state problem
for the specific systems it is worthwhile to summarize some basic
ideas of Discrete Light Cone Quantization ( for a complete review see
\cite{BPP}).

Let us consider a general relativistic system in two dimensions. Usual
canonical quantization of such a system means that one imposes certain
commutation relations between coordinates and momenta at equal time. However
as was pointed out by Dirac long ago \cite{Dirac} this is not the only
possibility. Another scheme of quantization treats the light like coordinate
$x^+=\frac{1}{\sqrt{2}}(x^0+x^1)$ as a
new time and then the system is quantized canonically. This scheme (called
light cone quantization) has both positive and negative sides. The main
disadvantage of light cone quantization is the presence of constraints,
even for systems as simple as free bosonic field. From the action
\be
S=\int d^2 x \d_+\phi \d_-\phi
\ee
one can derive the constraint relating coordinate and momentum:
\be
\pi=\d_-\phi.
\ee
For more complicated systems the constraints are also present and in general
they are hard to resolve.

The main advantage of the light cone is the decoupling between positive and
negative momentum modes. This property is crucial for DLCQ. In the
Discrete Light Cone Quantization one considers the theory on the finite circle
along the $x^-$ axis: $-L<x^-<L$. Then all the momenta become quantized and
the integer number measuring the total momentum in terms of "elementary
momentum" is called the harmonic resolution $K$. Due to the decoupling
property one may work only in the sector with positive momenta where there are
a finite number of states for any finite value of resolution. Of course the
full quantum field theory in the continuum corresponds to the limit
$L\rightarrow\infty$ and , in this limit the elementary bit of momentum goes
to zero, as the harmonic resolution goes to infinity and the infinite number
of degrees of freedom are restored. However it is believed that the "quantum
mechanical" approximation is suitable for describing the lowest states in the
spectrum.
Note that the problem of constraints in DLCQ is a quantum mechanical one and
thus it is easier to solve. Usually this problem can be reformulated in terms
of zero modes and the solution can be found for any value of the resolution.

DLCQ is mainly used in order to solve the bound state problem. Let us
formulate this problem for general two dimensional theory. The theory in the
continuum has the full Poincare symmetry, thus the states are naturally
labeled by the eigenvalues of Casimir operators of the Poincare algebra. One
such Casimir is the mass operator: $M^2=P^\mu P_\mu$. Another Casimir is
related to the spin of the particle and we will not use it. After compactifying
the $x^-$ direction one looses Lorenz symmetry, but not the translational
invariance in $x^+$ and $x^-$ directions. Thus $P^+$ and $P^-$ are still
conserved charges, but now the mass operator is not the only Casimir of the
symmetry group: the states are characterized by both $P^+$ and $P^-$. However
if we consider DLCQ as an approximation to the continuum theory we anticipate
that in the limit of infinite harmonic resolution (or $L\rightarrow\infty$)
the Poincare symmetry is restored and the mass will be the only quantity
having invariant meaning. Thus the aim would be to study the value of $M^2$
as function of $K$ and to extrapolate the results to the $K=\infty$.

The usual way to define $M^2$ in DLCQ is based on separate calculation of
$P^+$ and $P^-$ in matrix form and then bringing them together:
\be
M^2=2P^+P^-.
\ee
Usually one works in the sector with fixed $P^+$, and the calculation of light
cone Hamiltonian $P^-$ is the nontrivial problem. An important
simplifications occur
for supersymmetric theories \cite{sakai}.

Supersymmetry is the only
nontrivial extension of Poincare algebra compatible with the existence of the
S matrix \cite{WessBag}. Namely in addition to usual bosonic generators of
symmetries, fermionic ones are allowed and the full (super)algebra in
two dimensions reads:
\bea\label{SUSYalg}
\{Q^I_\alpha,Q^J_\beta\}&=&2\delta^{IJ}\gamma^\mu_{\alpha\beta}P_\mu+
   \varepsilon_{\alpha\beta}Z^{IJ},\\
\left[ P_\mu,P_\nu \right]&=&0,\qquad
\left[P_\mu,Q^I_\alpha \right]=0.
\eea
In this expression $\varepsilon$ is an antisymmetric $2\times 2$ matrix,
$\varepsilon_{12}=1$ and $Z^{IJ}$ is the set of c--numbers called the central
charges. In these lectures we will put them equal to zero.
It is convenient to choose two dimensional gamma matrices in the form:
$\gamma^0=\sigma^2$, $\gamma^1=i\sigma^1$, then one can rewrite
(\ref{SUSYalg}) in terms of light cone components:
\bea
\{Q^+_I,Q^+_J\}&=&2\sqrt{2}\delta^{IJ}P^+,\\
\{Q^-_I,Q^-_J\}&=&2\sqrt{2}\delta^{IJ}P^-,\\
\{Q^+_I,Q^-_J\}&=&2Z_{IJ}.
\eea
As we mentioned before, in DLCQ diagonalization of $P^+$ is trivial and
the construction of Hamiltonian is the main problem. The last set of
equations suggests an alternative way of dealing with this problem: one
can first construct the matrix representation for the supercharge $Q^-$
and then just square it. This version of DLCQ first suggested in
\cite{sakai} appeared to be very fruitful. First of all it preserves
supersymmetry at finite resolution, while the conventional DLCQ
applied to supersymmetric theories doesn't (we will consider the relation
between these two approaches in section \ref{ChExSUSY}).  The supersymmetric
version
of DLCQ (SDLCQ) also provides better numerical convergence.

To summarize, in this subsection we defined two procedures for studying the
bound state spectrum: DLCQ and SDLCQ. To implement the first one we
construct the light cone Hamiltonian and diagonalize it, while in the second
approach one constructions the supercharge and from it the Hamiltonian. Of
course the SDLCQ method is appropriate only for the theories with
supersymmetries, although it can be modified to study models with soft
supersymmetry breaking (see section
\ref{Ch3D}).

\subsection{Reduction from Three Dimensions.}

Let us start by the defining a simple supersymmetric system in two
dimensions. It can be constructed by dimensional reduction of SYM from
three dimensions to two dimension. The more general case can be found in
the next
subsection.

Our starting point is the action for SYM in $2+1$ dimensions:
\be\label{3daction}
S=\int d^3 x \mbox{tr} \left(-\frac{1}{4} F_{AB}F^{AB}
+i \bar{\Psi}\gamma^A D_{A}\Psi \right).
\ee
The system consists of gauge field $A_A$ and two--component Majorana fermion
$\Psi$, both transforming according to adjoint representation of gauge group.
We assume that this group is either $U(N)$ or $SU(N)$ and thus matrices
$A^A_{ij}$ and $\Psi_{ij}$ are hermitian.
Studying dimensional reduction of $SYM_{D}$ we introduce the following
conventions for the indices: the capital latin letters correspond to $D$
dimensional spacetime, greek indices label two dimensional coordinates and
the lower case letters are used as matrix indices.
According to this conventions the indexes in (\ref{3daction}) go from
zero to two, the field strength $F_{AB}$ and covariant derivative $D_{A}$ are
defined in the usual way:
\bea
F_{AB}=\d_A A_B-\d_B A_A +ig[A_A,A_B],\nonumber\\
D_A\Psi=\d_A\Psi +ig[A_A,\Psi].
\eea

Dimensional reduction to $1+1$ means that we require all fields to be
independent on coordinate $x^2$, in other words we place the system on
the cylinder with radius $L_\perp$ along the $x^2$ axis and consider only
zero modes of the fields. The possible improvement of this approximation
will be suggested in section \ref{Ch3D}, here we consider this reduction as a
formal
way of getting two dimensional matrix model. In the reduced theory it is
convenient to introduce two dimensional indices and treat $A^2$ component
of gauge field as two dimensional scalar $\phi$. The action for the reduced
theory has the form:
\begin{equation}\label{2dact3dfer}
S  =  \int d^2 x \hspace{1mm}
  \mbox{tr} \left(-\frac{1}{4} F_{\mu \nu}F^{\mu \nu}
+\frac{1}{2}D_\mu \phi D^\mu \phi +
i \bar{\Psi}\gamma^\mu D_{\mu}\Psi -2ig\phi
\bar{\Psi}\gamma_5\Psi \right),
\end{equation}
We also could choose the
special representation of three dimensional gamma matrices:
\be
\gamma^0=\sigma^2,\qquad \gamma^1=i\sigma^1, \qquad \gamma^2=i\sigma^3,
\ee
then it would be natural to write the spinor $\Psi$ in terms of its
components:
\be\label{3ddecomp}
\Psi=(\psi,\chi)^T.
\ee
Taking all these definitions into account one can rewrite the dimensional
reduction of (\ref{3daction}) as:
\bea\label{2daction}
S&=&L_\perp\int d^2x\left(\frac{1}{2}D_\mu\phi D^\mu\phi +i\sqrt{2}\psi D_+\psi
+i\sqrt{2}\chi D_-\chi+\right.\nonumber\\
&+&\left.2g\psi\{\psi,\chi\}-\frac{1}{4}F_{\mu\nu}F^{\mu\nu}\right).
\eea
The covariant derivatives here are taken with respect to the light cone
coordinates:
\be
x^\pm=\frac{x^0\pm x^1}{\sqrt{2}}.
\ee
Note that by rescaling the fields and coupling constant $g$ we can make the
constant $L_\perp$ to be equal to one, so below we simply omit this constant.

The bound state problem for the system (\ref{2daction}) was first studied in
\cite{sakai}. The supersymmetric version of the discrete light cone
quantization was used in order to find the mass spectrum. However the zero
modes were neglected by authors of \cite{sakai}, so we spend some time
studying this problem in the next section. As we will see, while zero modes
are not very important for calculations of massive spectrum, they play
crucial role in the description of the vacuum of the theory.

Let us consider (\ref{2daction}) as the theory in the continuum. In
this case one can choose the light cone gauge:
\be
A^+=0,
\ee
then equations of motion for $A^-$ and $\chi$ give constraints:
\bea
&&-\partial_-^2 A^- =gJ^+, \\
&&\sqrt{2} i \partial_- \chi=g[\phi, \psi],\\
&&J^+(x)=\frac{1}{i}[\phi(x), \partial_-\phi(x)]-
\frac{1}{\sqrt 2}\{\psi(x), \psi(x)\}.
\eea
Solving this constraints and substituting the result back into the action
one determines the Lagrangian as function of physical fields $\phi$ and
$\psi$ only. Then using the usual Noether technique we can construct the
conserved charges corresponding to the translational invariance:
\bea
P^+&=&\int dx^- \mbox{tr}\left((\partial_-\phi)^2+
     i\sqrt{2}\psi\partial_-\psi\right),\\
P^-&=&
  \int dx^-{\rm tr} \left(
-\frac{g^2}{2} J^+\frac{1}{\partial_-^2} J^+
+\frac{ig^2}{2\sqrt 2}[\phi, \psi]
\frac{1}{\partial_-}[\phi, \psi]\right).
\eea
We can also construct the Noether charges corresponding to the supersymmetry
transformation. However the naive SUSY transformations break the gauge fixing
condition $A^+=0$, so they should be accompanied by compensating gauge
transformation:
\bea\label{susytrans}
\delta A_\mu=\frac{i}{2} {\bar\varepsilon}\gamma_\mu \Psi-D_\mu
\frac{i}{2} {\bar\varepsilon}\gamma_-\frac{1}{\partial_-} {\Psi},\\
\delta\Psi=\frac{1}{4}F_{\mu\nu}\gamma^{\mu\nu}\varepsilon-\frac{g}{2}[
{\bar\varepsilon}\gamma_-\frac{1}{\partial_-} {\Psi},\Psi].\nonumber
\end{eqnarray}
The resulting supercharges are:
\bea
Q^+=2\int dx^-\mbox{tr}\left(\psi\partial_-\phi\right),\\
Q^-=-2g\int dx^-\mbox{tr}\left(J^+\frac{1}{\partial_-}\psi\right).
\eea

Finally we make a short comment on supersymmetry in the pure fermionic
system. As one can see the expression for $Q^-$ contains the term cubic
in fermions, so if we formally put $\phi=0$ this supercharge will not
vanish. One may ask what kind of supersymmetric system this supercharge
corresponds to. This answer was found by Kutasov \cite{kut93} who
discovered the supersymmetry in the system of adjoint fermions, namely
the square of supercharge including fermions only gives Hamiltonian:
\be
P^- =  \int
dx^- \mbox{tr}\left(-\frac{{\rm i}m^2}{2}\psi\frac{1}{\partial_-}\psi-
\frac{g^2}{2}J^+\frac{1}{\partial_-^2}J^+\right),
\ee
$m^2=g^2N/\pi$. This $P^-$ corresponds to the system of adjoint fermions
in two dimensions with the special value of mass $m$. We will consider
this system in details in section \ref{ChExSUSY}.

\subsection{Reduction from Higher Dimensions.}

In this subsection we consider the general reduction of SYM$_D$ to two
dimensions. By counting the fermionic and bosonic degrees of freedom
one can see that the SYM can be defined only in limited number of
spacetime dimensions, namely $D$ can be equal to 2, 3, 4, 6 or 10. The
last case is the most general one: all other system can be obtained by
dimensional reduction and appropriate truncation of degrees of freedom.
So in this subsection we will concentrate on reduction $10\rightarrow 2$, and
the comments on four and six dimensional cases will be made in the end.

As in the last subsection we start from ten dimensional action:
\be\label{10daction}
S=\int d^3 x \mbox{tr} \left(-\frac{1}{4} F_{AB}F^{AB}+
i \bar{\Psi}\gamma^A D_{A}\Psi \right).
\ee

According to our general conventions the indexes in (\ref{10daction}) go from
zero to nine, $\Psi$ is the ten dimensional Majorana--Weyl spinor. A general
spinor in ten dimensions has $2^{10/2}=32$ complex components, if the
appropriate basis of gamma matrices is chosen then Majorana condition makes
all the components real. Since all the matrices in such representation are
real, the Weyl condition
\begin{equation}
\Gamma_{11}\Psi=\Psi
\end{equation}
is compatible with the reality of $\Psi$ and thus it eliminates half of its
components. In the special representation of Dirac matrices:
\begin{eqnarray}
&& \Gamma^0=\sigma_2 \otimes {\bf 1}_{16}, \\
&& \Gamma^I={\rm i}\sigma_1 \otimes \gamma^I, \hspace{6mm} I=1,\dots,8;\\
% && \Gamma_{11}= \Gamma^0 \cdots \Gamma^9=\sigma_3\otimes {\bf 1}_{16} .
&& \Gamma^9= {\rm i}\sigma_1 \otimes \gamma^9,
\end{eqnarray}
the $\Gamma_{11}= \Gamma^0 \cdots \Gamma^9$ has very simple form:
$\Gamma_{11}=\sigma_3\otimes {\bf 1}_{16}$. Then the Majorana spinor of
positive chirality can be written in terms of 16--component real object
$\psi$:
\begin{equation}
\Psi= 2^{1/2} { \psi \choose 0}. \label{spin16}
\end{equation}

Let us return to the expressions for $\Gamma$ matrices. The ten dimensional
Dirac algebra
$$
\{\Gamma_\mu,\Gamma_\nu\}=2g_{\mu\nu}
$$
is equivalent to the spin(8) algebra for $\gamma$ matrices:
$\{\gamma_I,\gamma_J\}=2\delta_{IJ}$ and the ninth matrix can be chosen to
be $\gamma^9=\gamma^1\dots\gamma^8$. Note that the 16 dimensional
representation of spin(8) is the reducible one: it can be decomposed as
${\bf 8}_s+{\bf 8}_c$
\begin{equation}
\gamma^I=\left(\begin{array}{cc}
0 & \beta_I\\
\beta_I^T & 0
\end{array}\right), \hspace{7mm} I=1,\dots,8.
\end{equation}
The explicit expressions for the $\beta_I$ satisfying
$\{\beta_I,\beta_J\}=2\delta_{IJ}$ can be found in \cite{GSW}. Such choice
leads to the convenient form of $\gamma^9$:
\begin{equation}
\gamma^9=\left(\begin{array}{cc}
{\bf 1}_{8} & 0\\
0 & -{\bf 1}_{8} \end{array}\right). \label{gamma9}
\end{equation}

So far we have found nonzero components of the spinor given by (\ref{spin16}).
However as we saw in the last subsection not all such components are physical
in the light cone gauge, so it is useful to perform the analog of
decomposition (\ref{3ddecomp}). In ten dimension it is related with breaking
the sixteen component spinor $\psi$ on the left and right--moving components
using the projection operators
\begin{equation}
P_L=\frac{1}{2}(1-\gamma^9), \qquad P_R=\frac{1}{2}(1+\gamma^9).
\end{equation}
After introducing the light--cone coordinates
$x^\pm=\frac{1}{\sqrt{2}}(x^0\pm x^9)$ the action (\ref{10daction}) can be
rewritten as
\begin{eqnarray}
S_{9+1}^{LC} & = & \int dx^+ dx^- d{\bf x}^{\perp} \hspace{1mm}
  \mbox{tr} \left( \frac{1}{2}F_{+-}^2 + F_{+I}F_{-I} - \frac{1}{4}F_{IJ}^2
  \right. \nonumber \\
& & \hspace{20mm}
+ \hspace{1mm}
i\sqrt{2} \psi_R^T D_+ \psi_R + i\sqrt{2}\psi_L^T D_- \psi_L +
      2i\psi_L^T \gamma^I D_I \psi_R \left.
\frac{}{} \right),
\label{LCversion}
\end{eqnarray}
where the repeated indices $I,J$ are summed over $(1,\dots,8)$. After applying
the light--cone gauge $A^+=0$ one can eliminate nonphysical degrees of freedom
using the Euler--Lagrange equations for $\psi_L$ and $A^-$:
\begin{eqnarray}
\label{fermioncon}
  \partial_- \psi_L = -\frac{1}{\sqrt{2}}\gamma^I D_I \psi_R, \\
\label{apluscon}
\partial_{-}^2 A_{+}=\partial_{-}\partial_{I}A_{I}+gJ^+\\
J^+=i[A_{I},\partial_{-}A_{I}]+2\sqrt{2}\psi_{R}^T\psi_{R}.
\end{eqnarray}

Performing the reduction to two dimensions means that all fields are assumed
to be independent on the transverse coordinates: $\partial_{I}\Phi=0$. Then
as in previous subsection one can construct the conserved momenta $P^\pm$ in
terms of physical degrees of freedom:
\begin{eqnarray}
P^+&=&\int dx^- \mbox{tr}\left((\partial_-A_I)^2+
     i\sqrt{2}\psi_R\partial_-\psi_R\right),\\
P^-&=&
  \int dx^-{\rm tr} \left(
-\frac{g^2}{2} J^+\frac{1}{\partial_-^2} J^+
+\frac{ig^2}{2\sqrt 2}[A_I, \psi_R^T] \beta^T_I
\frac{1}{\partial_-}\beta_J[A_J, \psi_R]\right)-\nonumber\\
&-&\frac{1}{4}\int dx^-{\rm tr}\left([A_IA_J]^2\right).
\end{eqnarray}

We can also construct the Noether charges corresponding to the supersymmetry
transformation (\ref{susytrans}). As in the three dimensional case it is
convenient to decompose the supercharge in two components:
$$
Q^+=P_L Q,\qquad Q^-=P_R Q.
$$
The resulting eight component supercharges are given by
\begin{eqnarray}
Q^+=2\int dx^-\mbox{tr}\left(\beta^T_I\psi_R\partial_-A_I\right),\\
Q^-=-2g\int dx^-\mbox{tr}\left(J^+\frac{1}{\partial_-}\psi_R+
\frac{i}{4}[A_IA_J](\beta_I\beta^T_J-\beta_J\beta^T_I)\psi_R\right).
\end{eqnarray}

Finally we make a short comment on dimensional reduction of $SYM_{3+1}$ and
$SYM_{5+1}$. These systems can be constructed repeating the procedure just
described. However there is an easier way to construct the Hamiltonian and
supercharges for the dimensionally reduced theories, namely one has to
truncate the unwanted degrees of freedom in the ten dimensional expressions.
This is especially easy for the bosonic coordinates: one simply considers
indices $I$ and $J$ running from one to two (for $D=4$) or to four (for $D=6$).
The fermionic truncation can also be performed by requiring the spinor
$\psi_R$ to be 2-- or 4--component. Then the only problem is the choice of
$2\times 2$ or $4\times 4$ beta matrices satisfying
\begin{equation}
\{\beta_I,\beta_J\}=2\delta_{IJ},
\end{equation}
that can be done easily.

%%%%%%%%%%%%%%%%%%%%%%%%%%%%%%%%%%%%%%%%%%%%%%%%%%%%%
%%%%%%%%%%%%%%%%%%%%%%%%%%%%%%%%%%%%%%%
%%%%%%%%%%%%%%%%%%%%%%%%%%%%%%%%%%%%%%%
\section{Zero Modes and Light Cone Vacuum.}
\label{ChVac}
\renewcommand{\theequation}{2.\arabic{equation}}
\setcounter{equation}{0}

The results of this section are based on the paper \cite{alptzm}
%%%%%%%%%%%%%%%%%%%%%%%%%%%%%%%%%%%%%%%%%%%%%%%
\subsection{Gauge Fixing in DLCQ}
We consider the supersymmetric Yang-Mills theory in
1+1 dimensions \cite{fer65} which is described by
the action (\ref{2dact3dfer}):
\begin{equation}
S  =  \int d^2 x \hspace{1mm}
  \mbox{tr} \left(-\frac{1}{4} F_{\mu \nu}F^{\mu \nu}
+\frac{1}{2}D_\mu \phi D^\mu \phi +
i \bar{\Psi}\gamma^\mu D_{\mu}\Psi -2ig\phi
\bar{\Psi}\gamma_5\Psi \right).
\end{equation}
A convenient  representation of the gamma matrices is $\gamma^0=\sigma^2$,
$\gamma^1=i\sigma^1$ and $\gamma^5=\sigma_3$ where $\sigma^a$ are
the Pauli matrices. In this representation the Majorana spinor
is real. We use the matrix notation for $SU(N)$ so that $A^\mu_{ij}$ and
$\Psi_{ij}$ are $N\times N$ traceless matrices.

We now introduce the light-cone coordinates
$x^{\pm}=\frac{1}{\sqrt 2}(x^0 \pm x^1)$.
The longitudinal coordinate $x^-$ is compactified
on a finite interval $x^-\in [-L, L]$ \cite{MaYam,BP85}
and we impose periodic boundary conditions on all fields
to ensure unbroken supersymmetry.

The light-cone gauge $A^+=0$ can not be used in a finite
compactification radius,
but the modified condition $\partial_-A^+=0$ \cite{kpp94}
is consistent with the light-like  compactification.
We can make a
global rotation in color space so that
the zero mode is diagonalized $A^+_{ij}(x^+)=v_i(x^+)\delta_{ij}$ with
$\sum_i v_i =0$ \cite{kpp94}. The gauge zero modes correspond to a
(quantized) color
electric flux loops around the compactified space.
The modified light-cone gauge is not a complete gauge fixing. We still have
large gauge transformations preserving the gauge condition $\partial_-A^+=0$.
There are two types of such transformations \cite{lnt94,lst95}: displacements
$T_D$ and central conjugations $T_C$. Their actions on the physical fields of
the theory and complete gauge fixing will be discussed in the end of this
subsection. Now we just mention that being discrete transformations, $T_D$ and
$T_C$ don't affect quantization procedure.

The quantization in the light--cone gauge with or without dynamical $A^+$ is
widely explored in the literature \cite{pk96,mrp97,sakai,alp2,BPP},
here we provide  only the results which are useful for later purposes.
The quantization proceeds in two steps. First, we must resolve the
constraints to eliminate the redundant degrees of freedom.
There are two constraints in the theory,
\begin{eqnarray}\label{constr1}
&&-D_-^2 A^- =gJ^+, \\
\label{constr2}
&&\sqrt{2} i D_- \chi=g[\phi, \psi],
\end{eqnarray}
where $\Psi\equiv (\psi, \chi)^{{\rm T}}$ and
the current operator is
\begin{equation}\label{current}
J^+(x)=\frac{1}{i}[\phi(x), D_-\phi(x)]-
\frac{1}{\sqrt 2}\{\psi(x), \psi(x)\}.
\end{equation}

Different components of (\ref{constr1}), (\ref{constr2}) play different roles
in the theory. First we look at diagonal zero modes of these equations.
The diagonal zero
mode of (\ref{constr2}) gives us constraints on the physical fields:
\begin{equation}\label{fermConstr}
[\phi,\psi]^0_{ii}=0.
\end{equation}
There is no sum over $i$ in above expression. As one can see this constraint
leads to decoupling of $\stackrel{0}{\chi}_{ii}$, this field plays the role
of Lagrange multiplier for above condition. The same is true for
${\stackrel{0}{A}}^-_{ii}$, the corresponding constraint is
$\stackrel{0}{J}_{ii}=0$.
The reason we treated the diagonal zero modes of (\ref{constr1}) and
(\ref{constr2}) separately is that for all other modes the $D_-$ operator is
invertible and instead of constraints on physical fields $\psi$ and $\phi$
one gets expressions for non-dynamical ones:
\begin{equation}
A^-=-\frac{g}{D_-^2}J^+,\qquad
\chi=\frac{g}{i\sqrt{2}}\frac{1}{D_-}[\phi, \psi].
\end{equation}

The next step is to derive the commutation relations
for the physical degrees of freedom.
As in the ordinary quantum mechanics, the zero mode
$v_i$ has a  conjugate momentum  $p=2L\partial_+v_i$ and
the commutation relation is $[v_i, p_j]=i\delta_{ij}$.
The off--diagonal components of the scalar field
are complex valued operators with
$\phi_{ij}=(\phi_{ji})^{\dagger}$. The canonical momentum
conjugate to $\phi_{ij}$ is $\pi_{ij}=(D_-\phi)_{ji}$.
They satisfy the canonical commutation relations \cite{pk96,sakai}
\begin{equation}
[\phi_{ij}(x), \pi_{kl}(y)]_{x^+=y^+}=
[\phi_{ij}(x), D_-\phi_{lk}(y)]_{x^+=y^+}=\frac{i}{2}
(\delta_{ik}\delta_{jl}-\frac{1}{N}\delta_{ij}\delta_{kl})
\delta(x^--y^-).
\end{equation}
On the other hand, the quantization of the
diagonal component $\phi_{ii}$ needs care. As mentioned in
\cite{pk96}, the zero mode of $\phi_{ii}$, the mode independent
of $x^-$, is not an independent degree of freedom but
obeys a certain constrained equation \cite{MaYam,pk96,kall}.
Except the zero
mode, the commutation relation is canonical
\begin{equation}
[\phi_{ii}(x), \partial_-\phi_{jj}(y)]_{x^+=y^+}=
\frac{i}{2}(1-\frac{1}{N})\delta_{ij}
\left[\delta(x^--y^-)-\frac{1}{2L}\right].
\end{equation}
The commutator of diagonal and non-diagonal elements of $\phi$ vanishes.
The canonical anti-commutation relations for fermion fields are \cite{sakai}
\begin{equation}
\{\psi_{ij}(x), \psi_{kl}(y)\}_{x^+=y^+}=
\frac{1}{\sqrt 2}
\delta(x^--y^-)(\delta_{il}\delta_{jk}-\frac{1}{N}\delta_{ij}\delta_{kl}).
\end{equation}
There are two differences between this expression and one from \cite{sakai}.
First one is technical: we consider commutators for $SU(N)$ group, this gives
$1/N$ term. Second difference is that unlike \cite{sakai} we include zero
modes in the expansion of $\psi$, we also include such modes in non-diagonal
elements of $\phi$.

Finally we return to the problem of complete gauge fixing. The actions of
$T_D$ and $T_C$ on physical fields are given by \cite{lnt94,lst95}:
\begin{eqnarray}
T_D:&& v_i(x^+)\rightarrow v_i(x^+)+\frac{n_i\pi}{gL},\qquad
        n_i\in {\bf {\rm Z}}, \qquad \sum n_i=0,\\
  && \psi_{ij}\rightarrow \exp(\frac{\pi i(n_i-n_j)x^-}{L})\psi_{ij},\qquad
        \phi_{ij}\rightarrow \exp(\frac{\pi i(n_i-n_j)x^-}{L})\phi_{ij};
    \nonumber\\
T_C:&& v_i(x^+)\rightarrow v_i(x^+)+\frac{\nu_i\pi}{gL},\qquad
        \nu_i=n(1/N-\delta_{iN}), \\
  && \psi_{ij}\rightarrow \exp(\frac{\pi i(\nu_i-\nu_j)x^-}{L})\psi_{ij},\qquad
     \phi_{ij}\rightarrow \exp(\frac{\pi i(\nu_i-\nu_j)x^-}{L})\phi_{ij}.
    \nonumber
\end{eqnarray}
There are also permutations of the color basis $i\rightarrow P(i)$ which
leave the theory invariant.
These symmetries preserve the gauge condition $\partial_-A^+=0$, but two
configurations related by $T_D$, $T_C$ or $P$ are equivalent. To fix the gauge
completely one therefore considers $v_i$ only in the fundamental domain,
other regions related with this domain by $T_D$, $T_C$ or $P$ give gauge
``copies" of it \cite{gri78}. The easiest thing to do is to describe the
boundaries of fundamental domain imposed by displacements $T_D$:
$-\frac{\pi}{2gL}<v_i<\frac{\pi}{2gL}$. The invariance under $T_C$
limits this region even more, but since we will not need the explicit form of
fundamental domain, we do not discuss such limits for $SU(N)$. For the
simplest case of $SU(2)$ the fundamental domain is given by
$0<v_1=-v_2<\frac{\pi}{2gL}$, the result for $SU(3)$ can be found in
\cite{lst95}. The $P$ symmetries do not respect the fundamental domain,
so they are not symmetries of gauge fixed theory. However there is one
special transformation among $P$ which being accompanied with combination of
$T_D$ and $T_C$ leaves fundamental domain invariant. Namely if $R$ is cyclic
permutation of color indexes then there exists a combination $T$ of $T_D$ and
$T_C$ such that $S=TR$ is the symmetry of gauge fixed theory. The explicit
form of $T$ depends on the rank of the group, for $SU(2)$ and $SU(3)$ it may
be found in \cite{lst95}. The operator $S$ satisfies the condition $S^N=1$
and it was used in classifying the vacua \cite{lst95,pin97a}.

%%%%%%%%%%%%%%%%%%%%%%%%%%%%%%%%%%%%%%%%%%%%%%%
\subsection{Current Operators}
The resolution of the Gauss-law constraint  (\ref{constr1})
is a necessary step for obtaining
the light-cone Hamiltonian.
The expression for the current operator is,
however, ill--defined unless an appropriate definition
is specified,  since
the operator products are defined at the same point.
We shall use the point--splitting regularization
which respects the symmetry of the theory under the
large gauge transformation.

To simplify notation it is convenient to introduce the dimensionless
variables $z_i=Lgv_i/\pi$ instead of quantum mechanical coordinates $v_i$
describing $A^+$.
The mode--expanded fields at the light-cone time $x^+=0$ are
\begin{eqnarray}\label{2mode_expansion}
\phi_{ij}(x)&=&\frac{1}{\sqrt{4\pi}}
\left( \sum_{n=0}^{\infty} a_{ij}(n)u_{ij}(n) {\rm e}^{-ik_n x^-}
+\sum_{n=1}^{\infty} a^{\dagger}_{ji}(n) u_{ij}(-n)
{\rm e}^{ik_n x^-}\right),\ i\ne j,\nonumber \\
\phi_{ii}(x)&=&\frac{1}{\sqrt{4\pi}}
\sum_{n=1}^{\infty}\frac{1}{\sqrt n}\left(a_{ii}(n) {\rm e}^{-ik_n x^-}
+ a^{\dagger}_{ii}(n) {\rm e}^{ik_n x^-}\right), \nonumber \\
\psi_{ij}(x)&=&
\frac{1}{2^{\frac{1}{4}}\sqrt{2L}}
\left( \sum_{n=0}^{\infty}b_{ij}(n) {\rm e}^{-ik_n x^-}
+\sum_{n=1}^{\infty} b^{\dagger}_{ji}(n) {\rm e}^{ik_n x^-}\right),
\end{eqnarray}
where  $k_n=n\pi/L$, $u_{ij}(n)=1/\sqrt{\vert n-z_i+z_j \vert}$
\footnote{$u_{ij}(n)$ is
well-defined in the
fundamental domain. Similarly,
$(D_-)^2$ in the Gauss-law constraint
have no zero modes in this domain.}.
The (anti)commutation
relations for Fourier modes are found in \cite{pk96,mrp97}
and in our notation they take the form
\begin{eqnarray}\label{ccr}
&&[a_{ij}(n), a^{\dagger}_{kl}(m)]={\rm sgn}(n+z_j-z_i)\delta_{n, m}
(\delta_{ik}\delta_{jl}-\frac{1}{N}\delta_{ij}\delta_{kl}),\nonumber \\
&&\{b_{ij}(n), b^{\dagger}_{kl}(m)\}=\delta_{n, m}
(\delta_{ik}\delta_{jl}-\frac{1}{N}\delta_{ij}\delta_{kl})
\end{eqnarray}
The zero modes in above relations deserve special consideration. Although we
formally wrote them as $a_{ij}(0)$ and $b_{ij}(0)$, these modes also
act as creation operators because the conjugation of zero mode gives another
zero mode:
\begin{equation}
a_{ij}^\dagger(0)=a_{ji}(0), \qquad b_{ij}^\dagger(0)=b_{ji}(0).
\end{equation}
In particular the diagonal components of fermionic zero mode are real and we
will use them later to describe the degeneracy of vacua. Now we concentrate
our attention on non-diagonal zero modes. In the fundamental domain all $z_i$
are different, then one can always make take them to satisfy the inequality
$z_N<z_{N-1}<\dots<z_1$ in this domain. Such condition together with
(\ref{ccr}) leads to interpretation of $a_{ij}(0)$ as creation operator if
$i<j$ and as annihilation operator otherwise. The situation for fermions
is more ambiguous. One can consider $b_{ij}(0)$ as creation operator either
when $i<j$ or when $i>j$, both assumptions are consistent with (\ref{ccr}).
Later we will explore each of these situations.

Let us now discuss the definition of singular operator products
in the current (\ref{current}). We define the current operator
by point splitting:
\begin{equation}
J^+ \equiv \lim_{\epsilon \rightarrow 0}\left( {J^+}_{\phi}(x;
\epsilon)
+{J^+}_{\psi}(x; \epsilon) \right),
\label{current2}
\end{equation}
where the divided pieces are given by
\begin{eqnarray}
&& {J^+}_{\phi}(x; \epsilon)=\frac{1}{i}
\left[{\rm e}^{-i\frac{\pi\epsilon}{2L} M}
\phi(x^- -\epsilon){\rm e}^{i\frac{\pi\epsilon}{2L} M},
D_-\phi(x^-)\right] \label{bcurrent}\\
&& {J^+}_{\psi}(x; \epsilon)=-\frac{1}{\sqrt 2}
\left\{ {\rm e}^{-i\frac{\pi \epsilon}{2L} M}
\psi(x^--\epsilon){\rm e}^{i\frac{\pi\epsilon}{2L} M},
\psi(x^-)\right\}.
\label{fcurrent}
\end{eqnarray}
Here $M$ is diagonal matrix: $M={\mbox diag}(z_1,\dots,z_N)$.
An advantage of this regularization is that
the current transforms covariantly under the
large gauge transformation.

To evaluate (\ref{bcurrent}) and (\ref{fcurrent}) we will generalize
the approach used in \cite{pk96,mrp97} to the SU(N) case. First let
us calculate the vacuum average of bosonic current. Taking into account the
interpretation of zero modes as creation--annihilation operators we obtain:
\begin{eqnarray}
&&\langle 0|{J^+_{ij}}_{\phi}(x;\epsilon)|0\rangle=
\frac{1}{i}\langle 0|{\rm e}^{-i\frac{\pi\epsilon}{2L} (z_i-z_k)}
\phi_{ik}(x^- -\epsilon)D_-\phi(x^-)_{kj}-\nonumber\\
&&-{\rm e}^{-i\frac{\pi\epsilon}{2L} (z_k-z_j)}
\phi_{kj}(x^- -\epsilon)D_-\phi(x^-)_{ik}|0\rangle=\nonumber\\
&&=\frac{1}{4L}\sum_k\sum_{m>0}\left(
{\rm e}^{-i\frac{\pi\epsilon}{2L} (z_i-z_k)}-
{\rm e}^{-i\frac{\pi\epsilon}{2L} (z_k-z_j)}\right){\rm e}^{-ik_m\epsilon}
(\delta_{ij}-\frac{1}{N}\delta_{ik}\delta_{jk})+\nonumber\\
&&+\frac{1}{4L}\sum_{k<j}{\rm e}^{-i\frac{\pi\epsilon}{2L} (z_i-z_k)}
\delta_{ij}-
\frac{1}{4L}\sum_{k>i}{\rm e}^{-i\frac{\pi\epsilon}{2L} (z_k-z_j)}
\delta_{ij}.
\end{eqnarray}
Evaluating the sum and taking the limit one finds:
\begin{equation}
\lim_{\epsilon \rightarrow 0} {J^+_{ij}}_{\phi}(x;\epsilon)
=:{J^+_{ij}}_{\phi}(x): +\frac{1}{4L}\left(z_i-(N+1-2i)\right)\delta_{ij},
\label{bcurrent2}
\end{equation}
where $:{J^+}_{\phi}:$ is the naive normal ordered currents. To be more
precise, we have omitted the zero modes of the diagonal color sectors in
which the notorious constrained zero mode \cite{MaYam} appears.

The result for fermionic current depends on our interpretation of zero modes
as creation--annihilation operators and it is given by
\begin{equation}
\lim_{\epsilon \rightarrow 0} {J^+_{ij}}_{\psi}(x;\epsilon)
=:{J^+_{ij}}_{\psi}(x): -\frac{1}{4L}\left(z_i\mp(N+1-2i)\right)\delta_{ij}.
\label{fcurrent2}
\end{equation}
The minus sign here corresponds to the case where $b_{ij}(0)$ is a creation
operator if $i<j$ (i.e. the convention is the same as for the bosons) and
plus corresponds to the opposite situation.
As can be seen,
$ {J^+}_{\phi}$ and $ {J^+}_{\psi}$
acquire extra $z$ dependent terms, so called gauge corrections.
Integrating these charges over $x^-$, one finds that the charges are time
dependent. Of course this is an unacceptable situation, and
implies the need to impose
special conditions to single out `physical states' to form
a sensible theory. The
important simplification of the  supersymmetric model is that these time
dependent terms cancel, and the full current  (\ref{current2})
becomes
\begin{equation}
J^+_{ij}(x)=:{J^+_{ij}}_{\phi}:+ :{J^+_{ij}}_{\psi}:+C_i\delta_{ij}.
\label{nocurrent}
\end{equation}
Depending on the convention for fermionic zero modes the {\em z
independent} constants $C_i$ either vanish or they are given by
\begin{equation}
C_i=-\frac{1}{2L}(N+1-2i).
\end{equation}
The regularized current is thus equivalent to the naive
normal ordered current up to an irrelevant constant.
Similarly, one can show that $P^+$ picks up gauge correction when the
adjoint scalar or adjoint fermion are considered separately but in the
supersymmetric theory it is nothing more than  the expected normal ordered
contribution of the matter fields.

In one sense these results are a consequence of the well known fact that the
normal ordering constants in a supersymmetric theory cancel between
fermion and boson contributions. The important point here is that these normal
ordered constants are not actually constants, but rather quantum mechanical
degrees of freedom. It is therefore not obvious that they should
cancel. Of course, this property profoundly effects the
dynamics of the theory.

%%%%%%%%%%%%%%%%%%%%%%%%%%%%%%%%%%%%%%%%%%%%%%%
\subsection{Vacuum Energy}
The wave function of the vacuum state for the
supersymmetric Yang-Mills theory in 1+1 dimensions
has already been discussed in
the
equal-time formulation \cite{Oda97}.
An effective potential is computed in a weak
coupling region as a function of the gauge zero mode
by using the adiabatic  approximation.
Here we analyze the vacuum structure of the same theory
in the context of the DLCQ formulation.

The presence of zero modes renders the light-cone
vacuum quite nontrivial, but the advantage of the light-cone quantization
becomes  evident: the ground state is the Fock vacuum for a fixed
gauge zero mode and therefore our ground state may be
written in the tensor product form
\begin{equation}
\vert \Omega \rangle \equiv \Phi[z]\otimes
\vert 0 \rangle,
\label{vacuum}
\end{equation}
where we have taken the  Schr\"{o}dinger representation
for the quantum mechanical degree of freedom $z$
which is defined in the fundamental domain.
In contrast, to find the ground state of the fermion
and boson for a fixed value of the gauge zero mode
turns out to be a highly nontrivial task in the equal-time
formulation \cite{Oda97}.

Our next task is to derive an effective Hamiltonian
acting on $\Phi[z]$. The light-cone Hamiltonian
$H \equiv P^-$ is obtained from energy momentum tensors,
or through the canonical procedure:
\begin{eqnarray}\label{hamil}
H&=&-\frac{g^2L}{4\pi^2}\frac{1}{K(z)}\sum_i
\frac{\partial}{\partial z_i}
K(z)\frac{\partial}{\partial z_i}+ \nonumber\\
&+& \int_{-L}^L dx^-{\rm tr} \left(
-\frac{g^2}{2} J^+\frac{1}{D_-^2} J^+
+\frac{ig^2}{2\sqrt 2}[\phi, \psi]
\frac{1}{D_-}[\phi, \psi]\right),\\
K(z)&=&\prod_{i>j}{\rm sin}^2(\frac{\pi (z_i-z_j)}{2}),
\end{eqnarray}
where the first term is the kinetic energy of the
gauge zero mode,
and in the second term the zero modes of $D_-$ are understood
to be removed. Note that the kinetic term of the gauge
zero mode is not the standard form $-d^2/dz^2$ but acquires a nontrivial
Jacobian $K$ which is nothing but the Haar measure of SU(N).
The Jacobian originates from the unitary transformation of the variable from
$A^+$ to $v$, and can be derived  by explicit evaluation of a functional
determinant \cite{lnt94,lst95}. In the present context
it is found in \cite{kall}. Also we mention that Hamiltonian (\ref{hamil})
seems to contain terms quadratic in diagonal zero modes
$\stackrel{0}{\psi}_{ii}$. However using constraint equations one can show
that the total contribution of all such term vanishes. This also can be
seen by using the fact that Hamiltonian is proportional to the square
of supercharge (\ref{Qminus}).

Projecting the light-cone
Hamiltonian onto the Fock vacuum sector we obtain
the quantum mechanical Hamiltonian
\begin{equation}
H_0=-\frac{g^2L}{4\pi^2}\frac{1}{K(z)}\sum_i
\frac{\partial}{\partial z_i}
K(z)\frac{\partial}{\partial z_i}+V_{JJ}+V_{\phi\psi},
\end{equation}
where the reduced potentials are defined by
\begin{eqnarray}
&& V_{JJ}\equiv -\frac{g^2}{2}\int_{-L}^L dx^-
\langle {\rm tr} J^+\frac{1}{D_-^2} J^+  \rangle, \\
&&V_{\phi\psi}\equiv\frac{ig^2}{2\sqrt 2} \int_{-L}^L dx^-
\langle {\rm tr}
[\phi, \psi]
\frac{1}{D_-}[\phi, \psi]\rangle,
\end{eqnarray}
respectively. As stated in the previous subsection,
the gauge invariantly regularized current
turns out to be precisely the normal ordered current
in the absence of the zero modes.
It is now straightforward to evaluate $V_{JJ}$ and $V_{\phi\psi}$
in terms of modes. One finds that they cancel among themselves
as expected from the supersymmetry:
\begin{eqnarray}
V_{JJ}&=&-V_{\phi\psi}=\frac{g^2L}{16\pi^2}\left[\sum_{n, m=1}^\infty
\sum_{ijk}\frac{1}{(n-z_i+z_k)(m+z_j-z_k)}-\sum_{n, m=1}^\infty\frac{N}{mn}+
\right.\nonumber\\
&+&\sum_{n=1}^\infty\sum_{ij}\left(\sum_{k>j}\frac{1}{(n-z_i+z_k)(z_j-z_k)}+
\sum_{k<i}\frac{1}{(n+z_j-z_k)(z_k-z_i)}\right)+\nonumber\\
&+&\left.\sum_{ij} \sum_{i>k>j}\frac{1}{(z_k-z_i)(z_j-z_k)}\right].
\label{Vcanc}
\end{eqnarray}
This cancellation was found as the result of formal manipulations with
divergent series like ones in the right hand side of the last formula. Such
transformations are not well defined mathematically and as the result they
may lead to the finite "anomalous" contribution. The famous chiral
anomaly initially was found as the result of careful analysis of
transformations analogous to ones we just performed \cite{adler}. However if
one considers derivatives of $V_{JJ}$ or $V_{\phi\psi}$ with respect to any
$z_i$ then all the sums become convergent, the order of summations becomes
interchangeable and as the result the derivatives of $V_{JJ}+V_{\phi\psi}$
vanish. Thus if there is any anomaly in the expression above it is given by
$z$--independent constant. Such constant in the Hamiltonian would correspond
to the shift of energy levels and usually it is ignored. However in
supersymmetric case there is a natural choice for such constant: in order
for vacuum to be supersymmetric it should be zero. Below we assume that SUSY
is not broken, then we expect that (\ref{Vcanc}) is true.

Thus we arrive at
\begin{equation}
H_0=-\frac{g^2L}{4\pi^2}\frac{1}{K(z)}\sum_i\frac{\partial}{\partial z_i}
K(z)\frac{\partial}{\partial z_i}.
\label{h0}
\end{equation}
The relevant solutions of this equation should be finite in the fundamental
domain, this requirement leads to discrete spectrum due to the fact that
Jacobian vanishes on the boundary of this domain. However the operator $H_0$
is elliptic, and therefore it can't have negative eigenvalues.
If the eigenvalue problem
\begin{equation}
H_0\Phi(z)=E\Phi(z)
\end{equation}
has a solution for $E=0$, this solution corresponds to the ground state
of the theory. It is easy to see that such solution exists and it is given
by $\Phi(z)=const$ \footnote{some authors prefer to rewrite this to
include the measure in the definition of the wave function and then in SU(2)
for example the ground state wave function is a sin}.  We have thus found
that the ground state has a vanishing vacuum energy, suggesting that the
supersymmetry is not broken spontaneously.

\subsection{Supersymmetry and Degenerate Vacua.}

As we saw in the previous subsection supersymmetry leads to the cancellation of
the anomaly terms in current operator. However these terms played an
important role in the description of $Z_N$ degeneracy of vacua \cite{pin97a},
so we should find another explanation of this fact here. It appears that
fermionic zero modes give a natural framework for such treatment.

First we will generalize the supersymmetry transformation given in
\cite{sakai} to the present case, i.e. we include $A^+$ and the zero modes
of fermions. The naive SUSY transformations spoil the
gauge fixing condition, so we combine them with compensating gauge
transformation following \cite{sakai}. In three dimensional notation
(spinors have two components and indices go from 0 to 2) the result reads:
\begin{eqnarray}\label{SUSYtr}
\delta A_\mu=\frac{i}{2} {\bar\varepsilon}\gamma_\mu \Psi-D_\mu
\frac{i}{2} {\bar\varepsilon}\gamma_-\frac{1}{D_-} \tilde{\Psi},\\
\delta\Psi=\frac{1}{4}F_{\mu\nu}\gamma^{\mu\nu}\varepsilon-\frac{g}{2}[
{\bar\varepsilon}\gamma_-\frac{1}{D_-} \tilde{\Psi},\Psi].\nonumber
\end{eqnarray}
The difference between above expression and those in \cite{sakai} is that
we include the zero modes. Namely we defined $\Psi$ as the
complete field with all the zero modes included and $\tilde{\Psi}$ as fermion
without diagonal zero modes. The introducing of $\tilde{\Psi}$ is
necessary, because diagonal zero modes form the kernel of operator $D_-$, so
$\frac{1}{D_-}$ is not defined on this subspace.

In particular we are
interested in supersymmetry transformations for $A^+$ and fermionic zero modes.
Performing a mode expansion one can check that diagonal elements of matrix
$[\frac{1}{D_-}\psi,\psi]^0$ vanish, then from (\ref{SUSYtr}) we get:
\begin{eqnarray}\label{zmSUSY}
\delta A^+_{ii}=\frac{i}{\sqrt{2}} {\varepsilon_+^T}\stackrel{0}{\psi}_{ii},
\nonumber\\
\delta \stackrel{0}{\psi}_{ii}=-2\partial_+ A^+_{ii} \varepsilon_+.
\end{eqnarray}
This expression is written in two component notation and the decomposition of
spinor $\varepsilon$: $\varepsilon=(\varepsilon_+,\varepsilon_-)^T$ is used.
Note that since
$\bar{\varepsilon}Q=\sqrt{2}(\varepsilon_+Q^-+\varepsilon_-Q^+)$ the fields
involved in transformations (\ref{zmSUSY}) don't contribute to $Q^+$, this
is consistent to the fact that being $x^-$ independent they don't contribute
to $P^+$. The equations (\ref{zmSUSY}) look like supersymmetry transformation
for the quantum mechanical system built from free bosons and free fermions. In
fact as one can see the supercharge $Q^-$ is the sum of supercharge for the
quantum mechanical system and from the QFT without diagonal zero modes:
\begin{equation}\label{Qminus}
Q^-=-2g\int dx^- \mbox{tr}(J^+\frac{1}{D_-}\psi)+4L\mbox{tr}
     (\partial_+ A^+\stackrel{0}{\psi}).
\end{equation}
Calculating $(Q^-)^2$ and writing the momentum conjugate to $A^+$ as
differential operator \footnote{using Schr\"odinger coordinate
representation for quantum mechanical degree of freedom - note that the QFT
term has non-trivial dependence on the quantum mechanical coordinate.} we
reproduce Hamiltonian (\ref{hamil}). Note that
$\psi$ there has all the zero modes in it. The square of another supercharge
\begin{equation}
Q^+=2\int dx^-\mbox{tr}(\psi D_-\phi)
\end{equation}
gives $P^+$ while the anti-commutator of $Q^-$ with $Q^+$ is proportional to
the constraint (\ref{fermConstr}) and thus vanishes.

One can check that although $[\stackrel{0}{\psi}_{ii},H]$ does not vanish,
this commutator annihilates Fock vacuum $|0\rangle$, then it also
annihilates $|\Omega\rangle$. In subsection 1 we mentioned that
$\stackrel{0}{\chi}_{ii}$ decouples from the theory, and therefore it
commutes with
Hamiltonian. Thus acting on the vacuum state $|\Omega\rangle$ by diagonal
elements of either $\stackrel{0}{\psi}$ or $\stackrel{0}{\chi}$ we get states
annihilated by $P^-$ and $P^+$ (the latter statement is obvious since zero
modes commute with momentum). Not all such states however may be considered
as vacua. Although we fixed the gauge in subsection 1, the theory still has
residual symmetry
$P$, corresponding to permutations of the color basis. Physical states are
constructed from operator acting on the physical vacuum $|\Omega \rangle $
and both
the operators and the physical vacuum must be invariant under
$P$. Such objects can always be written as combinations of
traces. The candidates for the vacuum state may have any combination of
$\stackrel{0}{\psi}$ and $\stackrel{0}{\chi}$ inside the trace, here and
below we consider only diagonal components of zero modes. Since
$\stackrel{0}{\chi}$ is not dynamical we have the usual c--number relation
\begin{equation}
\{\stackrel{0}{\chi}_{ii},\stackrel{0}{\chi}_{jj}\}=0
\end{equation}
instead of canonical anti-commutator, so
$\stackrel{0}{\chi}\stackrel{0}{\chi}=0$. From the relations
(\ref{ccr}) one finds:
\begin{equation}
\stackrel{0}{\psi}\stackrel{0}{\psi}=\frac{1}{4L\sqrt{2}}(1-\frac{1}{N}),
\end{equation}
also we have $\stackrel{0}{\chi}\stackrel{0}{\psi}=-
\stackrel{0}{\psi}\stackrel{0}{\chi}$. Using all these relations and the
$SU(N)$ conditions $\mbox{tr}(\stackrel{0}{\psi})=0$ and
$\mbox{tr}(\stackrel{0}{\chi})=0$ we find that the only nontrivial trace
involving
only zero modes is
$\mbox{tr}(\stackrel{0}{\psi}
\stackrel{0}{\chi})$. Then the family of vacua is given by:
\begin{equation}
\left(\mbox{tr}(\stackrel{0}{\psi}\stackrel{0}{\chi})\right)^n|\Omega\rangle,
\qquad 0\le n\le N-1.
\end{equation}
The region for $n$ is determined taking into account the fact that
$\stackrel{0}{\chi}$ is anti-commuting field with $N-1$ independent
components. Thus we explained the $Z_N$ degeneracy of vacua first mentioned
in \cite{wit79}.

In addition to this discrete vacuum degeneracy supersymmetric theories also
have a continuum space of vacua called moduli space. In DLCQ approach the
moduli space is easy to understand. Let us suppose that scalar field $\phi$
developed a VEV. To have a consistent theory this VEV should commute with the
Wilson loop in the compact direction, which in our case happened to be
$\exp(i\int dx^- A^+)$. Since $A^+$ is a general diagonal matrix this leads to
the condition for the VEV: $\langle\phi_{ij}\rangle=w_i\delta_{ij}$. Now we
can make the substitution $\phi\rightarrow \phi+\langle\phi\rangle$ in the
supercharges (\ref{Qminus}) to find the correction in $Q^-$ due to the scalar
VEV:
\be
\delta Q^-=-2ig\mbox{tr}\left(w\int dx^-[\phi\psi]\right).
\ee
We used integration by part and the equation $D_-w=0$. Taking into account
the fermionic constraint (\ref{fermConstr}) we conclude that for any diagonal
$w$:
$\delta Q^-=0$, i.e. we can choose the state with arbitrary VEV
$\langle\phi_{ij}\rangle$ as the new vacuum. This is precisely the moduli
space of the theory: the models constructed starting from different vacua
are not coupled with each other.

\subsection{Solving for Massive Bound States.}

As we saw the zero modes play an important role in the description of the
vacuum. However solving for massive bound states one usually neglect the
zero mode contribution. Does this lead to errors in the mass spectrum? The
answer depends on the problem we are solving. If one is interested in the
spectrum of the theory at the finite value of resolution then zero modes
are important, but as we will show their contribution becomes smaller and
smaller as the resolution goes to infinity, so they may be neglected if one
is interested only in the large $K$ extrapolation.

First let us formulate the DLCQ problem with zero modes precisely. We will
use the Hamiltonian formulation, but the consideration for SDLCQ
formalism is the same. The space of states of the theory is the direct
product of Fock space and quantum mechanical Hilbert space for zero modes:
a general state can be written as
\be
|state\rangle =\Phi (z)\otimes |Fock State\rangle,
\ee
the Hamiltonian has the form:
\be
H=K(z)+V(z,a,a^\dagger,b,b^\dagger).
\ee
Here $K(z)$ is some differential operator, while $V$ is some function of
zero modes $z$ and creation--annihilation operators (see for example
(\ref{hamil})). In general one should solve the bound state problem
$H|\Psi\rangle=E|\Psi\rangle$ in two steps: first one should determine
the effective potential $\tilde{V}$:
\be
V(z,a,a^\dagger,b,b^\dagger)|Fock State\rangle =
\tilde{V}(z)\|Fock State\rangle
\ee
and then solve the Schr\"odinger equation for zero modes:
\be
(K(z)+\tilde{V}(z))\Psi(z)=E\Psi(z).
\ee
However in practice this is hard to carry out. Fortunately, solving the
Schr\"odinger equation is not important for calculating the continuum
limit of mass spectrum. The reason for this is the following.

Studying the large $L$ limit in DLCQ one is usually interested in situation
when the total momentum $P^+=\sum n_i/L$ is kept fixed. Then most of the terms
in $V$ (and thus in $\tilde{V}$) are of order $L^0$, while $K(z)$ scales
like $L$. Assume for a moment that the whole ${\tilde V}$ is of order one,
then one can consider ${\tilde V}$ as perturbation and use the standard
expression for the eigenvalue:
\be
E_i=E_i^{(0)}+\int dz \Psi^\dagger_i (z){\tilde V}(z)\Psi_i(z),
\ee
where $E_i^{(0)}$ and $\Psi_i(z)$ are eigenvalue and eigenfunction of
unperturbed system. To get finite masses in the continuum limit only the ground
state of $K(z)$ should be considered: $i=0$ and $E_0^{(0)}=0$ in the last
expression. Introducing the averaging procedure as
\be
\langle A \rangle =\int dz \Psi^\dagger_0 (z)A(z)\Psi_0(z)
\ee
we find: $E=\langle {\tilde V} \rangle$ and thus the continuum eigenvalues
are just solutions of the $z$--independent equation:
\be
\langle V(a,a^\dagger,b,b^\dagger)\rangle |Fock State\rangle =
E|Fock State\rangle.
\ee

The assumption of $L^0$ scaling for $\tilde{V}$ is not the trivial one.
Namely it is responsible for the difference in the constraint equations in
DLCQ and continuum cases. For example looking at the Hamiltonian (\ref{hamil})
  one can see that $V(z)$ includes a term linear in $L$:
\be
\frac{g^2L}{2}\frac{1}{(z_i-z_j)^2}{\tilde J}^+_{ij}(0){\tilde J}^+_{ji}(0),
\ee
so the assumption being false for $V$ may be satisfied for $\tilde{V}$ only
dynamically. One can make this specific term vanish if instead of DLCQ
constraint $\int dx^- {J}_{ii}(x)=0$ its continuum version
\be\label{/25constraint}
\int dx^- {J}_{ij}(x)=0
\ee
is used. Of course imposing this condition is not enough to make all the terms
in $\tilde{V}$ to be of order $L^0$, but following the usual path in DLCQ
calculations we choose not to impose other conditions explicitly. In our
numerical study we rather perform calculations with Hamiltonian
$\langle V(a,a^\dagger,b,b^\dagger)\rangle$ in the sector satisfying
(\ref{/25constraint}) and then concentrate our attention only on states whose
masses can be extrapolated to finite value. This way we make sure that our
assumption $\tilde{V}\sim 1$ holds and thus the $z$ dependence is not
important.

To summarize, we have shown that zero modes of gauge field and diagonal zero
modes of fermions play an important role in the description of vacuum
structure.
However if studying the bound state problem for the states with nonzero total
momentum $P^+$ one is interested only in the extrapolation to the continuum
limit, the zero mode of $A^+$ can be omitted from the theory. This fact leads
to significant simplifications in the numerical procedure. As soon as $A^+$
is excluded from the theory one also has to exclude the bosonic zero modes
(otherwise the  expression $1/0$ is encountered in the
(\ref{2mode_expansion})). What about the fermionic zero modes? In principle
we can either keep them or disregard them. However in the latter case one
should be very careful: as we will see in the next section such modes play an
important role in the ensuring of supersymmetry.

\section{ \bf Fermionic Zero Modes and Exact Supersymmetry.}
\label{ChExSUSY}
\renewcommand{\theequation}{3.\arabic{equation}}
\setcounter{equation}{0}

In this section we study the relation between conventional DLCQ
and its supersymmetric version. Since usual DLCQ is formulated for the
Hamiltonian we should rewrite SDLCQ in the same form. Here one encounters
the first difference between two schemes: in DLCQ the fermions can be chosen
to be either periodic or antiperiodic on $x^-$, but in the Hamiltonian
formulation of SDLCQ they must be periodic due to supersymmetry. Then one
encounters the problem of fermionic zero modes. However the boundary
conditions
is not the only difference between the two approaches. Even after we choose
periodic fermions, DLCQ still has an ambiguity emerging from the choice of
regularization scheme. Taking the simplest SUSY system as an example we will
show that supersymmetry dictates the unique regularization and we study the
relation between this prescription and the principal value scheme, which is
usually used in the DLCQ calculations. We show that fermionic zero modes play
an important role in deriving this relation.

As we already mentioned in section one the simplest
supersymmetric system in two dimension is the one involving
only gauge fields and adjoint fermions \cite{kut93}. We derive all the
relations for this particular system.

%%%%%%%%%%%%%%%%%%%%%%%%%%%%%%%%%%%%%%%%%%%%%%%%%%%%%%
%%%%%%%%%%%%%%%%%%%%%%%%%%%%%%%%%%%%%%%%%%%%%%%%%%%%%%
\subsection{Zero Modes and Supersymmetric Regularization.}

We consider the $1+1$ dimensional SU($N$) gauge theory
coupled to an adjoint
Majorana fermion.
The light-cone quantization of this model in the light-cone
gauge and large $N$ limit has been dealt with explicitly
before \cite{dak93,bdk93}.
The expressions for the light-cone momentum $P^+$
and light-cone Hamiltonian $P^-$ for this model are
\begin{eqnarray}
\label{momenta}
P^+ & = & \int dx^-
\mbox{tr}(i\sqrt{2}\psi\partial_-\psi),\\
P^- & = & \int
dx^- \mbox{tr}\left(-\frac{im^2}{\sqrt{2}}\psi\frac{1}{\partial_-}\psi-
\frac{g^2}{2}J^+\frac{1}{\partial_-^2}J^+\right).
\end{eqnarray}
Here
$J^+_{ij}=-\sqrt{2}\psi_{ik}\psi_{kj}$ is the longitudinal current.
It is well
known that at a special value of fermionic mass (namely
$m^2=g^2N/\pi$)
this system is supersymmetric \cite{kut93}.
This special value of the fermion mass will be denoted by $m_{SUSY}$.
At this supersymmetric point, the supercharge is given by
\begin{equation}
\label{sucharge}
Q^-=\sqrt{2}g\int dx^-\mbox{tr}(\psi\psi\frac{1}{\partial_-}\psi)
\end{equation}
which satisfies the supersymmetry relation $\{Q^-,Q^-\}=2\sqrt{2}P^-$.
This may
be checked explicitly by using the anticommutator at equal $x^+$:
\begin{equation}
\{\psi_{ij}(x^-),\psi_{kl}(y^-)\}=
\frac{1}{2}\delta_{il}\delta_{jk}\delta(x^- -y^-).
\end{equation}
In the DLCQ formulation, the theory is regularized  by
a light-like compactification,
and either periodic or antiperiodic boundary conditions
may be imposed for fermions.
If $P^+$ denotes the total light-cone momentum, light-like
compactification is equivalent to restricting
the light-cone momentum of partons to be non-negative
integer multiples of $P^+/K$, where $K$ is some positive
integer that is sent to infinity in the decompactified limit\footnote{
$K$ is sometimes called the `harmonic resolution', or just `resolution'.}.
Anti-periodic boundary conditions will in general explicitly break the
supersymmetry in the discretized theory,
although supersymmetry will be restored in the
decompactification limit $K\rightarrow
\infty$ \cite{bdk93}.
  If we wish to maintain supersymmetry
at any finite $K$, we must at least impose
periodic boundary conditions for the fermions.
This, however, leads to the notorious ``zero-mode problem''\footnote{
  For anti-periodic boundary conditions,
the light-cone momentum of partons is restricted to
{\em odd} integer multiples of
$P^+/K$, and so there are no zero-momentum modes in such a formulation.}.
 From a numerical perspective, omitting zero-momentum modes
in our analysis is absolutely necessary, since
it guarantees a {\em finite} Fock basis for each finite resolution $K$.
The mass spectrum of the continuum theory may be obtain
by extrapolating from a sequence of finite mass matrices $M^2=2P^+P^-$.
But are we really justified in omitting the zero-momentum modes?
To date, the general consensus is that omitting
zero momentum modes in a two dimensional interacting field
theory does not affect the spectrum of the decompactified
theory, where
$K \rightarrow \infty$.
Actually, the numerical results of the next subsection are
consistent with this viewpoint.

However, the goal of this work is to understand
the structure of a supersymmetric theory at finite resolution.
As we will see shortly, understanding why the DLCQ
and SDLCQ prescriptions differ involves studying certain
intermediate zero-momentum processes. But first,
we need to be more precise about the form of the
light-cone operators of the theory.
If we expand the fermion field $\psi_{ij}$ in terms of its
Fourier components, we may express the uncompactified
light-cone supercharge
and Hamiltonian in a momentum space representation
involving fermion creation and annihilation operators:
(\cite{kut93,dak93,bdk93}):
\begin{eqnarray}
\label{qminus1}
&&Q^-=\frac{i2^{-1/4}g}{\sqrt{\pi}}\int_0^\infty dk_1dk_2dk_3
\delta(k_1+k_2-k_3)
\left(\frac{1}{k_1}+\frac{1}{k_2}-\frac{1}{k_3}\right)\times\nonumber\\
&&\times\left(b^\dagger_{ik}(k_1)b^\dagger_{kj}(k_2)b_{ij}(k_3)+
b^\dagger_{ij}(k_3)b_{ik}(k_1)b_{kj}(k_2)\right),\nonumber\\
&&  \\
\label{pminus}
&&P^-=\frac{m^2}{2}\int_0^\infty\frac{dk}{k}b^\dagger_{ij}(k)b_{ij}(k)+
\frac{g^2N}{\pi}\int_0^\infty\frac{dk}{k}\int_0^k dp \frac{k}{(p-k)^2}
b^\dagger_{ij}(k)b_{ij}(k)+\nonumber\\
&&{+}\frac{g^2}{2\pi}\int_0^\infty
dk_1dk_2dk_3dk_4\left[ \frac{}{}\delta(k_1{+}k_2{-}k_3{-}k_4)
A(k)b^\dagger_{kj}(k_3)b^\dagger_{ji}(k_4)b_{kl}(k_1)b_{li}(k_2)+\right.
\nonumber\\
&&+\delta(k_1{+}k_2{+}k_3{-}k_4)B(k)\times\nonumber\\
&&\qquad\times\left.\left(
b^\dagger_{kj}(k_4)b_{kl}(k_1)b_{li}(k_2)b_{ij}(k_4)-
b^\dagger_{kj}(k_1)b^\dagger_{jl}(k_2)b^\dagger_{li}(k_3)b_{ki}(k_4)
\right)\frac{}{}\right]\nonumber
\\
& &
\end{eqnarray}
with
\begin{eqnarray}
A(k)=\frac{1}{(k_4-k_2)^2}-\frac{1}{(k_1+k_2)^2},\nonumber\\
B(k)=\frac{1}{(k_3+k_2)^2}-\frac{1}{(k_1+k_2)^2}.
\end{eqnarray}
As we mentioned earlier,
the continuum theory is supersymmetric
for a special value of fermion mass. We
will therefore consider only the case $m=m_{SUSY}$.
In the DLCQ formulation, one simply
restricts integration of the light-cone momenta $k_i$
in expression (\ref{pminus}) for $P^-$
above to be {\em positive} integer multiples of $P^+/K$.
i.e. one simply drops the zero-momentum mode.
The DLCQ mass spectrum is then obtained by diagonalizing
the mass operator $M^2=2P^+P^-$.
Similarly, in the SDLCQ formulation, the
light-cone momenta $k_i$
in expression (\ref{qminus1}) for $Q^-$
are restricted to {\em positive} integer multiples of $P^+/K$.
One then simply {\em defines} $P^-$ to be the square
of the supercharge: $2 \sqrt{2} P^- = \{Q^-,Q^- \}$.
The mass operator $M^2=2P^+P^-$ is then easily constructed
and diagonalized to obtain the
SDLCQ spectrum.

In general, the following observations are made; at finite
resolution, the DLCQ spectrum of a supersymmetric theory
is not supersymmetric. However, supersymmetry is restored
after extrapolating to the continuum limit $K \rightarrow \infty$
(see \cite{bdk93}, for example). In contrast, for any
finite resolution, the SDLCQ spectrum is supersymmetric.
The DLCQ and SDLCQ spectra agree only in the decompactified
limit $K \rightarrow \infty$.

Not surprisingly, the difference in the
DLCQ and SDLCQ prescriptions at finite resolution
may be understood as a zero-mode contribution.
What is surprising is that we can
encode the effect of these zero-mode contributions into
a simple well defined operator.
The main result here is the
the precise operator form of this contribution at finite $K$.

In order to motivate our argument, note that the
anticommutator  for the
supercharge $Q^-$ in the continuum theory involves
products of terms of the form
$b^\dagger(k)b^\dagger(0)b(k)$ and $b^\dagger(p)b(0)b(p)$, and these
provide contributions to $P^-$ that may be expressed in terms
of non-zero momentum modes.
The problem is
exacerbated by the fact that the coefficients of these terms behave
singularly. To
examine this more closely, we consider
the discretized theory
where the light-cone momenta $k_i$ in the expression
for $Q^-$ [eqn(\ref{qminus1})] are restricted to positive integer
multiples of $P^+/K$. We also include the
effects of zero-momentum modes by introducing
an `$\epsilon$ regulated zero mode', which are modes
with momentum $k_i = \epsilon$, where $\epsilon$ is
much less than $P^+/K$, and is sent to zero at the end of
the calculation.
Then the anticommutator of two $Q^-$
gives contributions
  of the following form:
\begin{eqnarray}
\left\{(\frac{1}{\epsilon}+\frac{\epsilon}{k(k+\epsilon)})
b^\dagger(k)b^\dagger(\epsilon)b(k+\epsilon),
(\frac{1}{\epsilon}+\frac{\epsilon}{p(p+\epsilon)})
b^\dagger(p+\epsilon)b(\epsilon)b(p)\right\}=\nonumber\\
=b^\dagger(k)b(k{+}\epsilon)b^\dagger(p{+}\epsilon)b(p)
\left[\frac{1}{\epsilon^2}+(\frac{1}{p(p{+}\epsilon)}+
\frac{1}{k(k{+}\epsilon)})
+\frac{\epsilon^2}{pk(p{+}\epsilon)(k{+}\epsilon)}\right],\nonumber\\
\
\end{eqnarray}
where any terms involving
an $\epsilon$ regularized zero mode on the right-hand-side are
dropped and zero modes are omitted from $P^-$.
We have suppressed all matrix indices in this expression. In the limit
$\epsilon\rightarrow 0$ the last term on the right-hand-side
in the brackets vanishes,
while the first term
is the pure momentum--independent divergence
that was identified in an earlier study of this model
\cite{bdk93}, and is canceled if we adopt
a principal value prescription for singular amplitudes
in the definition of $P^-$. The second term however,
is clearly a finite contribution to $P^-$,
although it arises from the $\epsilon$ regulated zero modes
in $Q^-$, which are not present in the SDLCQ prescription
for defining $Q^-$. Consequently, in order to ensure the
supersymmetry relation
$\{Q^-,Q^- \} = 2 \sqrt{2} P^-$
in the discretized formulation,
we must include an $\epsilon$ regularization of the
zero modes in the definition for $Q^-$,  and
then apply a principal value prescription in the presence
of any singular processes to eliminate $1/\epsilon$ divergences.

Stated slightly differently, we may decompose the supercharge into a part
without zero
modes $Q^-_{SDLCQ}$ (i.e. $k_i = nP^+/K, n=1,2,\dots$),
and terms with $\epsilon$ regularized zero modes,
$Q^-_\epsilon$.
The anti-commutator $\{Q^-_{SDLCQ},Q^-_{\epsilon}\}$
contains only terms with $\epsilon$
regulated zero-modes. Since
$Q^-=Q^-_{SDLCQ}+Q^-_{\epsilon}$ one finds
\begin{equation}
\label{twoQ}
\{Q^-_{SDLCQ},Q^-_{SDLCQ}\}=
2\sqrt{2}P^-_{SDLCQ}= 2\sqrt{2}P^-_{DLCQ}-\{Q^-_{\epsilon},
Q^-_{\epsilon}\}_{PV},
\end{equation}
after dropping any $\epsilon$ regulated zero-mode terms
in the calculated expression for $\{Q^-,Q^-\}$.
Note that the first equality above is just the definition
for the light-cone Hamiltonian $P^-$ in the SDLCQ prescription.
The $PV$ abbreviation on the right hand side indicates a principal
value
regularization prescription,  which is tantamount to
dropping all $1/\epsilon$ terms as $\epsilon\rightarrow 0$. The
procedure is well known in the context of the present model
\cite{bdk93}. It is clear that our definition for $P^-_{SDLCQ}$
gives rise to the supersymmetry relation $[Q^-_{SDLCQ},P^-_{SDLCQ}]=0$,
which yields a supersymmetric spectrum for any finite resolution
$K$. Moreover, we know that
$P^-_{SDLCQ}$ and $P^-_{DLCQ}$ yield the same spectrum
in the continuum limit $K\rightarrow \infty$,
so it remains to calculate the difference at finite resolution $K$.
  We will write this
difference in terms of their respective
mass operators: $M^2=2P^+P^-$. A straightforward
calculation of the
anticommutator on the right-hand-side of (\ref{twoQ}) leads to the result:
\begin{eqnarray}
\label{Mdiff}
&&M^2_{SDLCQ}-M^2_{DLCQ}=M^2_{\Delta}=-\frac{g^2NK}{\pi}\sum_n \frac{1}{n^2}
B^\dagger_{ij}(n)B_{ij}(n)\nonumber\\
&& -\frac{g^2NK}{\pi}\sum_{mn}
(\frac{1}{m^2}+\frac{1}{n^2})\frac{1}{N}
B^\dagger_{kj}(m)B^\dagger_{ji}(n)B_{kl}(m)B_{li}(n).
\end{eqnarray}
We also write down
the expression for $M^2_{DLCQ}$ in the theory with periodic fermions:
\begin{eqnarray}
&&M^2_{DLCQ}=\frac{g^2NK}{\pi}\sum_n B^\dagger_{ij}(n)B_{ij}(n)(\frac{x}{n}+
\sum_m^{n-1}\frac{2}{(n-m)^2})+\nonumber\\
&&\frac{g^2K}{\pi}{\sum_{n_i}}
'\left\{
\delta_{n_1+n_2}^{n_3+n_4}
\left[\frac{1}{(n_2{-}n_4)^2}-\frac{1}{(n_1{+}n_2)^2}\right]
B^\dagger_{kj}(n_3)B^\dagger_{ji}(n_4)B_{kl}(n_1)B_{li}(n_2) \right.\nonumber\\
&&+\delta_{n_1+n_2+n_3,n_4}
\left[\frac{1}{(n_2+n_3)^2}-\frac{1}{(n_1+n_2)^2}\right]\times\\
&&\left.\left(B^\dagger_{kj}(n_4)B_{kl}(n_1)B_{li}(n_2)B_{ij}(n_3)-
B^\dagger_{kj}(n_1)B^\dagger_{jl}(n_2)B^\dagger_{li}(n_3)B_{ki}(n_4)\right)
\frac{}{}\right\}.\nonumber
\end{eqnarray}
In this expression the variable $x=\frac{\pi m^2}{g^2N}$ is a dimensionless
mass parameter, and for
the supersymmetric point we have $x=1$. The sums are performed over positive
integers, $0<n_i<K$, and we employ a principal value prescription
in sums labeled as $\sum'$, which implies that terms of the form
$1/(k-k)^2$ are dropped. In the SDLCQ procedure we calculate $Q^-$ which is
non-singular
and requires no principal value prescription.

The term $M^2_\Delta$ appears to be non-trivial
due to the presence of
$B^\dagger B^\dagger BB$ terms on the right hand side of
(\ref{Mdiff}).
However, the action of this term on any
SU($N$) Fock state turns out to be equivalent
to the first term, although with opposite sign, and twice the
magnitude.
Thus the
action of the right hand side of
(\ref{Mdiff}) is equivalent to the single quadratic operator:
\begin{equation}
  M^2_{\Delta}=\frac{g^2NK}{\pi}\sum_n \frac{1}{n^2}
B^\dagger_{ij}(n)B_{ij}(n).
\label{olegterm}
\end{equation}
Fortunately, we are able to test this analytical result by performing
direct numerical simulations of this model using both prescriptions,
and comparing the differences observed with the above prediction.
Interestingly,
although this result was derived for large $N$,
agreement turns out to be perfect for both finite and large $N$,
which was verified using the finite $N$ DLCQ
algorithms developed in \cite{anp98}.
We discuss this further in the next subsection.

%%%%%%%%%%%%%%%%%%%%%%%%%%%%%%%%%%%%%%%%%%%%%%%%%%%%%%%%%%%%%
%%%%%%%%%%%%%%%%%%%%%%%%%%%%%%%%%%%%%%%%%%%%%%%%%%%%%%%%%%%%%
\subsection{Soft SUSY Breaking and Numerical Results.}

In this subsection we compare the numerical results for different
regularization
schemes. Although in the continuum limit both PV and SUSY prescription
ns
should give the same results, the convergence
of the masses as $K\rightarrow\infty$ might be different. So if at a given
value of $K$ one wants to get a better approximation to continuum masses, one
scheme might work better than other. In previous subsection we described two
regularization schemes and found the operator $M^2_{\Delta}$ describing the
difference between them. It is convenient to introduce the family of
regularizations labeled by parameter $Y$:
\begin{equation}
P^-_Y=P^-_{PV}+YM^2_{\Delta}.
\end{equation}
Then at two special values of $Y$ we get the PV and SUSY prescriptions:
$P^-_{PV}=P^-_{Y=0}$, $P^-_{SUSY}=P^-_{Y=1}$. Since $P^-_{PV}$ is defined
for arbitrary value of fermionic mass $m$ (not only for supersymmetric one)
the last equation also defines the family of regularizations beyond the SUSY
point. On the other hand shifting the fermionic mass from its supersymmetric
value is equivalent to introducing additional fermionic mass, i.e. to the soft
SUSY breaking. Below we will give a numerical results for bound state masses
in the theory with two new "coupling constants": $X=\frac{\pi m^2}{g^2N}$ and
$Y$ which determine the value of fermionic mass and regularization scheme
accordingly.

Our investigation of this theory indicates that at $X=1$ ( the
supersymmetric value of the fermion mass) the lightest
fermionic and bosonic bound states are degenerate with
continuum masses approximately $M^2= 26$ \cite{bdk93,alp2}.
Using $P^-_{SUSY}$ we arrive at the same conclusion
for any value of $Y$.

Boorstein and Kutasov \cite{kub94} have investigated `soft' supersymmetry
breaking for small values of
this difference,
$X-1$ and they found that the degeneracy between the fermion and boson
bound state masses is broken
according to
\begin{equation} M^2_F(X) - M^2_B(X)= (1-X) M_B(1)+O((X-1)^3).
\label{linear}
\end{equation}

They calculated these masses using the PV prescription ($Y=0$) with
anti-periodic BC and found
very good agreement with the theoretical prediction.  We have compared this
theoretical prediction
at $Y=1$ and we  find
  that eq (\ref{linear}) is  very well satisfied. At resolution $K=5$, for
example, the slope is 4.76
and the predicted slope
$M_B(1)$ is 4.76. The indication is that this result is true for any value
of $Y$.
%%%%%%%%%%%%%%%%%%%%%%%%%%%%%
\begin{figure}[h]
\begin{center}
%\epsfscale=1400
\epsfxsize=5.5in
\epsfbox{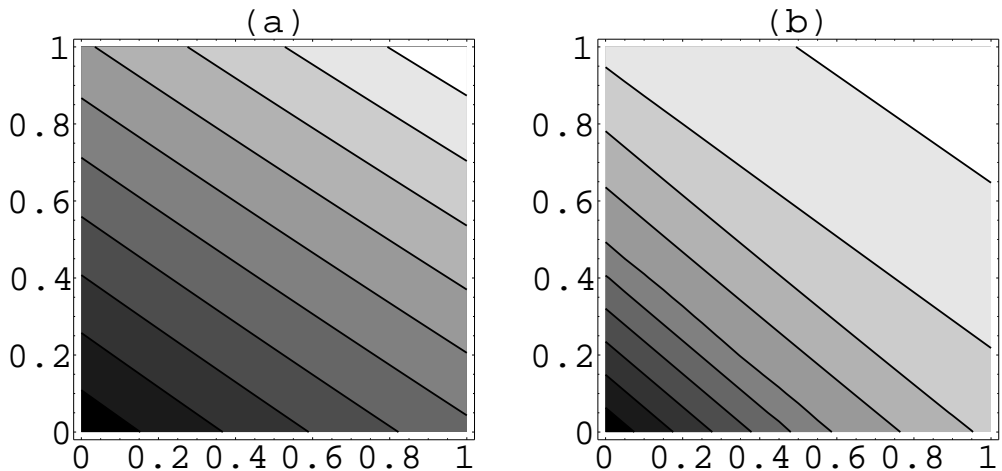}
%!!!!!!!!!!!!!!!!!!!!>>>>>>>>>>>>>
%\epsfig{file=twoplots.eps, width=14cm}
%<<<<<<<<<<<<<<!!!!!!!!!!!!!!!!!!!!!!
\end{center}
\caption{(a) The contour plots of
$Y=Y(X)$ for the mass squared of the lowest bound state in units of $g^2 N/
\pi$ as a function of
$X=m \pi /g^2 N$ and Y (b)The contour plots of
$Y=Y(X)$ for the mass squared of the second lowest bound state in units of
$g^2 N/
\pi$ as a function of
$X=m \pi /g^2 N$ and Y (b) }
\end{figure}
%%%%%%%%%%%%%%%%%%%%%%%%%%%%

In Fig. 1 we show the contour plots of the mass squared $M^2$
of the two lightest bosonic bound
states as a function of $X$
and $Y$ at resolution $K=10$. These contours
are lines of constant mass squared. Selecting a particular value of the
mass of the first bound
state then fixes a particular contour in Fig. 1a as a contour
of fixed mass, which we can write as
$Y=Y_p(X)$.

Interestingly, constructing the same contour plot
for the next to lightest bosonic bound state -- see
Fig. 1b -- yields contours that have approximately
the same functional dependence implied by Fig. 1a.
In fact, one obtains approximately the same contour
plots for the next twenty bound states (which is as far as we checked).
The simple conclusion is that the coupling $Y$ which
represents the strength of the additional operator
affects all bound state masses more or less equally.
This in turn suggests that at finite resolution, we can
smoothly interpolate between different values of fermion
mass $X$, and different prescriptions specified by the
coupling $Y$, without affecting too much the actual numerical spectrum.
Of course, in the decompactification limit $K \rightarrow \infty$,
such a dependence on $Y$ disappears, due to scheme independence.

Since the lightest bosonic bound state is primarily a
two particle state it is reasonable to truncate the
Fock basis to  two
particle states. This will permit
very high resolutions, which will be needed
to carefully scrutinize any possible discrepancies between
the two versions of
'soft' symmetry breaking presented here.
In fact, we are able to study the theory
for $K$ up to 800. The
mass of the lowest state as a function of the resolution for various
values of $X$ and $Y$ are
shown in Fig. 2. Each converging pair
of lines -- which extrapolate the actual data points -- in Fig. 2
corresponds to different
values of fermion mass $X$. The top
upper curve in each pair runs through data points
that were calculated via SDLCQ (i.e. $Y=1$), while the
lower corresponds to the PV (i.e. $Y=0$) prescription
commonly adopted in the literature. We
find that each pair of
curves converge to the same point at infinite resolution,
although this may not be completely obvious
for the lowest pair in the
figure (corresponding to the critical mass $X=0$).

Away from
$X=0$, the SDLCQ formulation is fitted with a linear function of $1/K$,
while the PV formulation is fit
with a polynomial of
$1/K^{2\beta}$, where $\beta$ is the solution of $1-X/2=\pi \beta
Cot(\pi
\beta)$ \cite{van96}. It
now appears that SDLCQ not only provides more rapid convergence
for supersymmetric models, but
also for the
massive t'Hooft model, which is not supersymmetric.
For the massless case, the situation is reversed;
the SDLCQ formulation
converges slower. It is fit by a polynomial in $1/Log(K)$ and gives the
same mass at infinite
resolution as the PV formulation. This behavior
may be understood from the observation that
the wave function of this
state does not vanish at
$x=0$. We have looked closely at `small' masses,
such as $X=.1$, and one finds that both PV and SDLCQ
vary as a polynomial
in $1/K^{2\beta}$ at large resolution. Thus careful extrapolation
schemes must be adopted at small masses.

We therefore conclude that the continuum of
regularization schemes that interpolate smoothly between
the SDLCQ and PV prescriptions -- which we characterized
by the parameter $Y$ -- yield the same continuum
bound state masses, although the rate of
convergence of the DLCQ spectrum may be altered significantly.
This implies that
the contour plots observed in Fig. 1 eventually
approach lines parallel to the $Y$ axis, and the
sole dependence on the parameter $X$ is recovered.

%%%%%%%%%%%%%%%%%%%%%%%%%%%%%
\begin{figure}[h]
\begin{center}
%\epsfscale=1200
\epsfbox{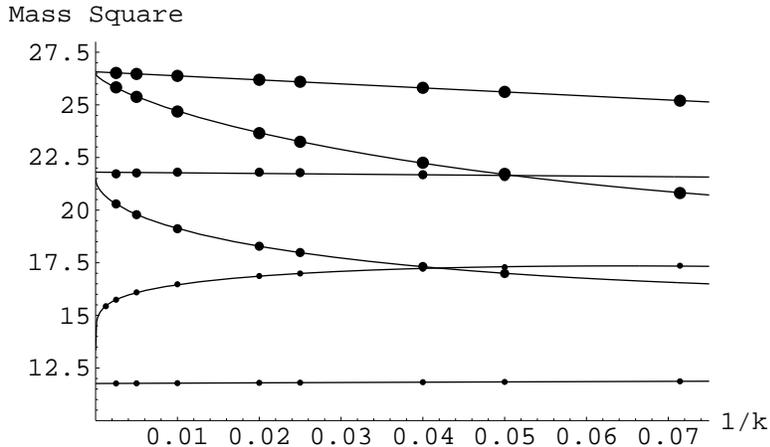}
%!!!!!!!!!!!!!!!!!!!!!!!>>>>>>>>>>>>>>>>>
%\epsfig{file=convergenceplot.eps}
%<<<<<<<<<<<<<<<<<!!!!!!!!!!!!!!!!!!!!!
\end{center}
\caption{Mass of the of the lowest bound state in units of $g^2 N/ \pi$
calculated in the t'Hooft
model. The top pair is at $X=1$, the second is at $X=.5$,
and the bottom pair is at $X=0$ }
\end{figure}
%%%%%%%%%%%%%%%%%%%%%%%%%%%%

Interestingly, since the
two-body equation studied here
for the adjoint fermion model
is simply the t'Hooft equation with
a rescaling of coupling constant,
we have arrived at an
alternative prescription
for regulating the Coulomb singularity
in the massive t'Hooft model
that improves the rate of convergence towards
the actual continuum mass.
Thus, a prescription that
arises naturally in the study of supersymmetric
theories is also applicable in the study of
a theory without supersymmetry.
We believe that this idea deserves to be exploited
further in a wider context of theories.
In particular, it is an open
question whether this procedure could provide a sensible approach to
regularizing softly broken gauge theories with bosonic degrees
of freedom, and in higher dimensions.

In any case, it appears that the special cancellations
afforded by supersymmetry -- especially in
the context of DLCQ bound state calculations --
might have scope beyond the domain
of supersymmetric field theory. This would be a crucial first step
towards a serious non-perturbative study of
theories with broken supersymmetry.

\section{\bf Massless States in Two Dimensional Models.}
\label{ChMsls}
\renewcommand{\theequation}{4.\arabic{equation}}
\setcounter{equation}{0}

In this section we will study the structure of bound states for two
dimensional supersymmetric models defined in section 1. We will
concentrate most of the attention on the model obtained by dimensional
reduction from SYM$_{2+1}$. For this theory we will prove that any
normalizable bound state in the continuum must include a contribution
with arbitrarily large number of partons. By generalizing this proof to the
theories with extended SUSY we show that this is the general property of
supersymmetric matrix models.
This scenario is to be contrasted with the simple bound states discovered in
a number of $1+1$ dimensional theories with complex fermions, such as the
Schwinger model, the t'Hooft model, and a dimensionally
reduced theory with complex adjoint fermions
\cite{anp97,pin97}.
We also study the massless states of SYM$_{2+1}$ in DLCQ. Some of them are
constructed explicitly and the general formula for the number of massless
states as function of harmonic resolution is derived for the large $N$ case.
This section is based in part on the results of \cite{alp98a}.
%%%%%%%%%%%%%%%%%%%%%%%%%%%%%%%%%%%%%%%%%%%%%%%%%%%%%%%%%%

\subsection{ Formulation of the bound state problem.}

The light-cone formulation of the
supersymmetric matrix model obtained by dimensionally
reducing ${\cal N} = 1$ $\mbox{SYM}_{2+1}$ to $1+1$ dimensions
was initially given in \cite{sakai}, and it was summarized in
the section 2 of these lectures. We simply note here that
the light-cone Hamiltonian $P^-$ is given in terms of the
supercharge $Q^-$ via the supersymmetry
relation $\{Q^-,Q^-\} = 2 \sqrt{2} P^-$, where
\begin{equation}
     Q^-  =  2^{3/4} g \int dx^- \mbox{tr} \left\{
          ({\rm i}[\phi,\partial_- \phi ] + 2 \psi \psi ) \frac{1}{
     \partial_-} \psi \right\}. \label{qminus2}
\end{equation}
In the above, $\phi_{ij} = \phi_{ij}(x^+,x^-)$ and
$\psi_{ij} = \psi_{ij}(x^+,x^-)$
are $N \times N$ Hermitian matrix fields representing the physical
boson and fermion degrees of freedom (respectively) of the theory,
and are remnants of the physical transverse degrees of freedom
of the original $2+1$ dimensional theory.
This is a special feature of light-cone quantization in light-cone
gauge: all unphysical degrees of freedom present in the original
Lagrangian may be explicitly eliminated. There are no ghosts.

In order to quantize $\phi$ and $\psi$ on the light-cone, we
first introduce the following
expansions at fixed light-cone time $x^{+}=0$ (the continuum counterpart of
(\ref{2mode_expansion}):
\begin{eqnarray}
\phi_{ij}(x^-,0)=\frac{1}{\sqrt{2\pi}}\int_0^{\infty}
\frac{dk^+}{\sqrt{2k^+}}\left(a_{ij}(k^+) e^{-ik^+ x^-}+a^\dagger_{ji}(k^+)
e^{ik^+ x^-}\right); \label{phiexp}\\
\psi_{ij}(x^-,0)=\frac{1}{2\sqrt{\pi}}\int_0^{\infty}
dk^+ \left(b_{ij}(k^+) e^{-ik^+ x^-}+b^\dagger_{ji}(k^+)
e^{ik^+ x^-}\right). \label{psiexp}
\end{eqnarray}
We then specify the commutation relations
\begin{equation}
\left[a_{ij}(p^+ ),a^\dagger_{lk}(q^+)\right]=\left\{ b_{ij}(p^+ ),
b^\dagger_{lk}(q^+)\right\}=\delta (p^+ -q^+)\delta_{il}\delta_{jk}
\end{equation}
for the gauge group U($N$), or
\begin{equation}
\left[a_{ij}(p^+ ),a^\dagger_{lk}(q^+)\right]=\left\{ b_{ij}(p^+ ),
b^\dagger_{lk}(q^+)\right\}=\delta (p^+ -q^+)\left(\delta_{il}\delta_{jk}-
\frac{1}{N}\delta_{ij}\delta_{kl}\right)
\end{equation}
for the gauge group SU($N$)\footnote{We assume the normalization
${\mbox tr}[T^a T^b] = \delta^{ab}$, where the $T^a$'s are the
generators of the Lie algebra of SU($N$).}.

For the bound state eigen-problem
$2P^+ P^- |\Psi> = M^2 |\Psi>$, we may restrict to the
subspace of states with fixed light-cone momentum $P^+$,
on which $P^+$ is diagonal, and so the bound state problem is
reduced to the diagonalization
of the light-cone Hamiltonian $P^-$.
Since $P^-$ is proportional to the square of the supercharge $Q^-$,
any eigenstate $|\Psi>$ of $P^-$ with mass squared $M^2$ gives
rise to a natural four-fold degeneracy in the spectrum because
of the supersymmetry algebra---all four states below have the same mass:
\begin{equation}
     |\Psi>, \hspace{4mm} Q^+ |\Psi>,\hspace{4mm}  Q^- |\Psi>,
\hspace{4mm}  Q^+ Q^- |\Psi>.
\end{equation}
Although this four-fold degeneracy is realized in the continuum
formulation of the theory, this property will not
necessarily survive if we choose to discretize the
theory in an arbitrary manner. However, a nice
feature of SDLCQ is that it does
preserve the supersymmetry (and hence the {\it exact} four-fold
degeneracy) for any resolution.

Focusing attention on zero mass eigenstates, we simply note
that a massless eigenstate of $P^-$ must also be annihilated by the
supercharge $Q^-$, since $P^-$ is proportional to $(Q^-)^2$. Thus
the relevant eigen-equation is $Q^- |\Psi> = 0$. We wish to study
this equation.
However, first we need to state the explicit equation for $Q^-$,
in the momentum representation,
which is obtained by substituting
the quantized field expressions (\ref{phiexp}) and (\ref{psiexp})
directly into the the definition of the supercharge (\ref{qminus2}).
The result is:
\begin{eqnarray}
\label{Qminus3}
Q^-&=& {{\rm i} 2^{-1/4} g \over \sqrt{\pi}}\int_0^\infty dk_1dk_2dk_3
\delta(k_1+k_2-k_3) \left\{ \frac{}{} \right.\nonumber\\
&&{1 \over 2\sqrt{k_1 k_2}} {k_2-k_1 \over k_3}
[a_{ik}^\dagger(k_1) a_{kj}^\dagger(k_2) b_{ij}(k_3)
-b_{ij}^\dagger(k_3)a_{ik}(k_1) a_{kj}(k_2) ]\nonumber\\
&&{1 \over 2\sqrt{k_1 k_3}} {k_1+k_3 \over k_2}
[a_{ik}^\dagger(k_3) a_{kj}(k_1) b_{ij}(k_2)
-a_{ik}^\dagger(k_1) b_{kj}^\dagger(k_2)a_{ij}(k_3) ]\nonumber\\
&&{1 \over 2\sqrt{k_2 k_3}} {k_2+k_3 \over k_1}
[b_{ik}^\dagger(k_1) a_{kj}^\dagger(k_2) a_{ij}(k_3)
-a_{ij}^\dagger(k_3)b_{ik}(k_1) a_{kj}(k_2) ]\nonumber\\
&& ({ 1\over k_1}+{1 \over k_2}-{1\over k_3})
[b_{ik}^\dagger(k_1) b_{kj}^\dagger(k_2) b_{ij}(k_3)
+b_{ij}^\dagger(k_3) b_{ik}(k_1) b_{kj}(k_2)]  \left. \frac{}{}\right\}.
\nonumber\\ \
\end{eqnarray}

In order to implement the DLCQ formulation \cite{BP85,MaYam}
of the theory, we simply restrict the
momenta $k_1,k_2$ and $k_3$ appearing in the above equation to
the following set of allowed momenta: $\{\frac{P^+}{K},\frac{2P^+}{K},
\frac{3P^+}{K},\dots \}$. Here, $K$ is some arbitrary positive integer,
and must be sent to infinity if we wish to recover the continuum
formulation of the theory. The integer $K$
is called the {\em harmonic resolution},
and $1/K$ measures the coarseness of our discretization.
Physically, $1/K$ represents the smallest unit of longitudinal
momentum fraction allowed for each parton.
As soon as we implement the DLCQ procedure, which is
specified unambiguously
by the harmonic resolution $K$, the integrals appearing
in the definition of $Q^-$ are replaced by finite sums,
and the eigen-equation $Q^- |\Psi> = 0$ is reduced to a finite matrix
problem. For sufficiently small values of $K$ (in this case
for $K \leq 4$)
this eigen-problem may be solved analytically.
For values $K > 5$, we may still compute
the DLCQ supercharge analytically as a function of $N$,
but the diagonalization procedure must be performed
numerically.

For now, we concentrate on the structure of the zero mass eigenstates
for the continuum theory. Firstly, note
that for the U($N$) bound state problem, massless states
appear automatically because of the decoupling of the U($1$)
and SU($N$) degrees of freedom that constitute U($N$). More explicitly,
we may introduce the U(1) operators
\begin{equation}
          \alpha (k^+)  =  \frac{1}{N}\mbox{tr} [a(k^+)]
\hspace{4mm} \mbox{and}
    \hspace{4mm} \beta (k^+)  =  \frac{1}{N}\mbox{tr} [b(k^+)],
\end{equation}
which allow us to decompose any U($N$) operator into a sum of
U(1) and SU($N$) operators:
\begin{equation}
             a(k^+) = \alpha (k^+)\cdot \mbox{${\bf 1}_{N\times N}$} +
{\tilde a}(k^+)
  \hspace{4mm} \mbox{and} \hspace{4mm}
       b(k^+) = \beta (k^+)\cdot \mbox{${\bf 1}_{N\times N}$}
  + {\tilde b}(k^+),
\end{equation}
where ${\tilde a}(k^+)$ and ${\tilde b}(k^+)$ are traceless $N \times N$
matrices. If we now substitute the operators above into the
expression for the supercharge (\ref{Qminus3}),
we find that
all terms involving the U(1) factors $\alpha(k^+), \beta(k^+)$ vanish -- only
the SU($N$) operators ${\tilde a}(k^+),{\tilde b}(k^+)$ survive.
i.e. starting with the definition of the U($N$)
supercharge, we end up with the definition of the SU($N$) supercharge.
In addition, the (anti)commutation relations
$[{\tilde a}_{ij}(k_1),\alpha^{\dagger}(k_2)] = 0$ and
$\{{\tilde b}_{ij}(k_1),\beta^{\dagger}(k_2)\} = 0$ imply
that this supercharge acts only on the SU($N$) creation operators
of a fock state - the U(1) creation operators only introduce
degeneracies in the SU($N$) spectrum. Clearly, since $Q^-$ has no
U(1) contribution, any fock state made up of only U(1)
creation operators must have zero mass.
  The non-trivial problem here is to determine whether there are
massless states for the SU($N$) sector. We will address
this topic next.

%%%%%%%%%%%%%%%%%%%%%%%%%%%%%%%%%%%%%%%%%%%%%%%%%%%%%%%%%%%%%%%%%%%%%%%
\subsection{The Proof for (1,1) Model.}
\label{proof}

It was pointed out in the previous subsection
that a zero mass eigenstate is annihilated by the
light-cone supercharge (\ref{Qminus3}):
\begin{equation}
\label{mslcond}
Q^-|\Psi \rangle =0
\end{equation}
We wish to show that if such an SU($N$)
eigenstate is normalizable, then
it must involve a superposition of an {\it infinite} number of Fock states.
The basic strategy is quite simple: normalizability will
impose certain conditions on the light-cone wave functions as one or
several momentum variables vanish. Moreover, if we assume a given
eigenstate $|\Psi \rangle$ has at most $n$ partons, then the terms in
$Q^- |\Psi \rangle$ consisting of $n+1$ partons must sum to zero,
providing relations between the $n$ parton wave functions only.
We then show these wave functions must all vanish by studying
various zero momentum limits of these relations. Interestingly, the
utility of studying light-cone wave functions at small momenta
also appears in the context of light-front
$\mbox{QCD}_{3+1}$ \cite{adb97}.

In order to proceed with a systematic presentation of
the proof, we start by considering the large $N$ limit case.
This simply means that we consider Fock states that
are made from a {\em single} trace of a product
of boson or fermion creation
operators acting on the light-cone Fock vacuum
$|0\rangle$. Multiple trace states correspond to $1/N$ corrections to
the theory, and are therefore ignored.
In this limit, a general state $|\Psi \rangle$ is
a superposition of Fock states of any length,
and may be written in the form
\begin{eqnarray}
\label{genfock}
\lefteqn{|\Psi \rangle =
   \sum_{n=2}^{\infty} \sum_{r=0}^{n}
    \sum_{P} \int_0^{P^+} {dq_1 \dots dq_n \over
    \sqrt{q_1...q_n}}\delta(q_1 + \cdots + q_n - P^+) \times} & &
\nonumber \\
& & \hspace{25mm}
f^{(n,r)}_P (q_1,\dots,q_n)
  \mbox{tr}[c^\dagger(q_1)\dots c^\dagger(q_n)]|0\rangle ,
\end{eqnarray}
where $c^{\dagger}(q^+)$ represents either a boson or fermion
creation operator
carrying light-cone momentum $q^+$, and $f^{(n,r)}_P$ denotes
the wave function of an $n$ parton Fock state containing $r$
fermions in a particular arrangement $P$. It is implied that
we sum over all such arrangements, which may not necessarily be distinct
with respect to cyclic symmetry of the trace.

At this point, we simply remark that normalizability of a
general state $|\Psi \rangle$ above implies
\begin{equation}
\int_0^{P^+} {dq_1 \dots dq_n \over q_1 \dots q_n}\delta(
q_1 + \cdots + q_n - P^+)|f^{(n,r)}_P(q_1,\dots,q_n )|^2 < \infty
\end{equation}
for any particular wave function $f^{(n,r)}_P$. Therefore,
any wave function vanishes
if one or several of its momenta are made to
vanish.

We are now ready to carry out the details of the proof. But first
a little notation. We will write $|\Psi_{(n,m)} \rangle$ to denote
a superposition of all Fock states -- as in (\ref{genfock}) --
with precisely $n$ partons, $m$ of which are fermions. Such a
Fock expansion involves only the wave functions $f^{(n,m)}_P$,
and the number of them is
enumerated by the index $P$.
For the special
case $|\Psi_{(n,0)} \rangle$ (i.e. no fermions), there
is only one wave function, which we denote by $f^{(n,0)}$
for brevity:
\begin{equation}
\label{stallbos}
|\Psi_{(n,0)} \rangle =
   \int_0^{P^+} {dq_1 \dots dq_n \over
    \sqrt{q_1...q_n}}\delta(q_1 + ... + q_n - P^+)\hspace{1mm}
f^{(n,0)}(q_1... q_n)
  \mbox{tr}[a^\dagger(q_1)... a^\dagger(q_n)]|0\rangle .
\end{equation}
There is another special case we wish to consider; namely,
the state $|\Psi_{(n,2)} \rangle$ consisting of $n$ parton
Fock states with precisely two fermions. If we place one of
the fermions at the beginning of the trace, then there are
$n-1$ ways of positioning the second fermion, yielding
$n-1$ possible wave functions. We will enumerate such
wave functions by the subscript index $k$, as in $f^{(n,2)}_k$,
where $k=2,3,\dots,n$.
The subscript $k$ denotes the location of the second fermion.
Explicitly, we have
\begin{eqnarray}
\label{twofermion}
\lefteqn{
|\Psi_{(n,2)} \rangle =
  \sum_{k=2}^{n} \int_0^{P^+} {dq_1 \dots dq_n \over
    \sqrt{q_1...q_n}}\delta(q_1 + \cdots + q_n - P^+)\times } & &\nonumber \\
& & \hspace{15mm}
f^{(n,2)}_k(q_1,\dots,q_k,\dots,q_n)
  \mbox{tr}[ b^\dagger(q_1)a^\dagger(q_2)\dots
    b^\dagger(q_k) \dots
a^\dagger(q_{n})]|0\rangle .\nonumber\\
\
\end{eqnarray}
Of course, depending upon the symmetry, the $n-1$
Fock states enumerated in this way need not be distinct
with respect to the cyclic properties of the trace.
This provides us with additional relations between wave functions -- a fact
we will make use of later on.

Now let us assume that $| \Psi \rangle$ is a
normalizable SU($N$) zero mass eigenstate with at most $n$ partons.
Glancing at the form of (\ref{Qminus3}), we see that
the $n+1$ parton Fock states containing a single
fermion in each of the combinations
  $Q^-|\Psi_{(n,0)} \rangle$ and  $Q^-|\Psi_{(n,2)} \rangle$
must cancel each other to guarantee a massless eigenstate.
This immediately
gives rise to the following wave function relation:
\begin{eqnarray}
\label{higheqnnorm}
\label{nn}
\frac{q_{1}+2q_2}{q_{1}+q_2} f^{(n,0)}(q_1+q_2,q_3,\dots,q_{n+1})
-\frac{q_1+2q_{n+1}}{q_1+q_{n+1}}f^{(n,0)}(q_{n+1}+q_1,q_2,\dots,q_n)
=\nonumber\\
=2 \frac{\sqrt{q_1}}{n}\sum_{k=2}^{n} \frac{q_{k+1}-q_k}{(q_{k+1}+q_k)^{3/2}}
f^{(n,2)}_k(q_1,\dots,q_{k-1},q_k+q_{k+1},q_{k+2},\dots,q_{n+1}).\nonumber\\
\
\end{eqnarray}
In the limit $q_i \rightarrow 0$, for $3 \leq i \leq n$, this last equation
is reduced to
\begin{eqnarray}
\lefteqn{
\frac{1}{\sqrt{q_{i+1}}}f^{(n,2)}_i(q_1,\dots,q_{i-1},q_{i+1},
\dots,q_{n+1}) } & & \nonumber \\
& & \hspace{25mm} -
\frac{1}{\sqrt{q_{i-1}}}f^{(n,2)}_{i-1}(q_1,\dots,q_{i-1},q_{i+1},
\dots,q_{n+1})=0.
\end{eqnarray}
An immediate consequence is that
any wave function $f^{(n,2)}_i$ for $i=3,4,\dots,n$, may
be expressed in terms of $f^{(n,2)}_2$. Explicitly, we have
\begin{equation}
f^{(n,2)}_i(q_1,q_2,\dots,q_n) = \sqrt{\frac{q_i}{q_2}}
     f^{(n,2)}_2(q_1,q_2,\dots,q_n), \hspace{8mm}i=3,4,\dots,n.
\label{chain}
\end{equation}
Moreover, the limit $q_2 \rightarrow 0$
of equation (\ref{higheqnnorm})
yields the further relation after a suitable change of variables:
\begin{eqnarray}
f^{(n,0)}(q_1,q_2,q_3,\dots,q_{n}) & = &\frac{2}{n}\sqrt{\frac{q_{1}}{q_{2
}}}
f^{(n,2)}_2(q_1,q_2,q_3,\dots,q_{n}). \label{rel1}
\end{eqnarray}
Finally, because of the cyclic properties of the trace, there
is an additional relation between wave functions:
\begin{equation}
   f^{(n,2)}_i(q_1,q_2,\dots,q_i,\dots,q_n)
  = -f^{(n,2)}_{n-i+2}(q_i,q_{i+1},\dots,q_n,q_1,q_2,\dots,q_{i-1}).
\end{equation}
Setting $i=2$ in the above equation, and $i=n$ in equation (\ref{chain}),
we deduce
\begin{equation}
        f^{(n,2)}_2(q_1,q_2,\dots,q_n) =
         -\sqrt{\frac{q_1}{q_2}} f^{(n,2)}_2(q_2,q_3,\dots,q_n,q_1).
\end{equation}
Combining this with equation (\ref{rel1}), we conclude
$(\frac{\sqrt{q_2}}{q_1} + \frac{\sqrt{q_3}}{q_2})
f^{(n,0)}(q_1,\dots,q_n) = 0$, where
we use the fact that the wave functions $f^{(n,0)}$ are
cyclically symmetric.  Thus $f^{(n,0)}$ must vanish.
It immediately follows that $f^{(n,2)}_i$ vanish for all $i$
as well.

To summarize, we have shown that if $|\Psi \rangle$ is a normalizable
zero mass eigenstate, where each Fock state
in its Fock state expansion has no more than $n$ partons,
the contributions
$|\Psi_{(n,0)}\rangle$ and $|\Psi_{(n,2)}\rangle$
in this Fock state expansion must vanish. Since we may
assume $|\Psi \rangle$ is bosonic, the only other
contributions involve Fock states with an even number of
fermions: $|\Psi_{(n,4)}\rangle$, $|\Psi_{(n,6)}\rangle$, and so on.
We claim that all such contributions vanish. To
see this, first observe that the $n+1$ parton Fock states
with three fermions
in the combinations $Q^- |\Psi_{(n,2)}\rangle$ and
$Q^- |\Psi_{(n,4)}\rangle$ must cancel each other, in order
to guarantee a zero eigenstate mass. But our previous analysis
demonstrated that $|\Psi_{(n,2)}\rangle  \equiv 0$, and thus
the $n+1$ parton Fock states with three fermions in
$Q^- |\Psi_{(n,4)}\rangle$ alone must sum to zero.

We are now ready to perform an
induction procedure. Namely,
we assume  that for some positive integer
$k$ the state $|\Psi_{(n,2[k-1])}\rangle$ vanishes.
Then the $n+1$ parton Fock states
in $Q^- |\Psi \rangle$ which contain $2k-1$ fermions
receive contributions only from  $Q^-|\Psi_{(n,2k)}\rangle$
in which a fermion is replaced by two bosons. This has to sum to zero.
We therefore obtain a relation among the wave functions
$f^{(n,2k)}_P$ by considering the action
of the supercharge (\ref{Qminus3}) in which a fermion is replaced
by two bosons. Keeping in mind that we are
  free to renormalize any wave function by a constant,
we end up with the following relation:
\begin{equation}
\label{indeqnnorm}
\sum_{P} {f}^{(n,2k)}_P(s_1,\dots,s_{i-1},s_i+s_{i+1},s_{i+2},
\dots, s_{n+1}) \frac{s_{i+1}-s_i}{(s_{i+1}+s_i)^{3/2}}=0.
\end{equation}
It is now an easy task to show that
  the wave functions ${f}^{(n,2k)}_P$ appearing
in equation (\ref{indeqnnorm}) must vanish; one simply
considers various limits $s_j \rightarrow 0$ as we did before.
This completes our proof by induction. Namely, there
can be no non-trivial normalizable massless state
with an upper limit on the number of allowed partons.
Of course, this proof is valid only in the large $N$ limit.
We now turn our attention to the finite $N$ case.

\medskip

Let us define $Q^-_{lead}$ to be that part of the
supercharge $Q^-$ that replaces a fermion with two bosons,
or replaces a boson with a boson and fermion pair.
As in the large $N$ case we begin
by assuming that we have a normalizable zero mass eigenstate
$|\Psi \rangle$ which is a sum of Fock states that have
at most $n$ partons.
The proof for finite $N$ consists of
two parts. First, we consider bosonic states
consisting of only $n$ parton Fock states
that have at most two fermions,
and show the wave functions must
vanish. We then invoke an induction argument to
consider $n$ parton wave functions involving an even number of fermions,
and show they must vanish as well.

The additional complication introduced by the assumption that $N$ is
finite is that a given Fock state may involve more than just
a single trace. However, note that
$Q^-_{lead}$
cannot decrease the number of traces; it can either increase the
number of traces by one, or leave the number unchanged.
Thus we have a natural induction
procedure in the number of traces as well.
Since the terms in $Q^-_{lead}$  have
  only one annihilation operator, it acts on
a given product of traces according to the Leibniz rule:
\begin{eqnarray}
Q^-_{lead}\left(\mbox{tr}[A]\mbox{tr}[B]\dots \right)|0\rangle =
\left(Q^-_{lead}\mbox{tr}[A]\right)\mbox{tr}[B]\dots|0\rangle+\nonumber\\
(-1)^{F(A)}\mbox{tr}[A]Q^-_{lead}\left(\mbox{tr}[B]\dots\right)|0\rangle.
\end{eqnarray}
Schematically, the general structure of an arbitrary Fock
state with $k$ traces has the form
\begin{equation}
f^{(n,i_1,i_2,\dots,i_k)}_P
\mbox{tr}[(b^\dagger)^{i_1} a^\dagger\dots a^\dagger]
\dots \mbox{tr}[(b^\dagger)^{i_k} a^\dagger \dots a^\dagger]|0\rangle,
\end{equation}
where  $n$ denotes
the total number of partons in each Fock state,
and the integers $i_1,i_2,\dots$
denote the number of fermions in the first trace, second trace, and so on.
We will always
order the traces so that the number of fermions
in each trace
decreases to the right. The index $P$ labels a particular arrangement
of fermions.

We now consider the $n+1$ parton Fock states of
$Q^-_{lead}|\Psi \rangle$ that have precisely one fermion.
The only possible contributions involve three types of wave functions;
$f^{(n,0)}$, $f^{(n,2)}_P$ and
$f^{(n,1,1)}$ (we only include the permutation index $P$
if there is more than one distinct arrangement).
If these three wave functions  contribute to the
same one fermion Fock state, then the distribution of
bosons in the Fock state corresponding to
$f^{(n,2)}_P$ determines the distribution of bosons for $f^{(n,0)}$ and
$f^{(n,1,1)}$. We allow
$Q^-_{lead}$ to act only on the first trace in both
$f^{(n,0)}$ and $f^{(n,2)}_P$,
and only on the second one in $f^{(n,1,1)}$. If
there are more than two traces in these states
they must be identical in all the
components, and so don't play a role in the calculation.
Thus, it is sufficient to
consider states with two traces only. Such a state has the form
\begin{eqnarray}
\label{finnphi}
&&|\Phi> =\int_0^{P^+} { d^{m+n}q \over \sqrt{q_1 \dots q_{n+m}}}
\delta(q_1+ \cdots +q_{n+m}
-P^+)
\nonumber \\
&&f^{(n+m,0)}(q_1,\dots,q_m|q_{m+1},\dots,q_{m+n})
\nonumber \\
&&\times \mbox{tr}\left[a^\dagger(q_1)\dots a^\dagger(q_m)\right]
\mbox{tr}\left[a^\dagger(q_{m+1})\dots a^\dagger(q_{m+n})\right]|0\rangle
\nonumber\\
&&\nonumber\\
&&+\int_0^{P^+} { d^{m+n-2}qdp_1dp_2 \over \sqrt{q_1\dots q_{n+m-2}p_1p_2}}
\delta(q_1+\cdots +q_{n+m-2}+p_1+p_2 -P^+) \left\{\frac{}{}\right.
\nonumber \\
&&f^{(n+m,1,1)}(p_1,q_1,\dots,q_m|p_2,q_{m+3},\dots,q_{m+n}) \times
\nonumber\\
&&\times \mbox{tr}\left[b^\dagger(p_1)a^\dagger(q_1)\dots a^\dagger(q_m)]
\mbox{tr}[b^\dagger(p_2)a^\dagger(q_{m+3})\dots a^\dagger(q_{m+n})\right]
  +
\nonumber\\
&&\nonumber\\
&&+\sum_P
f^{(n+m,2)}_P (p_1,P[q_1\dots q_{m-2};p_2]|q_{m+1}\dots q_{m+n}) \times
\nonumber \\
&&\left.\times \mbox{tr}\left(b^\dagger(p_1)
P[a^\dagger(q_1)\dots a^\dagger(q_{m-2});b^\dagger(p_2)]
\right)
\mbox{tr}\left[a^\dagger(q_{m+1})\dots a^\dagger(q_{m+n})\right]
\frac{}{}\right\}|0\rangle,\nonumber\\
&&\
\end{eqnarray}
\\
\noindent where we have  summed over the index
$P$ representing all possible permutation arrangements
  of bosons and fermions
that
contribute.
We then find:
\begin{eqnarray}
\lefteqn{{F}(p,q_1,\dots,q_m|q_{m+1},q_{m+2},\dots,q_{m+n})+}& &
\\
& & +\frac{q_{m+2}-q_{m+1}}{(q_{m+2}+q_{m+1})^{3/2}}
{f}^{(n+m,1,1)}(p,q_1\dots q_m|q_{m+1}+q_{m+2},q_{m+3}\dots q_{m+n})=0,
\nonumber
\end{eqnarray}
where $F$ is the contribution from $f^{(n+m,0)}$ and $f^{(n+m,2)}_P$.
Now we see that the limit $q_{m+1}\rightarrow 0$ gives:
$f^{(n+m,1,1)}\equiv 0$.
Thus if (\ref{finnphi}) represents
a contribution to the massless eigenstate state $|\Psi\rangle$,
then $|\Phi\rangle$ takes the form
\begin{eqnarray}
\label{finnphm}
|\Phi\rangle &=& \int_0^{P^+} {d^{m+n-2}q dK^+
  \over \sqrt{q_1\dots q_{n+m-2}}}
\delta(q_1+ \cdots + q_{n+m-2} - (P^+-K^+))\left[ \frac{}{} \right.
\nonumber \\
&&  \int_0^{P^+}
{dq_{m-1}dq_{m} \over \sqrt{q_{m-1}q_m}}
\delta(q_{m-1} +q_m -K^+)
\nonumber \\
& & f^{(n+m,0)}(q_1,\dots,q_m|q_{m+1},\dots,q_{m+n})
\mbox{tr}(a^\dagger(q_1)\dots a^\dagger(q_m))
\nonumber\\
&+&  \int_0^{P^+} {dp_1dp_1 \over \sqrt{p_1p_2}} \delta(p_1+p_2 -K^+)
\nonumber \\
&& \sum_P f^{(n+m,2)}_P(p_1,P[q_1,\dots,q_{m-2};p_2]|q_{m+1},\dots,q_{m+n})
\\
&&\left.
\mbox{tr}(b^\dagger(p_1)P[a^\dagger(q_1)\dots
a^\dagger(q_{m-2});b^\dagger(p_2)])
\mbox{tr}(a^\dagger(q_{m+1})\dots a^\dagger(q_{m+n}))\right]|0\rangle\nonumber
\end{eqnarray}
and $Q^-_{lead}$ acts only on the terms in the square brackets. All these
terms have only one trace, which is a scenario we already
encountered  in
the large $N$ limit case.
Using the results of that discussion, we find that the only
massless solution of the form (\ref{finnphm}) is the trivial one. This is the
starting point of the induction procedure for finite $N$.

As explained earlier, we look for $n$ parton
Fock states in the expansion for $|\Psi\rangle$ that have $2k$ fermions
($k>1$),
  To finish the proof we
need to
show that for any $k$ the only allowed wave function is the trivial one.
>From the
large $N$ result we know there are no such one trace states. We now
consider the state
with an arbitrary number of traces,
\begin{eqnarray}
&&|\Psi_{(n,2k)}\rangle=\sum_{P} \int_0^{P^+} {ds_1\dots ds_n \over
\sqrt{s_1\dots s_n}}
\delta  (s_1+\cdots + s_n -P^+)  \\
&&f^{(n,2k)}_P
(s_1\dots s_{i_1}|\dots |\dots s_n)
\mbox{tr}\left(c^\dagger(s_1)\dots c^\dagger(s_{i_1})\right)
\mbox{tr}\left(\dots \right)
\mbox{tr}\left(\dots c^\dagger(s_n)\right)|0\rangle,\nonumber
\end{eqnarray}
then the analog of (\ref{indeqnnorm}) for such states reads:
\begin{equation}
\label{finnind}
{\sum_{i}}'  f^{(n,2k)}_{P_i}
(s_1\dots |s_{j_a} \dots s_{i-1},s_i+s_{i+1},s_{i+2}\dots
s_{j_a+k_a}| \dots s_{n+1})
\frac{s_{i+1}-s_i}{(s_{i+1}+s_i)^{3/2}}=0.
\end{equation}
Here, $\sum_{i}'$ means that for each trace we should include one additional
term with $"i"=j_a+k_a$, $"i+1"=j_a$ if $c$ corresponding to both $j_a+k_a$
and $j_a$ is  $a$. If the number of traces is $a$, we introduce
$$
j_a =\sum_{b=1}^{a-1} k_b.
$$

If any of the blocks $\mbox{tr}(\dots)$ in the state for which
(\ref{finnind})
is written contains two or more fermions, then, as in the large $N$ case,
all the corresponding wave functions
$f^{(n,2k)}_{P}$ vanish. So we only need to consider the states
of the form:
\begin{eqnarray}
\label{st11b}
|\Psi_{(n,k_1+1,\dots)}\rangle
= \sum_P\int dpdq f^{(n,k_1+1,\dots)}_P
(p_1,q_1,\dots,q_{k_1}|p_2,q_{k_1+1},\dots,
q_{k_1+k_2}|\dots)  \times \nonumber \\
\mbox{tr}\left(b^\dagger(p_1)a^\dagger(q_1)\dots a^\dagger(q_{k_1})\right)
\mbox{tr}\left(b^\dagger(p_2)a^\dagger(q_{k_1+1})\dots
a^\dagger(q_{k_1+k_2})
\right)\dots |0\rangle.\nonumber
\end{eqnarray}
\begin{equation}
\end{equation}
Let ${\tilde Q}$ denote that part of the supercharge $Q^-$
which replaces a fermion with two bosons.
  Let us consider the result of such a change in the
first trace. Suppose there are $a$ traces having the same form as the
first trace. Then without loss of generality, we may assume
they are the first
$a$ traces. Then using the symmetries of the wave functions we find:
\begin{eqnarray}
&&{\tilde Q}|\Psi_{(n,k_1+1,\dots)}\rangle  =-\frac{1}{2\sqrt{2\pi}}
\sum_P
\int_0^{P^+} dkdpdq \nonumber\\
&&f^{(n,k_1+1,\dots)}_P
(p_1,q_1,\dots,q_{k_1}|p_2,q_{k_1+1},\dots,q_{2k_1}|\dots)
%\hspace{25mm}
\sum_{b=1}^{a} \frac{p_b-2k}{p_b}\frac{1}{\sqrt{k(p_b-k)}}(-1)^{b+1}
\times \nonumber\\
&&\mbox{tr}
\left(b^\dagger(p_1)a^\dagger(q_1)\dots a^\dagger(q_{k_1})\right)\dots
\mbox{tr}\left(a^\dagger(k)a^\dagger(p_b-k)a^\dagger(q_{(b-1)k_1+1})\dots
a^\dagger(q_{bk_1})\right)
\dots |0\rangle \nonumber\\
&&=-\frac{1}{2\sqrt{2\pi}}
\sum_P\int_0^{P^+} dkdpdq \frac{p_1-2k}{p_1}\frac{1}{\sqrt{k(p_1-k)}}
\mbox{tr}\left(a^\dagger(k)a^\dagger(p_1-k)a^\dagger(q_1)\dots
a^\dagger(q_{k_1})\right) \times \nonumber \\
&&\mbox{tr}\left(b^\dagger(p_2)a^\dagger(q_{k_1+1})\dots
a^\dagger(q_{k_1+k_2})\right)
\dots |0\rangle
  \sum_{b=1}^{a} (-1)^{b+1}(-1)^{b+1} \times \nonumber \\
&& \hspace{10mm}f_P^{(n,k_1+1,\dots)}
(p_1,q_1,\dots,q_{k_1}|p_2,q_{k_1+1},\dots,q_{k_1+k_2}|\dots).\nonumber
\end{eqnarray}
If the above expression vanishes then the only solution is the trivial
one in which all wave functions vanish. This
finishes the proof of the induction procedure for the finite $N$ case.

The extension of the proof to massive bound states is straightforward.
Firstly, assume $|\Psi\rangle$ is a normalizable
  eigenstate of $2P^+ P^-$
with mass squared $M^2 \neq 0$.
Then, since $P^- = \frac{1}{\sqrt{2}}(Q^-)^2$,
the state
\begin{equation}
           |{\tilde \Psi}\rangle \equiv |\Psi\rangle
                            + \alpha Q^- |\Psi\rangle
\end{equation}
for $\alpha^2 = \sqrt{2} P^+/M^2$ is a normalizable eigenstate
of the supercharge $Q^-$, with eigenvalue $1/\alpha$.
We therefore study the eigen-problem $Q^-|{\tilde \Psi}\rangle =
\frac{1}{\alpha} |{\tilde \Psi}\rangle$.
The resulting constraints on the wave functions
may be obtained by
modifying  our original expressions by including a
wave function multiplied by a finite constant. However,
in our analysis, we always need to
let some of the momenta vanish, and therefore
this additional contribution vanishes. The analysis (and therefore
the conclusions) remains unchanged.

We therefore conclude that any normalizable SU($N$)
bound state (massless or massive)
that exists in the model must be a superposition
of an infinite number of Fock states.

%%%%%%%%%%%%%%%%%%%%%%%%%%%%%%%%%%%%%%%%%%%%%%%%%%%%%%%%%%%%%%%%%%%%%%%%%%%%
\subsection{Higher Dimensional Theories.}

In this subsection we extend our theorem to the two--dimensional supersymmetric
theories obtained as the result of dimensional reduction from $D>3$
dimensions. The most important cases are $D=4$, 6 and 10 which have
$2$, $4$ and $8$ supersymmetries in two dimensions. Below we
consider only large N case, the generalization to arbitrary $SU(N)$ group is
trivial repetition of the arguments given in previous subsection.

Again our starting point is the fact that if there is normalizable
eigenstate of Hamiltonian having finite length than its main symbol
satisfies the condition:
\be\label{opereqn}
Q^-_{lead} |\Psi>=0,
\ee
where $Q^-_{lead}$ is the part of supercharge increasing the number of
partons. In three dimensional case we had only one supercharge $Q^-$, for
general $D$ dimensional SYM reduced in $1+1$ there are $D-2$ supercharges,
each of them squared gives $P^-$ and (\ref{opereqn}) should be true for all
of them. In general different supercharges are not anticommute with each
other, but since we consider quantization near trivial classical configuration
(with no monopoles and no external charges) then they do. It is easy to derive
the general form of supercharge:
\bea\label{Qminuscmp}
Q^-_{\alpha}&=&\int_o^\infty \frac{dk}{k} (b^{\alpha\dagger}_{ij}(k)
J_{ij}(-k)-( J_{ij}(-k))^\dagger b^\alpha_{ij}(k) )+\\
&+&\frac{\mu}{2\sqrt{2\pi}}\int_{-\infty}^{\infty} dk M^{\alpha\beta}_{IJ}
[A_I,A_J]_{ij} (k){\Psi}^{\beta}_{ji}(-k),\nonumber\\
J_{ij}(-k)&=&\frac{1}{2\sqrt{2\pi}}\int_o^\infty dp\frac{2p+k}{\sqrt{p(p+k)}}
\left( a^{I\dagger}_{ki}(p)a^I_{kj}(k+p)- a^{I\dagger}_{jk}(p)a^I_{ik}(k+p)
\right)+\\
&+&\frac{1}{2\sqrt{2\pi}}\int_o^k dp\frac{k-2p}{\sqrt{p(k-p)}}
a^I_{ik}(p)a^I_{kj}(k-p) +
\frac{1}{\sqrt{2\pi}}\int_o^k dp b^{\alpha}_{ik}(p)b^{\alpha}_{kj}(k-p)+
\nonumber\\
&+&\frac{1}{\sqrt{2\pi}}\int_o^\infty dp
\left( b^{\alpha\dagger}_{ki}(p)b^{\alpha}_{kj}(k+p)- b^{\alpha\dagger}_{jk}
(p)b^\alpha_{ik} (k+p)\right) .\nonumber
\eea
In the above expression we introduced $d=D-2$ kinds of bosons ($I=1,...,d$) and
$d$ kinds of fermions ($\alpha=1,...,d$) which we get as the result of
compactification. The $\mu$ is nonzero constant depending on $D$ and $M$
are combinations of $d$ dimensional Dirac matrices:
\be\label{Dirac}
M^{\alpha\beta}_{IJ}=(\gamma_I \gamma^T_J -\gamma_J \gamma^T_I )_{\alpha\beta}.
\ee

As before our proof is based on the induction on the number of fermionic
operators in the state. First we consider main symbol being superposition of
purely bosonic states and ones containing two fermionic operators. Now we
have $d$ types of bosons and
$d$ types of fermions so some additional indices should be included in the
wavefunctions. Defining bosonic indexes to be capital letters $A,\ B...$ and
fermionic ones to be Greek letters we write:
\bea
|\Psi,0>&=&\int_0^{P^+} \frac{dq_1...dq_n}{\sqrt{q_1...q_n}}
\delta (q_1+...+q_n-P^+)\sum_{A} f^{(0)}_{[A_1...A_n]}
(q_1...q_n)\times\nonumber\\
&\times &tr[a^\dagger_{A_1}(q_1)...a^\dagger_{A_n}(q_n)]|0>,\\
|\Psi,2>&=&\sum_{k=1}^{n-1}\int_0^{P^+} \frac{dq_1...dq_n}{\sqrt{q_1...q_n}}
\delta (q_1+...+q_n-P^+)\sum_{A,\alpha}
f^{(2)k}_{[A_1...A_{k-1}\alpha_1 A_k...A_{n-2}\alpha_2]} (q_1...q_n)\times
\nonumber\\
&\times &tr[a^\dagger_{A_1}(q_1)...a^\dagger_{A_{k-1}}(q_{k-1})
b^\dagger_{\alpha_1}(q_k) a^\dagger_{A_k}(q_{k+1})...
a^\dagger_{A_{n-2}}(q_{n-1}) b^\dagger_{\alpha_2}(q_n)]|0>.
\eea

It is now easy to find the one fermionic part of the result of action by
(\ref{Qminuscmp}) on the main symbol of the state. The vanishing of this
contribution leads to the generalization of the equation (\ref{higheqnnorm}):
\bea
\label{mstreqncmp}
&&\delta_{\alpha\beta}\frac{n}{p}\left(
\frac{2q_n+p}{q_n+p}f^{(0)}_{A_1...A_n}(q_1...q_{n-1},q_n+p)-
\frac{2q_{1}+p}{q_{1}+p} f^{(0)}_{A_1...A_n}(q_1+p...q_{n-1},q_n)\right)-
\nonumber\\
&&\frac{2}{\sqrt{p}}\sum_{k=1}^{n-1} \frac{q_{k+1}-q_k}{(q_{k+1}+q_k)^{3/2}}
\delta_{A_k A_{k+1}}f^{(2)k}_{[A_1...A_{k-1}\alpha A_{k+2}...A_n \beta]}
(q_1 ...q_{k-1},q_k+q_{k+1},q_{k+2}...q_n,p)+\nonumber\\
&&n\mu\left(\frac{M^{\alpha\beta}_{A_n B}}{q_n+p}
f^{(0)}_{A_1...A_{n-1} B}(q_1...q_{n-1},q_n+p)-
\frac{M^{\alpha\beta}_{A_1 B}}{q_{1}+p} f^{(0)}_{B A_2...A_n}(q_1+p...q_{n-1},
q_n) \right)+\nonumber\\
&&\frac{2\mu}{\sqrt{p}}\sum_{k=1}^{n-1}
\frac{M^{\alpha\gamma}_{A_{k+1}A_k}}{\sqrt{q_{k+1}+q_k}}
f^{(2)k}_{[A_1...A_{k-1}\gamma A_{k+2}...A_n \beta]}
(q_1 ...q_{k-1},q_k+q_{k+1},q_{k+2}...q_n,p)=0
\eea
This equation should be true for any possible $A_1...A_N$, $\alpha$ and
$\beta$. We
will show that the only solution of such system of equations is trivial one so
all the $f^{(0)}_{[...]}$ and $f^{(2)k}_{[...]}$ vanish. This will be proven
by induction. First we note that if $A_1=...=A_n$ and $\alpha=\beta$ then
equation
(\ref{mstreqncmp}) is reduced to (\ref{nn}) written for $f^{(0)}_{[A...A]}$
and $f^{(2)k}_{[A...A\alpha A...A\alpha]}$ and as we saw this leads to
\be
f^{(0)}_{[A...A]}=0,\qquad f^{(2)k}_{[A...A\alpha A...A\alpha]}=0
\ee
for arbitrary $A$ and $\alpha$. The next case to consider is $A_1=...=A_n$,
$\alpha\ne\beta$. Using relation just found the (\ref{mstreqncmp}) for this
case
again gives us (\ref{nn}), but this time correspondence reads:
\bea
f^{(0)}\rightarrow p\mu M^{\alpha\beta}_{AB}f^{(0)}_{[BA...A]},\nonumber\\
f^{(2)k}\rightarrow  f^{(0)}_{[A...A\alpha A...A\beta]}.
\eea
The proven property of (\ref{nn}) together with trivial identity
\be
\sum_{\alpha\beta} M^{\beta\alpha}_{CA} M^{\alpha\beta}_{AB}=4d\delta_{BC}(1-
\delta_{AC})
\ee
leads to
\be\label{start}
f^{(0)}_{[BA...A]}=0,\qquad f^{(2)k}_{[A...A\alpha A...A\beta]}=0
\ee
for any $A,B,\alpha,\beta$ ($A$ could be equal to $B$ and $\alpha$ to
$\beta$). We use this
equation as starting point of the induction procedure.

Let us introduce one useful function. For each set $\{A_1...A_k\}$ we define
${\bar n}(\{A_1...A_k\} )$ to be the maximal number of identical $A$ in the
set:
\be
{\bar n}(\{A_1...A_k\} )=\max_{I\le d}\left(\sum_{i=1}^k \theta(A_i-I)\right).
\ee
In terms of this new function our result (\ref{start}) can be rewritten as
\bea
f^{(0)}_{[A_1...A_n]}=0 \quad if \quad {\bar n}(\{A_1...A_n\} )\ge n-1
\nonumber\\
f^{(2)k}_{[A_1...A_{k-1}\alpha A_k...A_{n-2}\beta]}=0 \quad if \quad
{\bar n}(\{A_1...A_n\} )= n-2.
\eea
This condition will be used as starting point of induction then the assumption
of induction procedure is:
\bea\label{indforstart}
f^{(0)}_{[A_1...A_n]}=0 \quad if \quad {\bar n}(\{A_1...A_n\} )\ge m
\nonumber\\
f^{(2)k}_{[A_1...A_{k-1}\alpha A_k...A_{n-2}\beta]}=0 \quad if \quad
{\bar n}(\{A_1...A_n\} )\ge m-1
\eea
and we checked it for $m=n-1$. In our induction procedure we will
decrease parameter $m$ instead of increasing it. To perform the proof, we
start with writing (\ref{mstreqncmp}) for the set $\{A_1...A_k\}$ with
${\bar n}=m$:
\bea\label{ind1}
&&\frac{2}{\sqrt{p}}\sum_{k=1}^{n-1} \frac{q_{k+1}-q_k}{(q_{k+1}+q_k)^{3/2}}
\delta_{A_k A_{k+1}}f^{(2)k}_{[A_1...A_{k-1}\alpha A_{k+2}...A_n \beta]}
(q_1 ...q_{k-1},q_k+q_{k+1},q_{k+2}...q_n,p)=\nonumber\\
&&n\mu\left(\frac{M^{\alpha\beta}_{A_n B}}{q_n+p}
f^{(0)}_{A_1..A_{n-1} B}(q_1..q_{n-1},q_n+p)-
\frac{M^{\alpha\beta}_{A_1 B}}{q_{1}+p} f^{(0)}_{B
A_2..A_n}(q_1+p..q_{n-1},q_n) \right).
\eea
If $\alpha=\beta$ then the right hand side is zero and we have a recurrent
relations
for $f^{(2)k}_{[A_1...A_{k-1}\alpha A_{k+2}...A_n \alpha]}$ with different
$k$. Due
to the presence of $\delta$ symbol these relations would connect only
$f^{(2)k}$
inside some clusters (for $m<n$) and the boundary elements of such clusters
should be zero. Thus we deduce that if ${\bar n}(\{A_1...A_{n-2}\})=m-2$
then
\be
f^{(2)k}_{[A_1...A_{k-1}\alpha A_{k}...A_{n-2} \alpha]}=0.
\ee
For $\alpha\ne\beta$ we consider different limits $q_i\rightarrow 0$ in
(\ref{ind1}):
\be
\frac{2\sqrt{p}}{\sqrt{q_2}}
\delta_{A_1 A_2}f^{(2)1}_{[\alpha A_3...A_n \beta]}
(q_2 ...q_n,p)=
-n\mu M^{\alpha\beta}_{A_1 B}
f^{(0)}_{B A_2...A_n}(p...q_{n-1},q_n)
\ee
for $i=1$ and
\bea
\frac{1}{\sqrt{q_{i+1}}}
\delta_{A_i A_{i+1}}f^{(2)i}_{[A_1...A_{i-1}\alpha A_{i+2}...A_n \beta]}
(q_1 ...q_{i-1},q_{i+1}...q_n,p)=\nonumber\\
=\frac{1}{\sqrt{q_{i-1}}}
\delta_{A_i A_{i-1}}f^{(2)i}_{[A_1...A_{i-2}\alpha A_{i+1}...A_n \beta]}
(q_1 ...q_{i-1},q_{i+1}...q_n,p)
\eea
for $1<i<n$. Since ${\bar n}<n-1$ then there exist $i<n$:
$\delta_{A_i A_{i-1}}=0$ then from above equations we deduce for
$\alpha\ne\beta$:
\bea
f^{(0)}_{[A_1...A_n]}=0, \quad {\bar n}(\{A_1...A_n\} )=m-1 \nonumber\\
f^{(2)k}_{[A_1...A_{k-1}\alpha A_k...A_{n-2}\beta]}=0, \quad
{\bar n}(\{A_1...A_n\} )= m-2.
\eea
This finishes the proof by induction. Thus we have proven that equation
(\ref{mstreqncmp}) doesn't have any normalizable solutions.

To show that there are no finite length bound states we now turn to the
analog of equation (\ref{indeqnnorm}). This analog reads:
\bea
&&\sum_{i} A_i {f}^{(2k)}_{P_i} (s_1...s_{i-1},s_i+s_{i+1},s_{i+2}...
s_{n+1})\times\nonumber\\
&&\times\left(\delta_{A_i A_{i+1}}^{\alpha\alpha_i}\frac{s_{i+1}-s_i}{
(s_{i+1}+s_i)^
{3/2}}+\mu M_{A_i A_{i+1}}^{\alpha\alpha_i}\frac{1}{\sqrt{s_{i+1}+s_i}}\right)
=0,
\eea
where $P_i$ describes different permutations of $A$ and $\alpha$. This equation
gives linear relations between wavefunctions inside the block of $a$ in
\be
tr (...b^\dagger a^\dagger...a^\dagger b^\dagger...)|0>,
\ee
"boundary" elements (when index $i$ corresponds to fermions) vanish, so as in
the case $3\rightarrow 2$ all the $f^{(2k)}$ are zero.
This completes the proof for general compactification.

%%%%%%%%%%%%%%%%%%%%%%%%%%%%%%%%%%%%%%%%%%%%%%%%%%%%%%%%%%%%%%%%%%%
\subsection{Bound States in DLCQ.}
\label{dlcq}

In the previous subsection we proved that the continuum
formulation of the theory does not have any
normalizable bound states with a finite number of partons.
Our proof used the
behavior of wave functions  at small momenta
arising from the normalizability assumption. Neither
of these
properties can be used in DLCQ, however. Here we consider some simple
examples of
massless DLCQ solutions with $n$ bosons to help shed some light on the
relation between
DLCQ solutions and the solutions of the continuum theory.
For simplicity, we work in the large $N$ limit case.

We write the
momentum of a state in DLCQ in terms of the momentum fraction $q_i$ where
$q_i=\frac{r_i}{r}P^+$, and the $r_i$  are positive integers.
The wave function of
such a
state is $f^{(n,0)}(r_1,\dots,r_n)$. There  are two conditions that must be
satisfied
to show that it is massless. One is the that the coefficient of
the term with one additional fermion that is produced by the action of $Q^-$
is zero. This condition gives the relation,
\begin{equation}
\label{dscreqnorm}
\frac{2r_n+t}{r_n+t}f^{(n,0)}(r_1,\dots,r_{n-1},r_n+t)-
\frac{2r_{n-1}+t}{r_{n-1}+t} f^{(n,0)}(r_1,\dots,r_{n-1}+t,r_n)=0.
\end{equation}
where $t$ correspond to the momentum fraction of the one fermion. The second is
that the coefficient of the state with two fewer bosons and one additional
fermion which
is also produced by the action of $Q^-$ is zero. This condition
gives the relation,
\begin{equation}
\label{dcreqnxt}
\sum_{k,t}\frac{t-2k}{k(t-k)}f^{(n,0)}(r_1,\dots,r_{n-2} ,k,t-k)
\delta_{(r_{n-1}+r_{n},t)}=0.
\end{equation}

For the case where all $r_i =1$, and the total harmonic
resolution is $n$, it is
trivial that
eqn(\ref{dscreqnorm})  is satisfied since there is not enough resolution to
increase
the number of particles in the state. It is also easy to see from eqn
(\ref{dcreqnxt}) since the coefficient of the one term in the sum is zero.
Thus the
wave function $f^{(n,0)}(1,1,....1)$ is a massless state for every resolution

To discuss additional solutions it is useful to start by considering
eqn(\ref{dscreqnorm}). The case $t=1$, gives the equation
\begin{equation}
\label{discrcase1}
f^{(n,0)}(r_1,\dots,r_{n-2},r_{n-1},r_n+1)=\frac{2r_{n-1}+1}{2r_n+1}
\frac{r_n+1}{r_{n-1}+1} f^{(n,0)}(r_1,\dots,r_{n-2},r_{n-1}+1,r_n).
\end{equation}
This equation is trivial to satisfy if $r_i =1$ for all $i$.
The contributions in
eqn(\ref{dcreqnxt}) come from the two terms in the sum, $k=1$,
$t=3$ and $k=2$, $t=3$. Each term has the same coefficient but of opposite
sign and cancel. Therefore the state $f^{(n,0)}(1,...1,2)$ is a massless
state for
all resolutions

The next case  $t=2$ in eqn(\ref{dscreqnorm}) gives,
\begin{equation}
\label{discrcase2}
f^{(n,0)}(r_1,\dots,r_{n-2},r_{n-1},r_n+2)=\frac{2r_{n-1}+2}{2r_n+2}
\frac{r_n+2}{r_{n-1}+2} f^{(n,0)}(r_1,\dots,r_{n-2},r_{n-1}+2,r_n).
\end{equation}
Using (\ref{discrcase1}) twice we find:
\begin{eqnarray}
&&f^{(n,0)}(r_1...r_{n-2},r_{n-1},r_n+2)=\frac{2r_{n-1}+1}{2r_n+3}
\frac{r_n+2}{r_{n-1}+1} f^{(n.0)}(r_1...r_{n-2},r_{n-1}+1,
r_n+1)=\nonumber\\
&&=\frac{2r_{n-1}+1}{2r_n+3}\frac{r_n+2}{r_{n-1}+1}
\frac{2r_{n-1}+3}{2r_n+1}\frac{r_n+1}{r_{n-1}+2}
f^{(n,0)}(r_1...r_{n-2},r_{n-1}+2,r_n).
\end{eqnarray}
Comparing with (\ref{discrcase2}) we have:
\begin{equation}
\label{braceqn}
f^{(n,0)}(r_1...r_{n-2},r_{n-1}+2,r_n)\left(
\frac{(r_n+1)^2}{(2r_n+3)(2r_n+1)}-
\frac{(r_{n-1}+1)^2}{(2r_{n-1}+3)(2r_{n-1}+1)}\right)=0.
\end{equation}
Using relation (\ref{dscreqnorm}) several times we can always express
an arbitrary wave function in the following form:
\begin{equation}
\label{Leqn}
f^{(n,0)}(r_1...r_n)=C(r_1...r_n)f^{(n,0)}(1...1,L+1,1)
\end{equation}
where $L=r_1 +...+r_n -n$ and $C(r_1...r_n)$ is some nonzero
coefficient.The two
massless states we found above correspond to $L=0$ and $L=1$. Choosing
$r_1=...=r_{n-2}=r_n=1$ in (\ref{braceqn}) we find,
\begin{equation}\label{boundstate2}
f^{(n,0)}(1...1,(L-1)+2,1)=0 \quad for \quad L>2
\end{equation}
due to monotonic behavior of the function in the parenthesis. Then using
(\ref{Leqn}) we conclude that all the wave functions with $L>2$ vanish. So the
only case we need consider is $L=2$. In this case (\ref{dscreqnorm}) has
only two nontrivial cases: $t=1$ and $t=2$ which are given by
(\ref{discrcase1}) and (\ref{discrcase2}). In the second of these equations
  we
can only have $r_1=\dots=r_n=1$ so it is trivially satisfied. Equation
(\ref{discrcase1}) however gives a nontrivial relation for the wave function:
\bea
\label{boundstate3}
f^{(n,0)}(1,\dots,1,2,2)=f^{(n,0)}(1,\dots,2,1,2)=\dots=\nonumber\\
f^{(n,0)}(2,\dots,1,1,2)=
\frac{10}{9}f^{(n,0)}(1,\dots,1,3).
\eea
finally we must show that eqn(\ref{dcreqnxt}) is satisfied which is straight
forward.

These are only a few examples of massless states, and there are in fact many
more in DLCQ \cite{alp98a}. But the results of our numerical analysis show
that the states we just described are closely connected with the massless
states in the continuum. Let us formulate this relation for $N=\infty$. In
this case only single trace states should be kept in the spectrum and DLCQ
massless states have the following structure. The state first appear as
${\mbox tr}(a^\dagger (1)\dots a^\dagger (1))$ at resolution $P$, then one
can trace it to resolutions $P+1$ and $P+2$ as states with wavefunctions
(\ref{boundstate2}) and (\ref{boundstate3}). As we just proved at higher
resolutions there are no massless states containing exactly $K$ partons,
however at any resolution $K\ge P$ there is exactly one massless state whose
wavefunction is localized predominantly in the sector with $P$ partons
$a^\dagger$. So it is natural to collect all such states in the single
sequence and to call the limits of this sequence "continuum massless state
with $P$ bosons", although as we saw the wavefunction of continuum state has
contributions from sectors with different number of partons. The interesting
feature of this theory is that such "continuum massless states with $P$
bosons" are the only bosonic massless states seen by DLCQ (in principle
the theory in the continuum might have massless state whose wavefunction is
localized in sector with infinite number of partons, but we will ignore this
possibility). Thus one can easily count bosonic massless states at any
resolution $P$: they are just images of states with $P$ bosons for all
$P\le K$, thus there are $K-1$ such states. Acting on any of such states by
$Q^+$ we can get the fermionic massless state (then there are also $K-1$ of
them), while acting by $Q^-$ doesn't give any new state (the result is zero).
We will do the counting for finite $N$ case in the next subsection.

In the continuum limit we have proven that there are no massless
normalizable states with a finite number of particles.
However, at each finite value of
the harmonic resolution, one obtains an exactly massless
bound state, but as the harmonic resolution is sent to infinity, the number
of Fock states required to keep the bound state massless must
also be infinite.

\subsection{Counting of Massless States in DLCQ.}

Finally in this section we will count the massless states in DLCQ as function
of resolution, keeping the number of colors $N$ finite. However we will assume
that $N$ is not too small, so that the relation between $N$ and resolution $K$:
$K<N^2-1$ is satisfied. We will need this condition in order to insure all
states of the form ${\mbox tr}(c^\dagger\dots c^\dagger)\dots
{\mbox tr}(c^\dagger\dots c^\dagger)|0\rangle$ are linearly independent
unless they are related by either cyclic permutations in one of the traces or
permutations of traces themselves. The simplest example of violation of this
condition is the state
$$
{\mbox tr}(a^\dagger (1)a^\dagger (1)a^\dagger (1))|0\rangle=0
$$
for $SU(2)$ (here $3=4-1$). Although for $K>N^2-1$ some conclusions can be
made, the different $N$ and $K$ requires special consideration and we are not
going to proceed in this direction. From the numerical perspective we should
mention that in our calculation $K<11$, so the only excluded values of $N$
are $2$ and $3$.

As soon as the condition $K<N^2-1$ is satisfied the DLCQ Fock spaces for
$SU(N)$ and $SU(\infty)$ are the same if all multitrace states are taken into
account. Moreover our numerical analysis strongly suggests that the number of
massless states is the same for all $N>\sqrt{K+1}$, while wavefunctions depend
on $N$. However talking about $SU(\infty)$ Fock space one usually considers
only single trace states as fundamental ones, while multitrace states are
though of as the system of free bound states. Let us explain the reason for
this. The light--cone Hamiltonian can be written in the following schematic
form:
\begin{equation}
p^-=\frac{1}{N}P^-=\alpha c^\dagger_{ij}c_{ij}+
     \frac{1}{N}\beta c^\dagger_{ij}(ccc)_{ij}+
     \frac{1}{N}\beta (c^\dagger c^\dagger)_{ij}(cc)_{ij}+
     \frac{1}{N}\beta (c^\dagger c^\dagger c^\dagger)_{ij}(c)_{ij}.
\end{equation}
The $\frac{1}{N}$ is introduced in order to make the eigenvalues of $p^-$
finite as $N\rightarrow\infty$. Let us consider two eigenstates of $p^-$,
which are chosen to be combination of single traces in the large $N$ limit:
\begin{eqnarray}
p^- A|0\rangle=m_A A|0\rangle,\\
p^- B|0\rangle=m_A B|0\rangle.
\end{eqnarray}
This is equivalent to the following commutation relations:
\begin{eqnarray}\label{ntr_commut}
[p^-, A]=m_A A+\frac{1}{N}\sum \mu_A {\mbox{ntr}}(c^\dagger\dots c^\dagger c)+
          \frac{1}{N}\sum \nu_A {\mbox{tr}}(c^\dagger\dots c^\dagger cc)+
          O\left(\frac{1}{N^2}\right),\nonumber\\
\left[p^-, B\right]=m_B B+
\frac{1}{N}\sum \mu_B {\mbox{ntr}}(c^\dagger\dots c^\dagger c)+
          \frac{1}{N}\sum \nu_B {\mbox{ntr}}(c^\dagger\dots c^\dagger cc)+
          O\left(\frac{1}{N^2}\right).\nonumber\\
\end{eqnarray}
We introduced a convenient notation for the normalized trace here:
\begin{equation}
{\mbox{ntr}}(\underbrace{c^\dagger\dots c^\dagger c\dots c}_n)=
\frac{1}{N^{n/2}}(c^\dagger\dots c^\dagger)_{ij}(c\dots c)_{ij}.
\end{equation}
This way the state ${\mbox{ntr}}(c^\dagger\dots c^\dagger)|0\rangle$ has a
finite norm
in the large $N$ limit.
>From the equations (\ref{ntr_commut}) one can easily see that
\begin{equation}
p^- AB|0\rangle=(m_A+m_B)AB|0\rangle+O\left(\frac{1}{N}\right),
\end{equation}
i.e. we indeed have a combination of two free states in the large $N$ limit.
In particular this fact may be applied toward the classification of DLCQ
massless states at $N=\infty$: the multitrace state is massless if and only
if all of the traces involved correspond to massless states. We also mention
the trivial fact that if state $A$ has resolution $K_A$ and $B$ has $K_B$
then $AB$ is the state at resolution $K_A+K_B$.

Let us summarize what we have learned so far. The number of massless states in
$SU(N)$ theory at resolution $K<N^2-1$ is the same as one for $SU(\infty)$
theory if the multitrace states are included in the latter. On the other hand
due to the fact that multitrace massless states in $SU(\infty)$ have special
structure (namely any single trace in them is massless state itself), their
number can be calculated from the known number of massless states written as
linear combination of single traces. The remaining part of this subsection is
devoted to such calculation.

As we found in the end of the last subsection there are $2(K-1)$ single trace
massless states at resolution $K$. We will show how this information can be
used in order to count the total number of massless states. Let us introduce
some notation first. The value we want to calculate is $N_k$ --- the number
of massless states at resolution $k$. We also define $N^{(m)}_k$ as number
of such massless states at resolution $k$ that the resolution of any
single traces in them is greater or equal to $m$. Then for example
$N_k=N^{(2)}_k$ and $N^{(k)}_k=2(k-1)$. Finally we define $f_n(m)$ to be the
number of different massless states containing $n$ traces, each of which
corresponds has resolution {\it equal} to $m$. Then one can derive the
recurrent relation:
\begin{equation}\label{nmbst_rec1}
N^{(m)}_k=N^{(m+1)}_k+\sum_{n=1}^{[k/m]} f_n(m)
           \left(N^{(m+1)}_{k-mn}+\delta^{k}_{mn}\right).
\end{equation}
The starting point of the recurrent procedure are the relations
$N^{(k)}_k=2(k-1)$ and $N^{(m)}_k=0$ for $m>k$. In order to apply
(\ref{nmbst_rec1}) we only have to evaluate $f_n(m)$. This will
be our next task.

To calculate $f_n(m)$ we first assume that we have only bosonic traces at
our disposal. Let us count the states which contain $p$ such traces. After
combining identical traces together one can reduce the problem further
by considering only states with special trace structure. Namely we will
concentrate our attention on the massless states having $n_i$ traces of type
$i$, for different values of $i=1\dots r$.
Without the loss of generality we can require $n_1\ge\dots\ge n_r\ge 1$, one
can also see that the relation $\sum n_i =p$ holds. If all $n_i$ are different
then the number of massless states with fixed structure is given by simple
formula:
$$
g(m)\left(g(m)-1\right)\dots\left(g(m)-r+1\right),
$$
where $g(m)=m-1$ is the number of bosonic single trace massless states. In
general one gets additional combinatoric coefficient $C(n_1,\dots,n_r)$ in
the last expression, it is defined by the following rules:
\begin{eqnarray}\label{CoMs_conbin}
C(n_1,\dots,n_i,n_{i+1},\dots,n_r)&=&C(n_1,\dots,n_i)C(n_{i+1},\dots,n_r),
\quad {\mbox{if}} \quad n_i>n_{i+1},\nonumber\\
C(\underbrace{n,\dots,n}_a)&=&\frac{1}{a!}.
\end{eqnarray}
Now let us include fermionic traces in the picture. The only difference
between them and bosons is the Pauli principle, so considering the product
of $q$ fermionic traces one has to choose all of them to be different.
Thus the coefficient $C$ for this case is
\begin{equation}
C_F (q)=C(\underbrace{1,\dots,1}_q)=\frac{1}{q!}.
\end{equation}

Collecting all the information together we finally get:
\begin{eqnarray}\label{CoMs_coeff}
f_n(m)&=&\sum_{q=0}^{n}\frac{1}{q!}g(m)\left(g(m)-1\right)\dots
          \left(g(m)-q+1\right)\times\\
&\times&\sum_{r=0}^{n-q}{\cal F}(n-q,r)g(m)\left(g(m)-1\right)\dots
         \left(g(m)-r+1\right),\nonumber\\
\label{CoMs_coeF}
{\cal F}(p,r)&=&\sum_{\{n_1,\dots,n_r\}} C(n_1,\dots,n_r),\qquad
\begin{array}{c}
n_1+\dots+n_r=p\\
n_1\ge\dots\ge n_r\ge 1.\\
\end{array}
\end{eqnarray}

Now we can use the relations (\ref{CoMs_coeff}), (\ref{CoMs_coeF}) and
(\ref{CoMs_conbin}) to determine all the coefficients $f_n(m)$ and then
substituting them to the recurrent relation (\ref{nmbst_rec1}) one can find
all the $N^{(m)}_k$. This in turn leads to the results for $N_k=N^{(2)}_k$.
Although we were not able to find an analytic expression for $N_k$ as
function of $k$, the number of states can be evaluated numerically for
arbitrary $k$ using the procedure we just described. We performed such
calculations using Mathematica and the results for lowest resolutions are
summarized in the Table\ref{CoMs_table}. For instance one can see that up to
resolution $5$ $N_k=2^{k-1}$, but at higher resolutions this relations holds
only approximately.
\begin{table}[h!]
\begin{center}
\begin{tabular}{|c|c|c|c|c|c|c|c|c|c|c|c|c|}
\hline
$k$  &2&3&4&5 &6 &7 &8  &9  &10 &11 &12\\
\hline
$N_k$&2&4&8&16&32&60&114&212&384&692&1232\\
\hline
\end{tabular}
\caption{Number of multitrace massless states as function of
resolution.
\label{CoMs_table}}
\end{center}
\end{table}

\section{ \bf Correlation Functions in SYM and DLCQ.}
\label{ChMaldac}
\renewcommand{\theequation}{5.\arabic{equation}}
\setcounter{equation}{0}

The bound state problem we have studied so far is the traditional one
for DLCQ. However this is not the only calculation that can be done using
this method. The problem of computing of correlation functions, more
traditional for conventional quantum field theory, can also be addressed
in the light cone quantization. Unlike the usual methods of QFT the results
of DLCQ calculations are valid beyond the perturbation theory and thus they
can be used for the testing the duality between the gauge theory and
supergravity.

There has been a great deal of excitement during this past year
following the realization that certain field theories admit concrete
realizations as a string theory on a particular background
\cite{adscft}. By now many examples of this type of correspondence for
field theories in various dimensions with various field contents have
been reported in the literature (for a comprehensive review and list
of references, see \cite{agmoo}).  However, attempts to apply these
correspondences to study the details of these theories have only met
with limited success so far. The problem stems from the fact that our
understanding of both sides of the correspondence is limited. On the
field theory side, most of what we know comes from perturbation theory
where we assume that the coupling is weak. On the string theory side,
most of what we know comes from the supergravity approximation where
the curvature is small.  There are no known situations where both
approximations are simultaneously valid. At the present time,
comparisons between the dual gauge/string theories have been
restricted to either qualitative issues or quantities constrained by
symmetry. Any improvement in our understanding of field theories
beyond perturbation theory or string theories beyond the supergravity
approximation is therefore a welcome development.

We will study the field theory/string theory correspondence motivated
by considering the near-horizon decoupling limit of a D1-brane in type
IIB string theory \cite{IMSY}. The gauge theory corresponding to this
theory is the Yang-Mills theory in two dimensions with 16
supercharges.  Its SDLCQ formulation was recently reported in
\cite{alppt}. This is probably the simplest known example of a field
theory/string theory correspondence involving a field theory in two
dimensions with a concrete Lagrangian formulation.

A convenient quantity that can be computed on both sides of the
correspondence is the correlation function of gauge invariant
operators \cite{GKP,Wit}. We will focus on two point functions of
the stress-energy tensor.  This turns out to be a very convenient quantity
to compute for many reasons that we will explain along the way.  Some
aspects of this as it pertains to a consideration of black hole entropy
was recently discussed in \cite{akisunny}. There are other physical
quantities often reported in the literature. In the DLCQ literature,
the spectrum of hadrons is often reported.  This would be fine for
theories in a confining phase. However, we expect the SYM in two
dimension to flow to a non-trivial conformal fixed point in the
infra-red \cite{IMSY,DVV}.  The spectrum of states will therefore form
a continuum and will be cumbersome to handle.  On the string theory
side, entropy density \cite{BISY} and the quark anti-quark potential
\cite{BISY,RY,juanwilson} are frequently reported. The definition of
entropy density requires that we place the field theory in a
space-like box which seems incommensurate with the discretized light
cone.  Similarly, a static quark anti-quark configuration does not fit
very well inside a discretized light-cone geometry.  The correlation
function of point-like operators do not suffer from these problems. We
should mention that there exists interesting work on computing the QCD
string tension \cite{adi1,adi2} directly in the field theory. These
authors find that the QCD string tension vanishes in the
supersymmetric theories which is consistent with the power law quark
anti-quark potential found on the supergravity side. This section is based
on the results of paper \cite{alphash99}.

\subsection{Correlation Functions from Supergravity}

Let us begin by reviewing the computation of the correlation function
of stress energy tensors on the string theory side using the
supergravity approximation.  The computation is essentially a
generalization of \cite{GKP,Wit}.  The main conclusion on the
supergravity side was reported recently in \cite{akisunny} but we will
elaborate further on the details. The near horizon geometry of a
D1-brane in string frame takes the form
\begin{eqnarray}
ds^2& =& \alpha' \left( {U^3 \over  \sqrt{64 \pi^3 g_{YM} ^2 N}}
dx_\parallel^2 + { \sqrt{64 \pi^3 g_{YM}^2 N} \over U^{3}} dU^2 + \sqrt{64
\pi^3 g_{YM}^2 N} U d \Omega_{8-p}^2 \right) \nonumber \\
e^\phi & = & 2 \pi  g_{YM}^2 \left( {64 \pi^3 g_{YM}^2  N \over U^6}
\right)^{{1 \over 2}} .
\end{eqnarray}
In order to compute the two point function, we need to know the action
for the diagonal fluctuations around this background to the quadratic
order. What we need is an analogue of \cite{KRvN} for this background
which unfortunately is not currently available in the
literature. Fortunately, some diagonal fluctuating degrees of freedom
can be identified by following the early work on black hole absorption
cross-subsections \cite{krasnitz1,krasnitz2}. In particular, we can show
that the fluctuations parameterized according to
\begin{eqnarray}
ds^2 & = & \left(1 +  f(x^0,U) +  g(x^0,U) \right) g_{00} (dx^0)^2  +
\left(1 +5 f(x^0,U) +  g(x^0,U)\right) g_{11} (dx^1)^2  \nonumber \\
&& + \left(1 +  f(x^0,U) +  g(x^0,U)\right) g_{UU} dU^2 + \left(1 +
f(x^0,U) -    {5 \over 7} g(x^0,U)\right) g_{\Omega\Omega} d \Omega_7^2
\nonumber \\
e^\phi &=& \left(1 + 3 f(x^0,U) - g(x^0,U) \right) e^{\phi_0}
\end{eqnarray}
will satisfy the equations of motion
\begin{eqnarray}
f''(U)  + {7 \over U}  f'(U) - {64 \pi^3 g_{YM}^2  N k^2 \over U^{6}} f(U)
&=&  0 \nonumber \\
g''(U)  +  {7 \over U} g'(U)- {72 \over U^2} g(U)  - {64 \pi^3 g_{YM}^2  N
k^2 \over U^6} g(U)   &=&  0 \label{fgeq}
\end{eqnarray}
by direct substitution into the equations of motion in 10
dimensions. We have assumed without loss of generality that these
fluctuation vary only along the $x^0$ direction of the world volume
coordinates like a plane wave $e^{i k x^0}$. The fields $f(U)$ and
$g(U)$ are scalars when the D1-brane is viewed as a black hole in 9
dimensions; in fact there are the minimal and the fixed scalars in
this black hole geometry. In 10 dimensions, however, we see that they
are really part of the gravitational fluctuation. We expect therefore
that they are associated with the stress-energy tensor in the operator
field correspondence of \cite{GKP,Wit}. In the case of the
correspondence between ${\cal N} = 4$ SYM and $AdS_5 \times S_5$,
superconformal invariance allowed the identification of operators and
fields in short multiplets \cite{ferrara}. For the D1-brane, we do not
have superconformal invariance and this technique is not
applicable. In fact, we expect all fields of the theory consistent
with the symmetry of a given operator to mix.  The large distance
behavior should then be dominated by the contribution with the longest
range. The field $f(k^0,U)$ appears to be the one with the longest
range since it is the lightest field.

The  equation (\ref{fgeq}) for $f(U)$ can be solved explicitly in terms of the
Bessel's function
\begin{equation}
f(U) = U^{-3}  K_{3/2} (  \sqrt{16 \pi^3 g_{YM}^2 N}  U^{-2} k  ).
\end{equation}
By thinking of $f(U)$ in direct analogy with the minimally coupled
scalar as was done in \cite{GKP,Wit}, we can compute the flux factor
\begin{equation} {\cal F} = \lim_{U_0 \rightarrow \infty} \left. {1 \over 2
\kappa_{10}^2} \sqrt{g} g^{UU} e^{-2 (\phi - \phi_{\infty})} \partial_U
\log( f(U))  \right|_{U = U_0} = {N U_0^2 k^2\over 2 g_{YM}^2} - {N^{3/2}
k^3 \over 4 g_{YM}} + \ldots
\end{equation}
up to a numerical coefficient of order one which we have suppressed.
We see that the leading non-analytic (in $k^2$) contribution is due to
the $k^3$ term, whose Fourier transform scales according
to\footnote{It is not difficult to show that for a generic $p$-brane,
$\langle {\cal O}(x){\cal O}(0) \rangle = {{N^{{7-p} \over 5-p}}
g_{YM}^{-{2 (3-p) \over 5-p}} x^{-{19+2 p - p^2 \over 5-p}}}.$}
\begin{equation}
\langle {\cal O}(x) {\cal O} (0) \rangle = {N^{{3 \over 2}} \over
g_{YM} x^5}. \label{SG}
\end{equation}
This result passes the following important consistency test.  The SYM
in 2 dimensions with 16 supercharges have conformal fixed points in
both UV and IR with central charges of order $N^2$ and $N$,
respectively. Therefore, we expect the two point function of stress
energy tensors to scale like $N^2/x^4$ and $N/x^4$ in the deep UV and
IR, respectively. According to the analysis of \cite{IMSY}, we expect
to deviate from these conformal behavior and cross over to a regime
where supergravity calculation can be trusted. The cross over occurs
at $x = 1 / g_{YM} \sqrt{N}$ and $x = \sqrt{N} / g_{YM}$. At these
points, the $N$ scaling of (\ref{SG}) and the conformal result match
in the sense of the correspondence principle \cite{garyjoe}.

\subsection{Correlation functions from DLCQ}

The challenge then is to attempt to reproduce the scaling relation
(\ref{SG}), fix the numerical coefficient, and determine the detail
of the cross-over behavior using SDLCQ.  Ever since the original
proposal
\cite{Dirac}, the question of equivalence between quantizing on a
light-cone and on a space-like slice have been discussed
extensively. This question is especially critical whenever a
massless particle or a zero-mode in the quantization is present.  It
is generally believed that the massless theories can be described on
the light-cone as long as we take $m\rightarrow 0$ as a limit. The
issue of zero mode have been examined by many authors. Some recent
accounts can be found in
\cite{zm1,zm2,aptzm,alptzm,zm5}. Generally speaking, supersymmetry seems
to save SDLCQ from complicated zero-mode issues.  We will not
contribute much to these discussions. Instead, we will formulate the
computation of the correlation function of stress energy tensor in
naive DLCQ.  To check that these results are sensible, we will first
do the computation for the free fermions.
Extension to SYM with 16 supercharges will be essentially
straightforward, except for one caveat. In order to actually
evaluate the correlation functions, we must resort to numerical
analysis at the last stage of the computation. For the SYM with 16
supercharges, this problem grows too big too fast to be practical on
desk top computer where the current  calculations were performed.
We can only provide an algorithm,  which, when executed on an much
more powerful computer, should reproduce (\ref{SG}).
Nonetheless, the fact that we can define a concrete algorithm seems
to be a progress in the right direction.  One potential pit-fall is
the fact that the computation may not show any sign of convergence.
If this is the case, or if it converges to a result at odds with
(\ref{SG}), we must go back and re-examine the issue of equivalence
of forms and the issue of zero modes.

The technique of DLCQ is reviewed by many authors \cite{BPP,KresIgor}
so we will be brief here.  The basic idea of light-cone quantization
is to parameterize the space using light cone coordinates $x^+$ and
$x^-$ and to quantize the theory making $x^+$ play the role of time.
In the discrete light cone approach, we require the momentum $p_- =
p^+$ along the $x^-$ direction to take on discrete values in units of
$p^+/k$ where $p^+$ is the conserved total momentum of the system and
$k$ is an integer commonly referred to as the harmonic resolution.
One can think of this discretization as a consequence of compactifying
the $x^-$ coordinate on a circle with a period $2L = {2 \pi k /
p^+}$. The advantage of discretizing the light cone is the fact that
the dimension of the Hilbert space becomes finite.  Therefore, the
Hamiltonian is a finite dimensional matrix and its dynamics can be
solved explicitly.  In SDLCQ one makes the DLCQ approximation to the
supercharges and these discrete representations satisfy the
supersymmetry algebra. Therefore SDLCQ enjoys the improved
renormalization properties of supersymmetric theories.  Of course, to
recover the continuum result, we must send $k$ to infinity and as luck
would have it, we find that SDLCQ usually converges faster than the
naive DLCQ. Of course, in the process, the size of the matrices will
grow, making the computation harder and harder.

Let us now return to the problem at hand. We would like to compute a
general expression of the form
\begin{equation}
F(x^-,x^+) = \langle {\cal O}(x^-,x^+) {\cal O} (0,0) \rangle \ .
\end{equation}
In DLCQ, where we fix the total momentum in the $x^-$ direction, it is
more natural to compute its Fourier transform
\begin{equation}
\tilde{F}(P_-,x^+) = {1 \over 2 L} \langle {\cal O}(P_-,x^+) {\cal O}(-P_-,
0) \rangle\ .
\end{equation}
This can naturally be expressed in a spectrally decomposed form
\begin{equation}
\tilde{F}(P_-,x^+)= \sum_i {1 \over 2 L} \langle 0| {\cal O}(P_-) | i
\rangle e^{-i P_+^i x^+} \langle i|  {\cal O}(-P_-,0) |0 \rangle\ .
\label{master}
\end{equation}

\subsection{Correlator for Free Dirac Fermions}

Let us first consider evaluating this expression for the stress-energy
tensor in the theory of free Dirac fermions as a simple example. The
Lagrangian for this theory is
\begin{equation}
{\cal L} = i \bar{\Psi} \partial \!\!\!/\, \Psi - m \bar{\Psi} \Psi
\end{equation}
where for concreteness, we take $\gamma^0 = \sigma^2, \gamma^1 = i
\sigma^1$ and we take $\Psi = 2^{-1/4} ({\psi \atop \chi})$. In terms of
the spinor components, the Lagrangian takes the form
\begin{equation}
{\cal L} = i \psi^* \partial_+ \psi + i \chi^* \partial_- \chi - {i m \over
\sqrt{2}} (\chi^* \psi - \psi^* \chi) \ .
\end{equation}
Since we treat $x^+$ as time and since $\chi$ does not have any
derivatives with respect to $x^+$ in the Lagrangian, it can be
eliminated from the equation of motion, leaving a Lagrangian which
depends only on $\psi$:
\begin{equation}
{\cal L} = i \psi^* \partial_+ \psi + i {m^2  \over 2} \psi^* { 1 \over
\partial_- } \psi \ .
\end{equation}
We can therefore express the canonical momentum and energy  as
\begin{eqnarray}
P_- & = &  \int dx^-\,
  i \psi^* \partial_- \psi \nonumber \\
P_+ & = & \int dx^-\, -{i m^2 \over 2} \psi^* { 1 \over \partial_- } \psi \ .
\end{eqnarray}
In DLCQ, we compactify $x^-$ to have period $2L$. We can then expand
$\psi$ and $\psi^*$ in modes
\begin{eqnarray}
\psi ={ 1 \over \sqrt{2L}} \left( b(n) e^{-{i n \pi  \over L} x^-} + d(-n)
e^{ { i n \pi \over L} x^-} \right) \nonumber \\
\psi^* ={ 1 \over \sqrt{2L}} \left( b(-n) e^{{i n \pi  \over L} x^-} + d(n)
e^{ -{ i n \pi \over L} x^-} \right) \ .
\end{eqnarray}
Operators $b(n)$ and $d(n)$ with positive and negative $n$ are
interpreted as a destruction and creation operators, respectively. In
a theory with only fermions, it is customary to take anti-periodic
boundary condition in order to avoid zero-mode issues. Therefore, $n$
will take on half-integer values\footnote{In SDLCQ one must use
periodic boundary condition for all the fields to preserve the
supersymmetry.}. They satisfy the anticommutation relation
\begin{equation}
\{ b(n), b(-m) \} =  \{ d(n), d(-m) \} =  \delta_{n,m} \ .
\end{equation}
Now we are ready to evaluate (\ref{master}) in DLCQ. As a simple and
convenient choice, we take
\begin{equation}
{\cal O}(-k) = {1 \over 2}\int dx^- \, \left( i \psi^* \partial_- \psi - i
(\partial_- \psi^*) \psi \right)
  e^{- {i k \pi \over L} x^-} . \label{operator}
\end{equation}
which is the Fourier transform of the local expression for $P_-$ with
the total derivative contribution adjusted to make this operator
Hermitian. Therefore, this should be thought of as the $T^{++}$
component of the stress energy tensor. For reasons that will become
clear as we go on, this turns out to be one of the simplest things to
compute. When acted on the vacuum, this operator creates a state
\begin{equation}
T^{++}(-k) |0 \rangle =   {\pi \over L} \left( { k \over 2} - n \right)
b(-k+n) d(-n) |0 \rangle  \ . \label{state}
\end{equation}
Since the fermions in this theory are free, the plane wave states
\begin{equation}
|n \rangle = b(-k+n) d(-n) |0 \rangle
\end{equation}
constitute an eigenstate. The spectrum can easily be determined by
commuting these operators:
\begin{equation}
M^2_n | n \rangle = 2 P_- P_+ | n \rangle = {m^2 }  \left( {k \over n} + {k
\over k-n} \right) | n \rangle \label{mass}
\end{equation}
which is simply the discretized version of the spectrum of a two body
continuum. All that we have to do now is calculate eigenstates of the
actual theory we are interested in and to assemble these pieces into
(\ref{master}), but we can do a little more to make the result more
presentable. The point is that since (\ref{master}) is expressed in mixed
momentum/position space notation in Minkowski space, the answer is
inherently a complex quantity that is cumbersome to display.  For the
computation of two point function, however, we can go to position space
by Fourier transforming with respect to the $L$ variable. After Fourier
transforming, it is straight forward to Euclideanize and display the two
point function as a purely real function without loosing any information.
To see how this works, let us write (\ref{master}) in the form
\begin{equation}
\tilde{F}(P_-,x^+)=
\left|{L \over \pi} \langle 0 | T^{++}(k) |n \rangle \right|^2
{1 \over 2L} {\pi^2 \over L^2} e^{-i M^2_n \over 2 ({k \pi \over L})x^+} \ .
\end{equation}
The quantity inside the absolute value sign is designed to be
independent of $L$. Now, to recover the position space form of the
correlation function, we inverse Fourier transform with respect to $P_- = k
\pi/ L$.
\begin{equation}
F(x^-,x^+)=
\left|{L \over \pi} \langle 0 | T^{++}(k) |n \rangle \right|^2
\int {d \left({k \pi\over  L}\right) \over 2 \pi}
{1 \over 2L} {\pi^2 \over L^2} e^{-i{ M^2_n \over 2 ({k \pi \over L})}x^+ -
i {k \pi \over L} x^-} .
\end{equation}
The integral over $L$ can be done explicitly and gives
\begin{equation}
F(x^-,x^+)=
\left|{L \over \pi} \langle 0 | T^{++}(k) |n \rangle \right|^2
\left({x^+ \over x^-}\right)^2 {M_n^4 \over 8 \pi^2 k^3}
K_4\left(M_n\sqrt{2 x^+ x^-}\right)
\end{equation}
where $K_4(x)$ is the 4-th modified Bessel's function. We can now
continue to Euclidean space by taking $r^2 = 2 x^+ x^-$ to be real and
considering the quantity
\begin{equation}
\left({x^- \over x^+}\right)^2 F(x^-,x^+)=
\left|{L \over \pi} \langle 0 | T^{++}(k) |n \rangle \right|^2
{M_n^4 \over 8 \pi^2 k^3} K_4(M_n r) \label{general} \ .
\end{equation}
This is a fundamental result which we will refer to a number of times
in this paper.  It has explicit dependence on the harmonic resolution
parameter $k$, but all dependence on unphysical quantities such as
the size of the circle in the $x^-$ direction and the momentum along
that direction have been canceled. For the free fermion model,
(\ref{general}) evaluates to
\begin{equation}
\left({x^- \over x^+}\right)^2 F(x^-,x^+)
= {N \over k} \sum_n  {M_n^4 \over 32 \pi^2} {(k-2n)^2 \over k^2} K_4(M_n r)
\end{equation}
with $M_n^2$ given by (\ref{mass}). The large $k$ limit can be gotten
by replacing $n \rightarrow k x$ and ${1 \over k} \sum_n \rightarrow
\int_0^1 dx$. We recover the identical result using Feynman rules. For
$r \ll m^{-1}$, this behaves like
\begin{equation}
\left({x^- \over x^+}\right)^2 F(x^-,x^+) = {N \over k} \sum_n {3 (k-2n)^2
\over 2  \pi^2 k^2 r^4} \rightarrow {N \over 2 \pi^2 r^4} \ .
\end{equation}

\subsection{Correlator for Supersymmetric Yang-Mills Theory with 16
          Supercharges.}

Finally, let us turn to the problem of computing the two point
function of the $T^{++}$ operator for the SYM with 16 supercharges.
Adopting light-cone coordinates and
choosing the light-cone gauge will eliminate the gauge boson and half
of the fermion degrees of freedom.  The most significant change comes
from the fact that the fields in this theory are in the adjoint rather
than the fundamental representations and the theory is
supersymmetric. This does not cause any fundamental problem in the
DLCQ formulation of these theories. Indeed, the SDLCQ formulation of
this \cite{alppt} as well as many other related models with adjoint
fields have been studied in the literature. The main difficulty comes
from the fact that in supersymmetric theories low mass states such as
${\rm tr} [b(-n_1) b(-n_2) b(-k+n_1 + n_2)] | 0 \rangle$ with an
arbitrary number of excited quanta, or ``bits,'' appear in the
spectrum. This means that for a given harmonic resolution $k$, the
dimension of the Hilbert space grows like $\exp(\sqrt{k})$, which is roughly
the number of ways to partition $k$ into sums of integers.

The fact that the size of the problem grows  very fast is somewhat
discouraging from a numerical perspective.  Nevertheless, it is
interesting to note that DLCQ provides a well defined algorithm for
computing a physical quantity like the two point function of $T^{++}$
that can be compared with the prediction from supergravity. In the
following, we will show that this can be computed for the SYM theory
by a straight forward application of (\ref{general}).

As we found in section \ref{ChLCGauge} the momentum operator $P^+$
is given by
\begin{equation}
P^+ = \int dx^- {\rm tr} \left[ (\partial_- X_I)^2 + i u_\alpha \partial_-
u_\alpha \right].
\end{equation}
The local Hermitian form of this operator is given by
\begin{equation}
T^{++}(x) =  {\rm tr} \left[ (\partial_- X^I)^2 + {1 \over 2} \left(i
u^\alpha \partial_- u^\alpha  - i  (\partial_- u^\alpha) u^\alpha
\right)\right], \qquad I = 1 \ldots 8, \quad \alpha = 1 \ldots 8
\end{equation}
where $X$ and $u$ are the physical adjoint scalars and fermions
respectively, following the notation of \cite{alppt}.  When
discretized, these operators have the mode expansion
\begin{eqnarray}
X_{i,j}^I & = & {1 \over \sqrt{4 \pi}} \sum_{n=1}^{\infty} {1 \over
\sqrt{n}} \left[ A^I_{ij} (n) e^{-i \pi n x^-/L} + A^I_{ji}(-n) e^{i \pi n
x^-/L} \right]\nonumber \\
u_{i,j}^\alpha & = & {1 \over \sqrt{4 L}} \sum_{n=1}^{\infty} \left[
B^\alpha_{ij} (n) e^{-i \pi n x^-/L} + B^\alpha_{ji}(-n) e^{i \pi n x^-/L}
\right] .
\end{eqnarray}
In terms of these mode operators, we find
\bea
\label{susycorr}
&&T^{++}(-k) | 0 \rangle = \\
&&{\pi \over 2 L} \sum_{n=1}^{k-1}
\left[-  \sqrt{n ( k-n)}  A_{ij}(-k+n) A_{ji} (-n)
+  \left({k \over 2}
- n\right) B_{ij}(-k+n) B_{ji}(-n) \right] | 0 \rangle  .\nonumber
\eea
Therefore, $(L/\pi) \langle 0 | T^{++}(-k) | n \rangle$ is independent
of $L$ and can be substituted directly into (\ref{general}) to give an
explicit expression for the two point function.

We see immediately that (\ref{susycorr}) has the correct small $r$
behavior, for in that limit, (\ref{susycorr}) asymptotes to (assuming
$n_b = n_f$)
\bea
&&\left({x^- \over x^+ }\right)^2 F(x^-,x^+) =\\
&&{N^2 \over k} \sum_n \left( {3 (k-2 n)^2 n_f \over 4 \pi^2 k^2 r^4}
+ {3 n (k-n) n_b \over  \pi^2 k^2 r^4}\right) = {N^2(2 n_b + n_f) \over 4
\pi^2 r^4}
\left(1 - {1 \over k}\right)\nonumber
\eea

which is what we expect for the theory of $n_b$ free bosons and $n_f$
free fermions in the large $k$ limit.

\begin{figure}
%!!!!!!!!!!!!!!!!!!!!!!!!!!>>>>
\begin{tabular}{cc}
\epsfxsize=2.8in  \epsfbox{8.8.mass.epsi} &
\epsfxsize=2.8in
\epsfbox{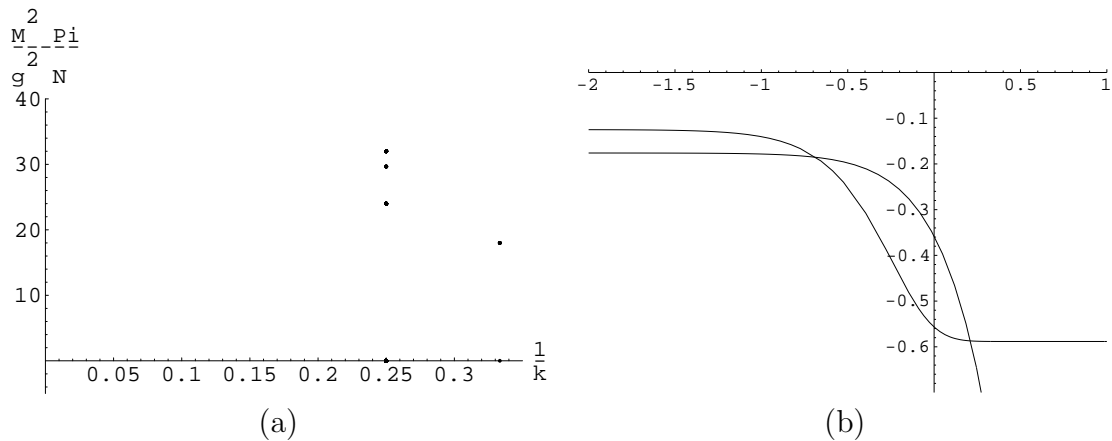} \\
(a) & (b)
\end{tabular}
%<<<<!!!!!!!!!!!!!!!!!!!!!!!!!
\caption{(a) The spectrum as a function of $1/k$ and (b) the Log-Log
plot of the correlation function $\langle T^{++}(x) T^{++}(0) \rangle
\left({x^- \over x^+} \right)^2 {4 \pi^2 r^4 \over N^2 (2 n_b +n_f)}$
v.s. $r$ in the units where $g_{YM}^2 N /\pi = 1$ for $k=3$ and $k=4$.
\label{figb}}
\end{figure}

Computing this quantity beyond the small $r$ asymptotics, however,
represents a formidable technical challenge. In \cite{alppt} we
constructed the mass matrix explicitly and
compute the spectrum for $k=2$, $k=3$, and $k=4$. Even for these
modest values of the harmonic resolution, the dimension of the Hilbert
space was as big as 256, 1632, and 29056 respectively (the symmetries
of the theory can be used to reduce the size of the calculation
somewhat). In figure \ref{figb}, we display the results with
the currently available values of $k$, except for the fact that we
display the correlation function multiplied by a factor of $ 4 \pi^2
r^4 / N^2 (2 n_b + n_f)$, so that it now asymptotes to 1 (or 0 in the
logarithmic scale) in the $k \rightarrow \infty$ limit. In this way
any deviation from the asymptotic behavior $1/r^4$ is made more
transparent. Note that with the values of the harmonic resolution $k$
obtained at present, the spectrum in figure \ref{figb}.a is far from
resembling a dense continuum near $M=0$. Clearly, we must probe much
higher values of $k$ before we can sensibly compare our field theory
results with the prediction from supergravity.

\subsection{Supersymmetric Yang-Mills Theories with Less Than 16 Supercharges}

The computation of the correlator for the stress energy tensor in the
(8,8) model is limited by our inability to carry out the computation
for large enough harmonic resolution. It is the (8,8) model which we
are ultimately interested in solving in order to compare against the
prediction of Maldacena's conjecture in the supergravity
limit. Nevertheless, the computation of the correlation function can
just as well be applied to models with less supersymmetry.  We will
conclude by reporting the results of such a computation.

First, let us consider the theory with supercharges (1,1).  This
theory is argued not to exhibit dynamical supersymmetry breaking in
\cite{Li95,Oda97}. We can also provide a physicist's proof that
supersymmetry is not spontaneously broken for this theory by adopting
the argument of Witten for the 2+1 dimensional SYM with Chern-Simons
interaction \cite{Witten99}. In \cite{Witten99}, the index of 2+1
dimensional SYM with gauge group $SU(N)$ and 2 supercharges on $R
\times T^2$ was computed and was found to be non-vanishing for
Chern-Simons coupling $k_3 > N/2$.  If the period $L$ of one of the
circles in $T^2$ is sufficiently small, this theory is approximately
the 2-dimensional SYM with (1,1) supersymmetry with gauge coupling
$g_2^2 = g_3^2 /L$ and BF coupling $k_2 = k_3 L$ \cite{BF}. Imagine
approaching this theory by taking $L \rightarrow 0$ keeping $g_2$ and
$k_3$ fixed.  In this limit, $k_2 \rightarrow 0$ in the units of $g_2$
so the limiting theory is that of pure SYM with (1,1) supersymmetry
and a vanishing BF coupling. Choosing different values of $k_3$
corresponds to a different choice in the path of approach to this
limit. If we chose $k_3 > N/2$, we are guaranteed to have a non-zero
index for finite $L$. This means that there will be a state with zero
mass in the $L \rightarrow 0$ limit also, indicating that
supersymmetry is not spontaneously broken in this limit. On the other
hand, the index is not a well defined quantity in the $L\rightarrow 0$
limit, as a different choice of $k_3$ will lead to a different value
of the index in the $L \rightarrow 0$ limit.  In fact, the index can
be made arbitrarily large by taking $k_3$ to be also arbitrarily
large.  This suggests that there are infinitely many states forming a
continuum near $m=0$.  The index is therefore an ill defined quantity,
akin to counting the number of exactly zero energy states on a
periodic box as one takes the volume to infinity.

This theory is also believed not to be confining \cite{adi1,adi2} and
is therefore expected to exhibit non-trivial infra-red dynamics.

\begin{figure}
\begin{tabular}{cc}
\epsfxsize=2.8in  \epsfbox{1.1.mass.epsi} &
\epsfxsize=2.8in  \epsfbox{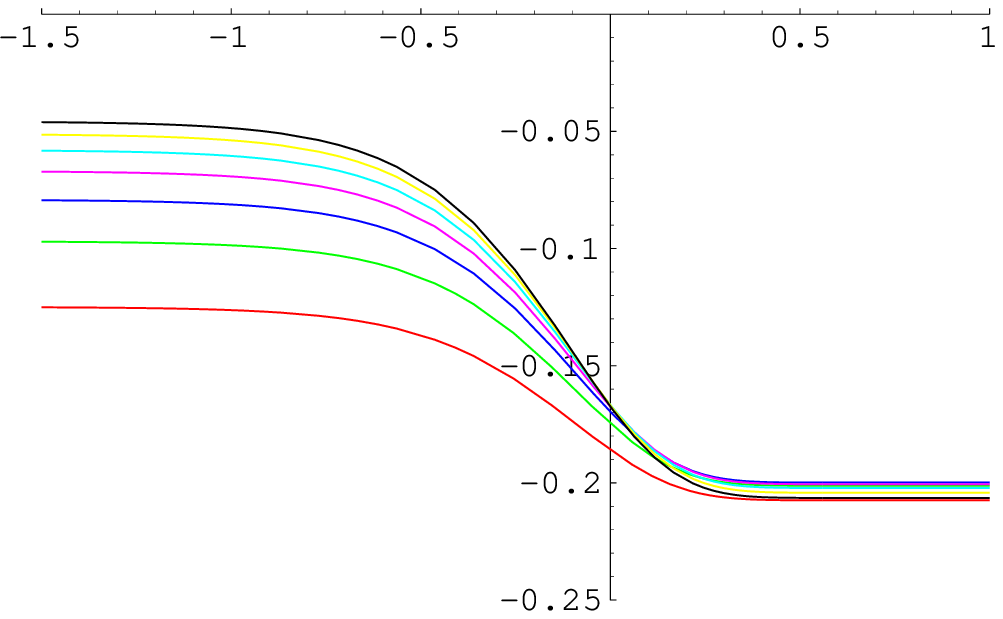} \\
(a) & (b)
\end{tabular}
%<<<<!!!!!!!!!!!!!!!!!!!!!!!!!!!!!!!!!!!!!!!!
\caption{(a) The spectrum as a function of $1/k$ and (b) the Log-Log
plot of the correlation function $\langle T^{++}(x) T^{++}(0) \rangle
\left({x^- \over x^+} \right)^2 {4 \pi^2 r^4 \over N^2 (2 n_b +n_f)}$
v.s. $r$ in the units where $g_{YM}^2 N /\pi = 1$ for $k=4 \ldots 10$.
\label{figc}}
\end{figure}

The SDLCQ of the 1+1 dimensional model with (1,1) supersymmetry
was solved in \cite{sakai,alp2}, and we apply these results
directly in order to compute (\ref{general}).  For simplicity, we
work to leading order in the large $N$ expansion. The spectrum of
this theory for various values of $k$, and the subsequent
computation of (\ref{general}) is illustrated in figure
\ref{figc}.a.

%!!!!!!!!!!!!!!!!!!!!!!!!!!!!!!!!>>>

The spectrum of this theory at finite $k$, illustrated in figure
\ref{figc}.a, consists of $2k-2$ exactly massless states\footnote{
i.e. $k-1$ massless bosons, and their superpartners.}, accompanied
by large numbers of massive states separated by a gap. The gap appears
to be closing in the limit of large $k$ however.  We have tried
extrapolating the mass of the lightest massive state as a function of
$1/k$ by performing a least square fit to a line and a parabola, giving the
extrapolated value of $M^2 \pi / g_{YM}^2 N = 1.7$ and $M^2 \pi/g_{YM}^2 N =
-0.6$, suggesting indeed that at large $k$, the gap is closed. This is
consistent with the expectation that the spectrum is that of a
continuum starting at $M=0$ discussed earlier, although one must be
careful when the order of large $N$ and large $k$ limits are
exchanged. At finite $N$, we expect the degeneracy of $2k-2$ exactly
massless states to be broken, giving rise to precisely a continuum
of states starting at $M=0$ as expected.

%!!!!!!!!!!!!!!!!!!!!!!!!!!!!!!!!>>>
\begin{figure}
%\centerline{\psfig{file=massless.1.1.epsi,width=3in}}
\begin{center}
\epsfxsize=3in  \epsfbox{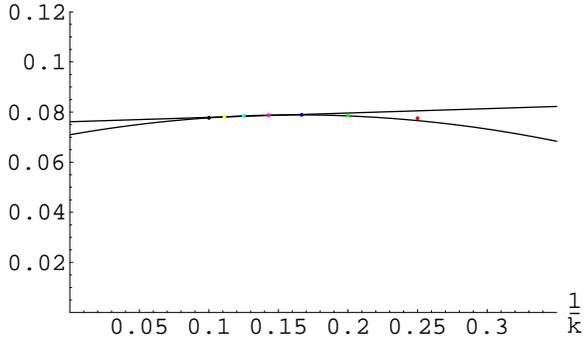}
\end{center}
\caption{$ {1 \over  k^3} \sum_{n} |{L \over \pi} \langle 0 | T^{++}(k) | n
\rangle
|^2$~v.s.~$k$ from states with $M|n\rangle = 0$.  This quantity
determines the coefficient of the $1/r^4$ asymptotic tail of the
correlation function in the large $r$ limit for the (1,1) model.
\label{figd}}
\end{figure}
%<<<!!!!!!!!!!!!!!!!!!!!!!!!!!!!!!!!!!!!

In the computation of the correlation function illustrated in figure
\ref{figc}.b, we find a curious feature that it asymptotes to the inverse
power law $c/r^4$ for large $r$. This behavior comes about due to the
coupling $\langle 0 | T^{++} | n \rangle$ with exactly massless states
$|n \rangle$. The contribution to (\ref{general}) from strictly massless
states are given by
\newpage
\bea
&&\left( x^- \over x^+ \right)^2 F(x^-,x^+) = \left. \left| {L \over \pi}
\langle 0 | T^{++}(k) | n \rangle \right|^2 {M_n^4 \over 8 \pi^2 k^3 }
K_4(M_n r) \right|_{M_n =0} =\\
&&\left| {L \over \pi} \langle 0 | T^{++}(k) | n \rangle \right|^2_{M_n=0}
{6 \over k^3 \pi^2 r^4} .\nonumber
\eea
We have computed this quantity as a function of $1/k$ and extrapolated
to $1/k \rightarrow 0$ by fitting a line and a parabola to the
computed values for finite $1/k$. The result of this extrapolation is
illustrated in figure \ref{figd}.  The data currently available
suggests that the non-zero contribution from these massless states
persists in the large $k$ limit.

\begin{figure}
\begin{tabular}{cc}
\epsfxsize=2.8in  \epsfbox{2.2.mass.epsi} &
\epsfxsize=2.8in  \epsfbox{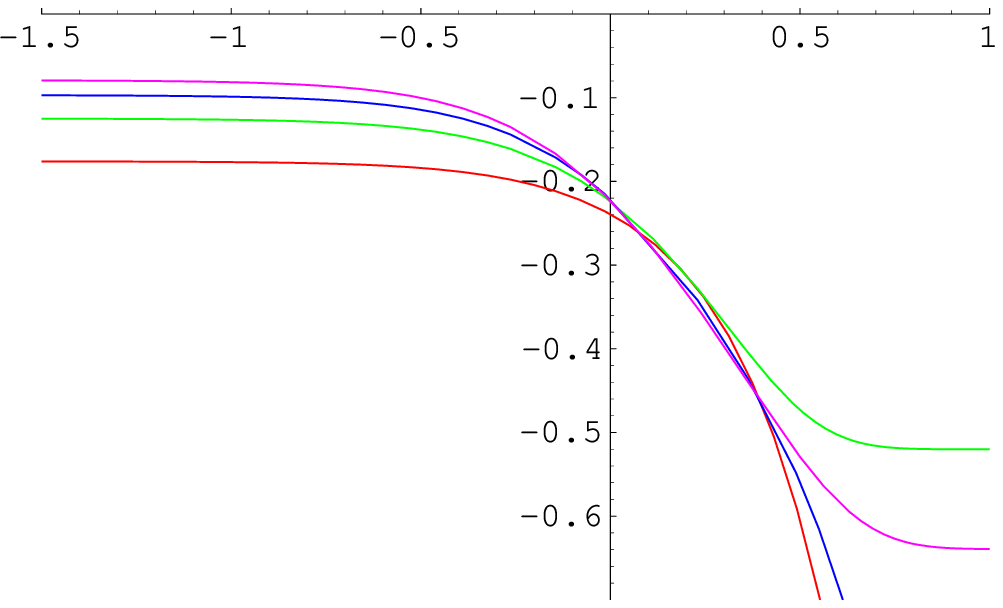} \\
(a) & (b)
\end{tabular}
\caption{(a) The spectrum as a function of $1/k$ and (b) the Log-Log
plot of the correlation function $\langle T^{++}(x) T^{++}(0) \rangle
\left({x^- \over x^+} \right)^2 {4 \pi^2 r^4 \over N^2 (2 n_b +n_f)}$
v.s. $r$ in the units where $g_{YM}^2 N / \pi = 1$ for $k=3 \ldots 6$.
\label{fige}}
\end{figure}

Let us now turn to the model with (2,2) supersymmetry. The SDLCQ
version of this model was solved in \cite{appt}. The result of this
computation can be applied to (\ref{general}). The result is
summarized in figure \ref{fige}. This model appears to exhibit the
onset of a gapless continuum of states more rapidly than the (1,1)
model as the harmonic resolution $k$ is increased. Just as we found in
the (1,1) model, this theory contains exactly massless states in the
spectrum.  These massless states appear to couple to $T^{++} | 0
\rangle$ only for $k$ even, and the overlap appears to be decreasing
as $k$ is increased. We believe that this model is likely to exhibit a
power law behavior $c/r^\gamma$ for $\gamma > 4$ for the $T^{++}$
correlator for $r \gg g_{YM} \sqrt{N}$ in the large $N$
limit. Unfortunately, the existing numerical data do not permit the
reliable computation of the exponent $\gamma$.

%!!!!!!!!!!!!!!!!!!!!!!!!!!!!!!!>>>
%<<<!!!!!!!!!!!!!!!!!!!!!!!!

\section{ \bf Bound States of Three Dimensional Supersymmetric Theory.}
\label{Ch3D}
\renewcommand{\theequation}{6.\arabic{equation}}
\setcounter{equation}{0}

Recently, there has been considerable
progress in understanding the properties of strongly coupled
gauge theories with supersymmetry
\cite{seibergwitten,seiberg}. In particular, there
are a number of supersymmetric gauge theories that are believed
to be inter-connected through a web of strong-weak coupling dualities.
Although these dualities provide a deep insight into the dynamics of
gauge theory at strong and weak couplings, they do not usually give much
information about the spectrum of bound states at intermediate values
of coupling constant $g$. The prominent exception is so called BPS states
whose mass is protected by supersymmetry and thus stays the same for all
values of $g$. An interesting new possibility for analytical treatment of
bound state problem in SYM$_{D}$ is based on the duality between SYM and
supergravity proposed by Maldacena \cite{adscft} as discussed in the previous
section. This idea was also exploited in \cite{coo98} to get the glueball
spectrum
of three dimensional theory and the results agree with lattice calculations.

However it would be interesting to solve the bound state problem for SYM
theory directly, starting from the first principles of quantum field theory.
As we have seen in the previous sections the solution can be found for the
various two dimensional theories by means of applying discrete light cone
quantization.
Evidently, it would be desirable to extend
these DLCQ/SDLCQ algorithms to solve higher dimensional theories.
One important difference between two dimensional and higher
dimensional theories is the phase diagram induced by
variations in the gauge coupling. The
spectrum of a $1+1$ dimensional gauge theory
scales trivially with
respect to the gauge coupling, while a theory in higher dimensions
has the potential of exhibiting a complex phase structure,
which may include a strong-weak coupling duality.
It is therefore interesting to study
the phase diagram of gauge theories in $D \geq 3$ dimensions.

Towards this end, we consider
three dimensional SU($N$) ${\cal N}=1$ super-Yang-Mills
compactified on the space-time ${\bf R} \times S^1 \times S^1$.
In particular, we compactify the light-cone coordinate
$x^-$ on a light-like circle via DLCQ, and wrap the remaining
transverse coordinate $x^{\perp}$ on a spatial circle.
By retaining only the first few excited modes in the
transverse direction, we are able to solve for bound state
wave functions and masses numerically by diagonalizing the
discretized light-cone
supercharge.
We show that the supersymmetric formulation
of the DLCQ procedure -- which was
studied in the context of two dimensional theories
extends naturally in $2+1$ dimensions, resulting in an exactly
supersymmetric spectrum.

%%%%%%%%%%%%%%%%%%%%%%%%%%%%%%%%%%%%%%%%%%%%%%%
\subsection{Light-Cone Quantization and SDLCQ}
\label{formulation}
We wish to study the bound states of ${\cal N} =1$
super-Yang-Mills in $2+1$ dimensions.
Any numerical approach necessarily involves introducing a
momentum lattice -- i.e.
parton momenta can only take on discretized values.
The usual space--time lattice explicitly breaks supersymmetry,
so if we wish to discretize our theory {\em and} preserve supersymmetry,
then a judicious choice of lattice is required.

In $1+1$ dimensions, it is well known that the light-cone
momentum lattice
induced by the DLCQ procedure
preserves supersymmetry if the supercharge rather than the
Hamiltonian is discretized \cite{sakai,alp99}.
In $2+1$ dimensions, a supersymmetric prescription
is also possible. We begin by introducing light-cone
coordinates $x^{\pm} = (x^0 \pm x^1)/\sqrt{2}$, and compactifying
the $x^-$ coordinate on a light-like circle.
In this way, the conjugate light-cone momentum $k^+$ is discretized.
To discretize the remaining (transverse) momentum $k^{\perp} = k^2$,
we may compactify $x^{\perp} = x^2$ on a spatial circle.
Of course, there is a significant
difference between the discretized light-cone momenta $k^+$,
and discretized transverse momenta $k_{\perp}$; namely,
the light-cone momentum $k^+$ is always positive\footnote{Since we wish
to consider the decompactified limit in the end, we omit zero
modes. This is a necessary technical constraint in
numerical calculations.}, while $k_{\perp}$ may take on positive
or negative values. The positivity of $k^+$ is a key property
that is exploited in DLCQ calculations;
for any given light-cone compactification,
there are only a finite number of choices for $k^+$ -- the
total number depending on how finely we discretize the
momenta\footnote{The `resolution' of the discretization is
usually characterized by a positive integer $K$, which is called
the `harmonic resolution' \cite{BP85,MaYam}; for a given choice of $K$,
the light-cone momenta $k^+$ are restricted to positive integer
multiples of $P^+/K$, where $P^+$ is the total light-cone momentum
of a state}.
In the context of two dimensional theories, this implies a finite
number of Fock states \cite{BP85}.

In the case
we are interested in -- in which there is an additional transverse
dimension -- the number of Fock states is no longer finite,
since there are an arbitrarily large number of transverse momentum
modes defined on the transverse spatial circle.
Thus, an additional truncation of the transverse momentum
modes is required to render the total number of Fock states
finite, and the problem numerically tractable\footnote{
This truncation procedure, which is characterized by some
integer upper bound, is analogous to the truncation of $k^+$
imposed by the `harmonic resolution' $K$.}.
In this work, we choose the simplest truncation procedure
beyond retaining the zero mode; namely,
only partons with transverse momentum $k_\perp=0,\pm\frac{2 \pi}{L}$
will be allowed, where $L$ is the size of the transverse circle.

Let us now apply these ideas in the context of a specific
super-Yang-Mills theory.
We start with $2+1$ dimensional ${\cal N}=1$ super-Yang-Mills theory
defined on a space-time with one transverse dimension
compactified on a circle:
\be
S=\int d^2 x \int_0^L dx_\perp \mbox{tr}(-\frac{1}{4}F^{\mu\nu}F_{\mu\nu}+
{\rm i}{\bar\Psi}\gamma^\mu D_\mu\Psi).
\ee
After introducing the light--cone coordinates
$x^\pm=\frac{1}{\sqrt{2}}(x^0\pm x^1)$, decomposing the spinor $\Psi$
in terms of chiral projections --
\be
\psi=\frac{1+\gamma^5}{2^{1/4}}\Psi,\qquad
\chi=\frac{1-\gamma^5}{2^{1/4}}\Psi
\ee
and choosing the light--cone gauge $A^+=0$, the action becomes
\bea\label{action}
S&=&\int dx^+dx^- \int_0^L dx_\perp \mbox{tr}\left[\frac{1}{2}(\d_-A^-)^2+
D_+\phi\d_-\phi+ {\rm i}\psi D_+\psi+ \right.\nonumber \\
& &
\left.
  \hspace{15mm} +{\rm i}\chi\d_-\chi+\frac{{\rm i}}{\sqrt{2}}\psi D_\perp\phi+
\frac{{\rm i}}{\sqrt{2}}\phi D_\perp\psi \right].
\eea
A simplification of the
light--cone gauge is that the
non-dynamical fields $A^-$ and $\chi$ may be explicitly
solved from their Euler-Lagrange equations of motion:
\bea
A^-&=&\frac{g}{\d_-^2}J=
\frac{g}{\d_-^2}\left(i[\phi,\d_-\phi]+2\psi\psi\right),\\
\chi&=&-\frac{1}{\sqrt{2}\d_-}D_\perp\psi.\nonumber
\eea

These expressions may be used to express any operator
in terms of the physical degrees of freedom only.
In particular, the light-cone energy, $P^-$, and momentum
operators, $P^+$,$P^{\perp}$,
corresponding to  translation
invariance in each of the coordinates
$x^\pm$ and $x_\perp$ may be calculated explicitly:
\bea\label{moment}
P^+&=&\int dx^-\int_0^L dx_\perp\mbox{tr}\left[(\d_-\phi)^2+
{\rm i}\psi\d_-\psi\right],\\
P^-&=&\int dx^-\int_0^L dx_\perp\mbox{tr}
\left[-\frac{g^2}{2}J\frac{1}{\d_-^2}J-
     \frac{{\rm i}}{2}D_\perp\psi\frac{1}{\d_-}D_\perp\psi\right],\\
P_\perp &=&\int dx^-\int_0^L dx_\perp\mbox{tr}\left[\d_-\phi\d_\perp\phi+
     {\rm i}\psi\d_\perp\psi\right].
\eea
The light-cone supercharge in this theory
is a two component Majorana spinor, and may be conveniently
decomposed in terms of its chiral projections:
\bea\label{sucharge2}
Q^+&=&2^{1/4}\int dx^-\int_0^L dx_\perp\mbox{tr}\left[\phi\d_-\psi-\psi\d_-
            \phi\right],\\
Q^-&=&2^{3/4}\int dx^-\int_0^L dx_\perp\mbox{tr}\left[2\d_\perp\phi\psi+
     g\left({\rm i}[\phi,\d_-\phi]+2\psi\psi\right)\frac{1}{\d_-}\psi\right].
\eea
The action (\ref{action}) gives the following canonical
(anti)commutation relations for
propagating fields at equal $x^+$:
\begin{eqnarray}
\left[\phi_{ij}(x^-,x_\perp),\d_-\phi_{kl}(y^-,y_\perp)\right]
&=&
\frac{1}{2}{\rm i}\delta(x^- -y^-)\delta(x_\perp -y_\perp)
\left( \delta_{il}\delta_{jk} - \frac{1}{N}\delta_{ij}\delta_{kl} \right),
\\
\left\{\psi_{ij}(x^-,x_\perp),\psi_{kl}(y^-,y_\perp)\right\}
&=&
\frac{1}{2}\delta(x^- -y^-)\delta(x_\perp -y_\perp)
\left( \delta_{il}\delta_{jk} - \frac{1}{N}\delta_{ij}\delta_{kl} \right).
\label{comm}
\end{eqnarray}

Using these relations one can check the supersymmetry algebra:
\be
\{Q^+,Q^+\}=2\sqrt{2}P^+,\qquad \{Q^-,Q^-\}=2\sqrt{2}P^-,\qquad
\{Q^+,Q^-\}=-4P_\perp.
\label{superr}
\ee

We will consider only states which have vanishing transverse momentum,
which is possible since the total transverse momentum operator
is kinematical\footnote{Strictly speaking, on a transverse
cylinder, there are separate sectors with total
transverse momenta $2\pi n/L$; we consider only one of them, $n=0$.}.
On such states, the light-cone supercharges
$Q^+$ and $Q^-$ anti-commute with each other, and the supersymmetry algebra
is equivalent to the ${\cal N}=(1,1)$ supersymmetry
of the dimensionally reduced (i.e. two dimensional) theory \cite{sakai}.
Moreover, in the $P_{\perp} = 0$ sector,
the mass squared operator $M^2$ is given by
$M^2=2P^+P^-$.

As we mentioned earlier, in order to render the bound state equations
numerically tractable, the transverse
momentum of partons must be truncated.
First, we introduce the Fourier expansion for the fields $\phi$ and $\psi$,
where the transverse space-time coordinate $x^{\perp}$ is periodically
identified:
\bea
\lefteqn{
\phi_{ij}(0,x^-,x_\perp) =} & & \nonumber \\
& &
\frac{1}{\sqrt{2\pi L}}\sum_{n^{\perp} = -\infty}^{\infty}
\int_0^\infty
  \frac{dk^+}{\sqrt{2k^+}}\left[
  a_{ij}(k^+,n^{\perp})e^{-{\rm i}k^+x^- -{\rm i}
\frac{2 \pi n^{\perp}}{L} x_\perp}+
  a^\dagger_{ji}(k^+,n^{\perp})e^{{\rm i}k^+x^- +
{\rm i}\frac{2 \pi n^{\perp}}{L}  x_\perp}\right]
\nonumber\\
\lefteqn{
\psi_{ij}(0,x^-,x_\perp) =} & & \nonumber \\
& & \frac{1}{2\sqrt{\pi L}}\sum_{n^{\perp}=-\infty}^{\infty}\int_0^\infty
  dk^+\left[b_{ij}(k^+,n^{\perp})e^{-{\rm i}k^+x^- -
{\rm i}\frac{2 \pi n^{\perp}}{L} x_\perp}+
  b^\dagger_{ji}(k^+,n^\perp)e^{{\rm i}k^+x^- +{\rm i}
\frac{2 \pi n^{\perp}}{L} x_\perp}\right]
\nonumber
\eea
Substituting these into the (anti)commutators (\ref{comm}),
one finds:
\begin{eqnarray}
\left[a_{ij}(p^+,n_\perp),a^\dagger_{lk}(q^+,m_\perp)\right]
&=&
\delta(p^+ -q^+)\delta_{n_\perp,m_\perp}
\left( \delta_{il}\delta_{jk} - \frac{1}{N}\delta_{ij} \delta_{lk} \right)
\\
\left\{b_{ij}(p^+,n_\perp),b^\dagger_{lk}(q^+,m_\perp)\right\}
&=&
\delta(p^+ -q^+)\delta_{n_\perp,m_\perp}
\left( \delta_{il}\delta_{jk} - \frac{1}{N}\delta_{ij} \delta_{lk} \right).
\end{eqnarray}
The supercharges now take the following form:
\bea\label{TruncSch}
&&Q^+={\rm i}2^{1/4}\sum_{n^{\perp}\in {\bf Z}}\int_0^\infty dk\sqrt{k}\left[
b_{ij}^\dagger(k,n^\perp) a_{ij}(k,n^\perp)-
a_{ij}^\dagger(k,n^\perp) b_{ij}(k,n^\perp)\right],\\
\label{Qminus5}
&&Q^-=\frac{2^{7/4}\pi {\rm i}}{L}\sum_{n^{\perp}\in {\bf Z}}\int_0^\infty dk
\frac{n^{\perp}}{\sqrt{k}}\left[
a_{ij}^\dagger(k,n^\perp) b_{ij}(k,n^\perp)-
b_{ij}^\dagger(k,n^\perp) a_{ij}(k,n^\perp)\right]+\nonumber\\
&&+ {{\rm i} 2^{-1/4} {g} \over \sqrt{L\pi}}
\sum_{n^{\perp}_{i} \in {\bf Z}} \int_0^\infty dk_1dk_2dk_3
\delta(k_1+k_2-k_3) \delta_{n^\perp_1+n^\perp_2,n^\perp_3}
\left\{ \frac{}{} \right.\nonumber\\
&&{1 \over 2\sqrt{k_1 k_2}} {k_2-k_1 \over k_3}
[a_{ik}^\dagger(k_1,n^\perp_1) a_{kj}^\dagger(k_2,n^\perp_2)
b_{ij}(k_3,n^\perp_3)
-b_{ij}^\dagger(k_3,n^\perp_3)a_{ik}(k_1,n^\perp_1)
a_{kj}(k_2,n^\perp_2) ]\nonumber\\
&&{1 \over 2\sqrt{k_1 k_3}} {k_1+k_3 \over k_2}
[a_{ik}^\dagger(k_3,n^\perp_3) a_{kj}(k_1,n^\perp_1) b_{ij}(k_2,n^\perp_2)
-a_{ik}^\dagger(k_1,n^\perp_1) b_{kj}^\dagger(k_2,n^\perp_2)
a_{ij}(k_3,n^\perp_3) ]\nonumber\\
&&{1 \over 2\sqrt{k_2 k_3}} {k_2+k_3 \over k_1}
[b_{ik}^\dagger(k_1,n^\perp_1) a_{kj}^\dagger(k_2,n^\perp_2)
a_{ij}(k_3,n^\perp_3)
-a_{ij}^\dagger(k_3,n^\perp_3)b_{ik}(k_1) a_{kj}(k_2,n^\perp_2) ]\nonumber\\
&& ({ 1\over k_1}+{1 \over k_2}-{1\over k_3})
[b_{ik}^\dagger(k_1,n^\perp_1) b_{kj}^\dagger(k_2,n^\perp_2)
b_{ij}(k_3,n^\perp_3)
+b_{ij}^\dagger(k_3,n^\perp_3) b_{ik}(k_1,n^\perp_1) b_{kj}(k_2,n^\perp_2)]
  \left. \frac{}{}\right\}. \nonumber \\
\eea
We now perform the truncation procedure; namely,
in all sums over the transverse momenta $n^{\perp}_{i}$
appearing in the above expressions for the supercharges, we
restrict summation to the following allowed momentum
modes: $n^{\perp}_{i}=0,\pm 1$. More generally, the truncation procedure
may be defined by $|n^{\perp}_{i}|\le N_{max}$, where $N_{max}$ is
some positive integer. In this work, we consider the simple
case $N_{max}=1$.
Note that this prescription is
symmetric, in the sense that there are as many positive modes as
there are negative ones. In this way we
retain parity symmetry in the transverse
direction.

How does such a truncation affect the
supersymmetry properties of the theory?
Note first that an operator relation $[A,B]=C$ in the
full theory is not expected to hold in the truncated formulation.
However, if A is
quadratic in terms of fields (or in terms of creation and
annihilation operators),
one can show that the relation $[A,B]=C$ implies
$$
[A_{tr},B_{tr}]=C_{tr}
$$
for the truncated operators $A_{tr}$,$B_{tr}$, and $C_{tr}$.
In our case, $Q^+$ is quadratic, and so the relations
$\{Q_{tr}^+,Q_{tr}^+\}=2\sqrt{2}P_{tr}^+$ and $\{Q_{tr}^+,Q_{tr}^-\}=0$ are
true in the $P_\perp=0$ sector of the truncated theory.
The $\{Q_{tr}^-,Q_{tr}^-\}$
however is not equal to $2\sqrt{2}P_{tr}^-$. So the diagonalization of
$\{Q_{tr}^-,Q_{tr}^-\}$ will yield a different bound state spectrum
than the one obtained after diagonalizing $2\sqrt{2}P_{tr}^-$.
Of course the two spectra should agree in the limit
$N_{max}\rightarrow\infty$. At any finite truncation,
however, we have the liberty to
diagonalize any one of these operators.
This choice specifies our regularization scheme.

Choosing to diagonalize the light-cone
supercharge, however, has an important advantage:
{\em the spectrum is exactly supersymmetric for
any truncation}. In contrast, the spectrum of the Hamiltonian becomes
supersymmetric only in the $N_{max}\rightarrow\infty$
limit\footnote{If one chooses
anti-periodic boundary conditions in the $x^-$ coordinate
for fermions, then there is no choice; one can only
diagonalize the light-cone Hamiltonian.
See \cite{dak93} for more details on this approach.}.

To summarize, we have introduced a truncation procedure
that facilitates a numerical study of the bound state problem, and
preserves supersymmetry.
The interesting property of the light-cone supercharge $Q^-$
[Eqn(\ref{Qminus5})] is the
presence of a gauge coupling constant as an independent variable,
which does not appear in the study of two dimensional theories.
In the next subsection, we will study how
variations in this coupling affects the bound states
in the theory.

%%%%%%%%%%%%%%%%%%%%%%%%%%%%%%%%%%%%%%%%%%%%%%%%%%%
\subsection{Numerical Results: Bound State Solutions}
\label{numerical}
In order to implement the DLCQ formulation of the bound state problem -- which
is tantamount to imposing periodic boundary conditions
$x^- = x^- + 2 \pi R$ \cite{MaYam} -- we simply restrict the
light-cone momentum variables $k_i$ appearing in the expressions for $Q^+$
and $Q^-$ to the following discretized set of momenta:
$\left\{ \frac{1}{K} P^+, \frac{2}{K} P^+, \frac{3}{K} P^+,\dots,
\right\}$. Here, $P^+$ denotes the total light-cone momentum of a
state, and may be thought of as a fixed constant, since it is
easy to form a Fock basis that is already diagonal with respect to the
operator $P^+$ \cite{BP85}. The integer $K$ is called the `harmonic
resolution', and $1/K$ measures the coarseness of our discretization.
The continuum limit is then recovered by taking the limit
$K \rightarrow \infty$. Physically, $1/K$ represents the smallest
positive unit of longitudinal momentum-fraction allowed for each
parton in a Fock state.

%%%%%%%%%%%%%%%%%%%%%%%%%%
%!!!!!!!!!!!!!!!!!!!!!!!!!!!>>>
\begin{figure}[h]
\begin{center}
\epsfbox{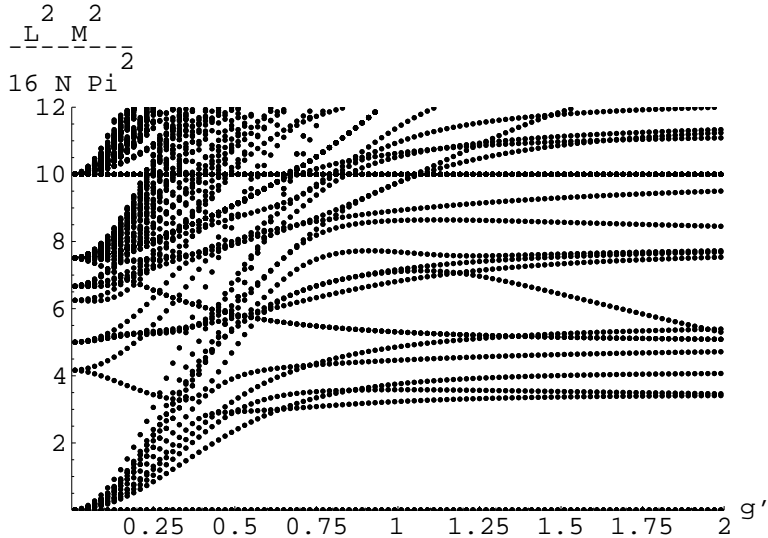}
\end{center}
\caption{{\small Plot of bound state mass squared $M^2$
in units $16 \pi^2 N/ L^2$
as a function of the dimensionless coupling $0 \leq g' \leq 2$,
defined by $(g')^2 = g^2 N L/16 \pi^3$,
at $N=1000$ and $K=5$. Boson and fermion masses are identical.
} \label{mass1}}
\end{figure}
%%%%%%%%%%%%%%%%%%%%%%%%%%

Of course, as soon as we implement the DLCQ procedure, which is
specified unambiguously by the harmonic resolution $K$,
and cut-off transverse momentum modes via the constraint
$|n_i^{\perp}| \leq N_{max}$,
the integrals
appearing in the definitions for $Q^+$ and $Q^-$ are replaced by finite sums,
and so the eigen-equation $2P^+P^-|\Psi\rangle = M^2 |\Psi\rangle$
is reduced to a finite matrix diagonalization problem.
In this last step we use the fact that $P^-$ is proportional
to the square of the light-cone supercharge\footnote{
Strictly speaking, $P^- = \frac{1}{\sqrt{2}}(Q^-)^2$ is an identity
in the continuum theory, and a {\em definition} in the compactified
theory, corresponding to the SDLCQ prescription \cite{sakai,alp99}.}
$Q^-$. In the present work, we are able to perform numerical
diagonalizations for $K=2,3,4$ and $5$ with the help of Mathematica and a
desktop PC.
%%%%%%%%%%%%%%%%%%%%%%%%%%
%!!!!!!!!!!!!!!!!!!>>>
\begin{figure}[ht]
\begin{center}
\epsfbox{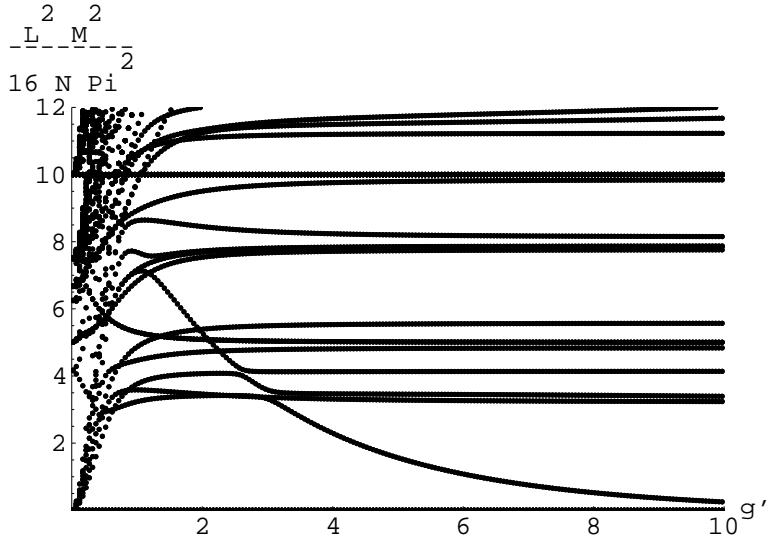}
\end{center}
\caption{{\small Plot of bound state mass squared $M^2$
in units $16 \pi^2 N/ L^2$
as a function of the dimensionless coupling $0 \leq g' \leq 10$,
defined by $(g')^2 = g^2 N L/16 \pi^3$,
at $N=1000$ and $K=5$. Note the appearance of a new massless
state at strong coupling.} \label{mass2}}
\end{figure}
%%%%%%%%%%%%%%%%%%%%%%%%%%
In Figure \ref{mass1}, we plot the bound state mass squared
$M^2$, in units $16 \pi^2 N/L^2$,
as a function of the dimensionless coupling
$g' = g \sqrt{N L}/4 \pi^{3/2}$, in the range $0 \leq g' \leq 2$.
We consider the specific case $N=1000$, although our algorithm
can calculate masses for any choice of
$N$, since it enters our calculations as an
algebraic variable.
Since there is an exact boson-fermion mass degeneracy,
one needs only diagonalize the mass matrix $M^2$
corresponding to the bosons. For $K=5$, there are precisely
600 bosons and 600 fermions in the truncated light-cone Fock space,
so the mass matrix that needs to be diagonalized has dimensions
$600 \times 600$. At $K=4$, there are $92$ bosons and $92$ fermions,
while at $K=3$, one finds $16$ bosons and $16$ fermions.

In Figure \ref{mass2}, we plot the bound state spectrum in
the range $0 \leq g' \leq 10$. It is apparent now that
new massless states appear in the strong coupling limit
$g' \rightarrow \infty$.
%%%%%%%%%%%%%%%%%%%%%%%%%%
%!!!!!!!!!!!!!!!!!!!!!!!>>>
\begin{figure}[h]
\begin{center}
\epsfbox{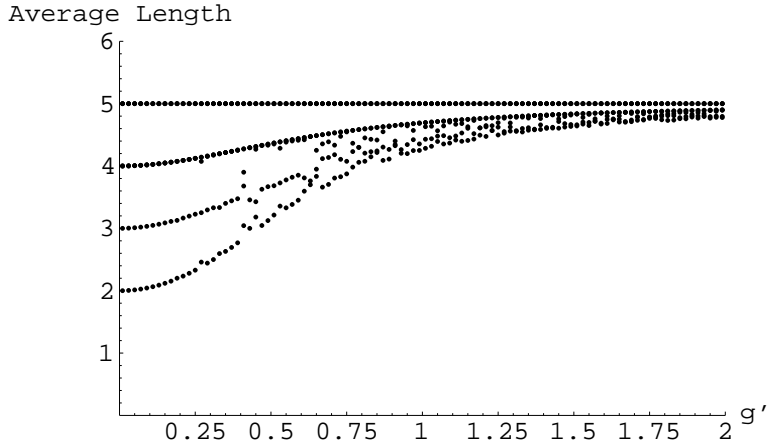}
\end{center}
\caption{{\small Plot of average length for the eight massless
bosonic states
as a function of the dimensionless coupling $g'$,
defined by $(g')^2 = g^2 N L/16 \pi^3$,
at $N=1000$ and $K=5$. Note that the states attain the maximum possible
length allowed by the resolution $K=5$
in the limit of strong coupling.} \label{lengths}}
\end{figure}
%%%%%%%%%%%%%%%%%%%%%%%%%%

An interesting property of the spectrum is the
presence of exactly massless states
that persist for all values of the coupling $g'$.
For $K=5$, there are 16 such states (8 bosons and 8 fermions).
At $K=4$, one finds 8 states (4 bosons and 4 fermions)
that are exactly massless
for any coupling, while for $K=3$, there are 4 states
(two bosons and two fermions) with
this property.
We will have more to say regarding these states in the next subsection,
but here we note that the structure of these states become
`string-like' in the strong coupling limit. This is illustrated
in Figure \ref{lengths}, where we plot the `average length'
(or average number of partons)
of each of these massless states\footnote{The `noisiness' in this
plot for larger values of $g'$ reflects the ambiguity of choosing
a basis for the eigen-space, due to the exact mass degeneracy
of the massless states.}.
This quantity is obtained
by counting the  number of partons in each  Fock state
that comprises a massless bound state, appropriately
weighted by the modulus of the wave function squared.
Clearly, at strong coupling, the average number of partons saturates
the maximum possible value allowed by the resolution -- in this
case 5 partons. The same behavior is observed at lower resolutions.
Thus, in the continuum limit $K \rightarrow \infty$, we expect
the massless states in this theory to become string-like
at strong coupling.
%%%%%%%%%%%%%%%%%%%%%%%%%%%%%%%%%%%%%%%%%%%%%%%%%%%%%%%
\begin{table}[h]
\begin{center}
\begin{tabular}{|c|c|c|c|c|}
\hline
\multicolumn{5}{|c|}{Comparison Between $1+1$ and $2+1$ Spectra} \\
\hline
- &
$1+1$ Model &
\multicolumn{3}{|c|}{Rescaled $2+1$ Model}
  \\
\hline
$K$ & - & $g'=.01$ & $g'=.1$ & $g'=1.0$ \\
\hline
$K=5$ & 15.63 & 15.5 & 15.17 & 3.7 \\
  & 18.23 & 17.6 & 17.9 & 3.5 \\
  & 21.8 & 21.3 & 21.7 & 3.2 \\
\hline
$K=4$ & - & - & - & - \\
  & 18.0 & 17.99 & 17.6 & 3.56 \\
  & 21.3 & 21.3 & 21.0 & 3.1 \\
\hline
$K=3$ & - & - & - & - \\
   & - & - & - & - \\
  & 20.2 & 20.2 & 19.8 & 3.1 \\
\hline
\end{tabular}
\caption{{\small Values for the mass squared $M^2$, in units
${\tilde g}^2 N/\pi$, with ${\tilde g}^2 = g^2/L$,
for bound states in the dimensionally
reduced ${\cal N}=(1,1)$ model, and
the $2+1$ model studied here.
The quantity ${\tilde g}$ is identified as the gauge
coupling in the $1+1$ model. We set $K=3,4$ and $5$, and $N=1000$.
Note that the comparison of masses between
the $1+1$ model, and the (re-scaled) $2+1$ model
is good only at weak coupling $g'$.}
\label{masspredictions}}
\end{center}
\end{table}
%%%%%%%%%%%%%%%%%%%%%%%%%%%%%%%%%%%%%%%%%%%%%%%%%

One interesting property of the model studied here is
the manifest ${\cal N}=(1,1)$
supersymmetry in the $P^{\perp}=0$ momentum sector, by virtue
of the supersymmetry relations (\ref{superr}).
Moreover, if we consider retaining only the zero mode $n_i^{\perp}=0$,
then the light-cone supercharge $Q^-$ for the $2+1$ model is identical
to the $1+1$ dimensional ${\cal N}=(1,1)$ supersymmetric
Yang-Mills theory studied in \cite{sakai,alp98a,alp2}, after a rescaling
by the factor $1/g'$.
(This is equivalent to expressing the mass squared
$M^2$ in units ${\tilde g}^2 N/\pi$, where ${\tilde g}=g/\sqrt{L}$.
The quantity ${\tilde g}$ is then identified as the gauge coupling
in the $1+1$ theory.)
  We may therefore think of the
additional transverse degrees of freedom in the $2+1$
model, represented by the modes $n^{\perp} = \pm 1$, as
a modification of the $1+1$ model. A natural question
that follows from this viewpoint is: How well does the
$1+1$ spectrum approximate the $2+1$ spectrum after
performing this rescaling? Before discussing the numerical
results summarized in Table \ref{masspredictions},
let us first attempt to predict what will happen
at small coupling $g'$. In this case, the coefficients
of terms in the rescaled Hamiltonian $P^-$ that
correspond to summing the
transverse momentum squared $|k^{\perp}|^2$ of partons in a state
will be large. So the low energy sector will be dominated
by states with $n^{\perp}=0$. i.e. those states that appear in the
Fock space of the ${\cal N}=(1,1)$ model in $1+1$ dimensions.
This is indeed supported by the results in Table \ref{masspredictions}.

For large coupling $g'$, however, it is clear that the
approximation breaks down. In fact, one can show that
the tabulated masses in the rescaled $2+1$ model
tend to zero in the strong
coupling limit, which  eliminates any scope for making
comparisons between the two and three dimensional models.

Thus, the non-perturbative problem of solving dimensionally
reduced models in $1+1$ dimensions can only provide information
about bound state masses in the corresponding {\em weakly
coupled} higher
dimensional theory.
%%%%%%%%%%%%%%%%%%%%%%%%%%%%%%%%%%%%%%%%%%%%%%%%%%%%%%%%%%%%%
\subsection{Analytical Results: The Massless Sector}
\label{analytical}

In the previous subsection we presented the results of studying the bound state
problem using numerical methods. In performing such a study we
conveniently chose the simplest nontrivial truncation
of the transverse momentum modes; namely,  $n_\perp=0,\pm 1$.
Surprisingly, such a simple scheme provided many interesting
insights concerning the massless and massive sector. In particular
we see that there are three types of massless states; those that are massless
only at $g=0$ or $g=\infty$ (but not both), and those
that are massless for any value of the coupling. In this subsection, we will
analyze only the massless sector of the theory, and show that the observed
properties of the
spectrum  with the truncation $n^{\perp}=0,\pm 1$ also persists
if we include higher modes: $n^{\perp} =0,\pm 1,\pm2,\dots,\pm N_{max}$.
We therefore consider
the model with supercharges given by (\ref{TruncSch}) and
(\ref{Qminus5}),
and restrict summation of transverse momentum modes
via the constraint $|n^{\perp}|\leq N_{max}$.

For states carrying positive light-cone momentum,
$P^+$ is never zero, and so massless states must satisfy
the equation $P^-|\Psi \rangle = 0$, which, using
the relation $P^-=\frac{1}{\sqrt{2}}(Q^-)^2$, and hermiticity of $Q^-$,
reduces to
\be\label{msls}
Q^-|\Psi\rangle=0.
\ee
This is the equation we wish to study in detail.

We begin with an analysis of the weak coupling limit of the theory.
This limit means that the dimensionless coupling constant
is small: i.e. $g \sqrt{L} \ll 1$.
We will
consider the strong--weak coupling behavior of the theory on a cylinder with
fixed circumference $L$ so it is convenient to choose the units in which
$L=1$ for this discussion.
The supercharge (\ref{Qminus5}) consists of two
parts: one is proportional to the coupling and the other is
coupling--independent:
\be\label{QminG}
Q^-=Q_{\perp}+g{\tilde Q}.
\ee
So at $g=0$, the equation (\ref{msls}) reduces to
$Q_{\perp}|\Psi\rangle=0$,
which means that $|\Psi\rangle$ may be viewed as a state
in the Fock space of the two dimensional
${\cal N}=(1,1)$ super Yang-Mills theory,
which may be obtained by dimensional reduction of the $2+1$ theory.
Thus the massless states at $g=0$ are states with any combination of
$a^\dagger(k,0)$ and $b^\dagger(k,0)$ modes, and no partons with nonzero
transverse
momentum.

What happens with these massless states when one switches on
the coupling? To
answer this question, we need some information about the
behavior of states
as functions of the coupling. We assume that wave functions are analytic
in terms of $g$ at least in the vicinity of $g=0$.
This means that in this region any
massless state $|\Psi\rangle$ may be written in the form:
\be
|\Psi\rangle=\sum_{n=0}^{\infty} g^n |n\rangle,
\ee
where states $|n\rangle$ are coupling independent.
Then using relation (\ref{QminG}),
the $g$--dependent equation (\ref{msls}) may be written as an
infinite system of
relations between different $|n\rangle$:
\bea
Q_\perp |0\rangle=0,\\
\label{InfSyst}
Q_\perp |n\rangle+{\tilde Q}|n-1\rangle=0, \qquad n>0.
\eea
The first of these equations was already used to exclude partons
carrying non-zero transverse momentum, which is a property
of the massless bound states at zero coupling.
The second equation is non-trivial. Let us
consider two different subspaces in the theory.
The first of these subspaces consists of
states with no creation operators for transverse modes which we will label
$1$. The other is the
complement of this space in which the operator $Q_\perp$ is invertible and
we label this space 2.
Equation (\ref{InfSyst}) defines the recurrence relation when ${\tilde Q}
|n-1\rangle$ is in subspace $2$:
\be\label{recur}
|n\rangle=-Q_\perp^{-1}\left(\left.{\tilde Q}
|n-1\rangle\right|_2\right),
\ee
The consistency condition is that  projection of ${\tilde Q}
|n-1\rangle$ in subspace $1$ is zero,
\be\label{consist}
\left.{\tilde Q}|n-1\rangle\right|_1=0.
\ee
This condition implies that not all states of the two dimension theory, $g=0$ ,
may be extended to such states at arbitrary $g$ using (\ref{recur}).
Taking $n=1$, (\ref{consist}) implies that $|0\rangle$
is a massless state of the  dimensionally reduced theory.  The numerical
solutions, of course, show
this correspondence between the  $2+1$ and
$1+1 $\cite{sakai,alp98a,alp2} massless bound states.  Starting from a
massless state of the two dimensional theory, and we construct states
$|n\rangle$ using (\ref{recur}), and for which (\ref{consist}) is always
satisfied.
Then $|\Psi\rangle$
may be found from summing a geometric series:
\be\label{3dmslsst}
|\Psi\rangle=\sum_{n=0}^\infty (-gQ_\perp^{-1}{\tilde Q})^n|0\rangle=
\frac{1}{1+gQ_\perp^{-1}{\tilde Q}}|0\rangle.
\ee
So, starting from the massless state of the two dimensional
${\cal N}=(1,1)$ model, one can always
construct unique massless states in the three dimensional theory
at least in the vicinity of $g=0$.

The state (\ref{3dmslsst}) turns out to be massless for any value of the
coupling:
\be
Q^-|\Psi\rangle=Q_{\perp}(1+gQ_\perp^{-1}{\tilde Q})
   \frac{1}{1+gQ_\perp^{-1}{\tilde Q}}|0\rangle=Q_{\perp}|0\rangle=0,
\ee
though the state itself is  dependent on g.
Thus, we have shown that massless states of the three
dimensional theory, at nonzero
coupling, can be constructed from massless states of
the corresponding model in two
dimensions. All other states containing only two dimensional modes can also be
extended to the eigenstates of the full theory. But such eigenstates are
massless only at zero coupling. Assuming analyticity, one can then
show that their masses grow linearly at $g$ in the vicinity of zero.
Such behavior also agrees with our numerical results.

To illustrate the general construction explained above we consider one simple
example. Working in DLCQ at resolution $K=3$ we choose a special
two dimensional massless state\footnote{
The state $|0 \rangle$ denotes a massless state,
while $|vac \rangle$ represents the light-cone vacuum.}
\cite{sakai,alp98a,alp2}:
\be
|0\rangle=\mbox{tr} (a^\dagger (1,0)a^\dagger (2,0))|vac\rangle.
\ee
Then in the SU($N$) theory we find:
\bea
{\tilde Q}|0\rangle&=&\frac{3}{2\sqrt{2}}\left[\mbox{tr}\left(a^\dagger (1,0)
(b^\dagger (1,-1)a^\dagger (1,1)-a^\dagger (1,1)b^\dagger (1,-1)+\right.\right.
\nonumber\\
&+&\left.\left.b^\dagger (1,1)a^\dagger (1,-1)-a^\dagger (1,-1)b^\dagger (1,1))
\right)
\right]|vac\rangle,\\
|1\rangle&=&-Q_\perp^{-1}{\tilde Q}|0>=
-\frac{\sqrt{L}}{4\pi^{3/2}}
\frac{3}{2\sqrt{2}}
\left(a^\dagger (1,0)a^\dagger (1,-1)a^\dagger (1,1)-\right.\nonumber\\
&-&\left. a^\dagger (1,0)a^\dagger (1,1)a^\dagger (1,-1)\right)
|vac\rangle \\
{\tilde Q}|1\rangle&=&0.
\eea
The last equation provides the consistency condition (\ref{consist}) for
$n=2$, and it also shows that for this special example we have only
two states
$|0\rangle$ and $|1\rangle$,
instead of a general infinite set.
The matrix form of the
operator $1+gQ_\perp^{-1}{\tilde Q}$ in the $|0\rangle , |1\rangle$
basis is
\be
1+gQ_\perp^{-1}{\tilde Q}=
\left(\begin{array}{cc}
1&-g\\0&1\end{array}\right)=
\left(\begin{array}{cc}
1&g\\0&1\end{array}\right)^{-1}.
\ee
Then the solution of (\ref{3dmslsst}) is
\bea
&&|\Psi\rangle=|0\rangle+g|1\rangle=
\mbox{tr} (a^\dagger (1,0)a^\dagger (2,0))|vac\rangle+\\
&&+\frac{g\sqrt{L}}{4\pi^{3/2}}\frac{3}{2\sqrt{2}}
\left(a^\dagger (1,0)a^\dagger (1,1)a^\dagger (1,-1)-
a^\dagger (1,0)a^\dagger (1,-1)a^\dagger (1,1)\right)|vac\rangle.
\nonumber
\eea
This state was observed numerically, and the dependence of the wave
function on the coupling constant is precisely the
one given by the last formula.

In principle, we can determine the wave functions of all massless states using
this formalism. Our procedure has an important advantage over a direct
diagonalization of the three dimensional supercharge. Firstly,
in order to find two dimensional massless states,
one needs to diagonalize the corresponding supercharge
\cite{sakai}.
However, the dimension of the relevant Fock space is
much less than the three dimensional theory
(at large resolution $K$, the ratio of these
dimensions is of order $(N_{max}+1)^{\alpha K}$, $\alpha\sim 1/4$). The
extension of the two dimensional massless solution
into a massless solution of the three dimensional theory
requires diagonalizing a matrix which has a smaller
dimension than the original problem in three dimensions.
Thus, if one is only interested in the massless sector
of the three dimensional theory, the most efficient
way to proceed in DLCQ calculations is to
solve the two dimensional theory, and then to
upgrade the massless states to massless solutions in three dimensions.

Finally, we will make some comments on
bound states at very strong coupling. Of
course, we have states (\ref{3dmslsst}) which are massless at any coupling,
but our numerical calculation show there are additional states which become
massless at $g=\infty$ (see Figure \ref{mass2}).
To discuss these state it is convenient to consider
\be
{\bar Q}^-=\frac{1}{g}Q_{\perp}+{\tilde Q}
\ee
instead of $Q^-$, and perform the strong coupling expansion. Since we are
interested only in massless states, the absolute
normalization doesn't matter.  We
repeat all the arguments used in the weak coupling case: first,
we introduce the space $1^*$  where ${\tilde Q}$ can not be inverted,
and its orthogonal complement $2^*$.
Then any state from $1^*$ is massless at $g=\infty$, but assuming the
expansion
\be
|\Psi\rangle=\sum_{n=0}^{\infty} \frac{1}{g^n} |n\rangle^*
\ee
at large enough $g$, one finds the analogs of (\ref{recur}) and
(\ref{consist}):
\bea
|n\rangle^*=-{\tilde Q}^{-1}\left(\left.Q_\perp|n-1\rangle^*
\right|_{2^*}\right),\\
\left.Q_\perp|n-1\rangle^*\right|_{1^*}=0.
\eea
As in the small coupling case, there are two possibilities: either
one can construct all states $|n\rangle^*$
satisfying the consistency conditions, or
at least one of these conditions fails.
The former case corresponds to the massless state in the vicinity
of $g=\infty$, which can be extended to the
massless states at all couplings.
The states constructed in this way -- and ones given by
(\ref{3dmslsst}) -- define the same subspace.
In the latter case,
the state is massless at $g=\infty$, but it acquires
a mass at finite coupling.
There is a big difference,
however, between the weak and strong coupling cases.
While the kernel of $Q_\perp$
consists of "two dimensional" states,
the description of the states annihilated by
${\tilde Q}$ is a nontrivial dynamical problem. Since the massless states
can be constructed starting from either $g=0$ or $g=\infty$,
we don't have to solve this problem to build them.
If, however, one wishes to show that
massless states become long in the strong coupling limit (there is numerical
evidence for such behavior -- see Figure \ref{lengths}),
the structure of $1^*$ space becomes important,
and we leave this question for future investigation.

\subsection*{Conclusion.}

In these lectures we have reviewed some of the progress in the application of
discrete light cone quantization to supersymmetric systems. Studying such
systems is especially interesting because the cancellation between bosonic
and fermionic loops make these theories much easier to renormalize than
the models without supersymmetry. Although we didn't need this advantage when
considering
two dimensional systems, it becomes crucial in higher dimensions. From
this point of view it is desirable to have exact SUSY  in
discretized theories to simplify the renormalization in DLCQ.

while we are still far from the point of solving the bound
state problem in three and four dimensional theories, we can
already  make some statements about these theories. For example in section
\ref{ChVac} we described the vacuum structure of
$SYM_{2+1}$ on a cylinder. The reason for this is that only zero
modes contribute to such structure, thus studying the theory dimensionally
reduced to $1+1$ provide all the necessary information. As we saw in section
\ref{Ch3D}, two dimensional models can also be used to determine the behavior
of bound states at weak coupling in three dimensions and to count the exact
massless
states. We performed such counting only for (1,1) theory, the case of (2,2)
supersymmetry \cite{appt} and the even more interesting case of the (8,8)
theory
\cite{alppt} which is known to have a mass gap have not been addressed.

Let us now mention a few of the immediate challenges for SDLCQ. First of all
is the extention of the numerical results of section
\ref{ChMaldac} to higher resolution and thus to test the Maldacena's
conjecture. The only problem here is the limits in one's computing resources
and the the speed of the algorithems one uses. Impovements here will also
help use to extend our analysis of higher dimensional system  and to larger
values of transverse truncation. Unfortunately the  transverse truncation
that we have achieved so far does not provide much information about behavior
of the spectrum as a function of transverse resolution and  was used mainly
for illustration of the general concepts. Another consideration is that our
approach studies theories on a light like cylinder and thus our theories may
have as aspects in which they are different than theories in infinite
spacetime or on a space like cylinder.

\section{Acknowledgments}
This work was supported in part by the US Department of Energy. All of the work
reported in these lectures was done in collaboration with Francesco
Antonuccio. We
would like to acknowledge the other members of the SDLCQ project, S.
Tsujimaru, C.
Pauli, J. Hiller, Uwe Trittmann, and Igor Filippos.


\begin{thebibliography}{99}

\bibitem{adler}  S.~Adler,
  ``Axial vector vertex in spinor electrodynamics,''
  {\em Phys. Rev.} {\bf 177} (1969) 2426.

\bibitem{agmoo}
O.~Aharony, S.~S. Gubser, J.~Maldacena, H.~Ooguri, and Y.~Oz, ``Large N field
   theories, string theory and gravity,''
   {{\tt hep-th/9905111}}.

\bibitem{adb97} F.~Antonuccio, S.J.~Brodsky, S.~Dalley,
  "Light-Cone Wavefunctions at Small $x$,"
   {\em Phys.Lett.} {\bf B412} (1997) 104.

\bibitem{alphash99} F.~Antonuccio, A.~Hashimoto, O.~Lunin, S.~Pinsky,
  ``Can DLCQ test the Maldacena Conjecture?'',
   {\em JHEP} {\bf 9907} (1999) 029, {{\tt hep-th/9906087}}.

\bibitem{alp98a}
F.~Antonuccio, O.~Lunin, S.~Pinsky,
``Bound States of Dimensionally Reduced $\mbox{SYM}_{2+1}$ at Finite $N$'',
   {\em Phys.Lett.} {\bf B429} (1998) 327-335, {{\tt hep-th/9803027}}.

\bibitem{alp2}
F.~Antonuccio, O.~Lunin, and S.~Pinsky, ``Nonperturbative spectrum of
   two-dimensional (1,1) superYang-Mills at finite and large N,'' {\em Phys.
   Rev.} {\bf D58} (1998) 085009,
   {{\tt hep-th/9803170}}.

\bibitem{alp99} F.~Antonuccio, O.~Lunin, and S.~Pinsky,
  ``On Exact Supersymmetry in DLCQ," {\em Phys.Lett.} {\bf B442} (1998) 173,
   {{\tt hep-th/9809165}}.

\bibitem{alp3} F.~Antonuccio, O.~Lunin, and S.~Pinsky,
``Super Yang-Mills at Weak, Intermediate and Strong Coupling'',
   {\em Phys.Rev. D} {\bf D59} (1999) 085001, {{\tt hep-th/9811083}}.

\bibitem{alppt}
F.~Antonuccio, O.~Lunin, S.~Pinsky, H.~C. Pauli, and S.~Tsujimaru, ``The DLCQ
   spectrum of N=(8,8) superYang-Mills,'' {\em Phys. Rev.} {\bf D58} (1998)
   105024, {{\tt hep-th/9806133}}.

\bibitem{alptzm}
F.~Antonuccio, O.~Lunin, S.~Pinsky, and S.~Tsujimaru, ``The Light cone vacuum
   in (1+1)-dimensional superYang-Mills theory,''
   {{\tt hep-th/9811254}}.

\bibitem{appt}
F.~Antonuccio, H.~C. Pauli, S.~Pinsky, and S.~Tsujimaru, ``DLCQ bound states of
   N=(2,2) super Yang-Mills at finite and large N,'' {\em Phys. Rev.} {\bf D58}
   (1998) 125006, {{\tt hep-th/9808120}}.

\bibitem{anp97} F.~Antonuccio, S.S.~Pinsky,
  "Matrix theories from reduced $SU(N)$ Yang--Mills with adjoint fermions,"
  {\em Phys.Lett} {\bf B397} (1997) 42, {{\tt hep-th/9612021}}.

\bibitem{anp98} F.~Antonuccio, S.~Pinsky,
  "On the transition from confinement to screening in QCD$_{1+1}$ coupled to
   adjoint fermions at finite N, " {\em Phys.Lett.} {\bf B439} (1998) 142,
   {{\tt hep-th/9805188}}.

\bibitem{aptzm}
F.~Antonuccio, S.~Pinsky, and S.~Tsujimaru, ``A Comment on the light cone
   vacuum in (1+1)-dimensional superYang-Mills theory,''
   {{\tt hep-th/9810158}}.

\bibitem{adi1}
A.~Armoni, Y.~Frishman, and J.~Sonnenschein, ``Screening in supersymmetric
   gauge theories in two- dimensions,'' {\em Phys. Lett.} {\bf B449} (1999) 76,
   {{\tt hep-th/9807022}}.

\bibitem{adi2}
A.~Armoni, Y.~Frishman, and J.~Sonnenschein, ``The String tension in
   two-dimensional gauge theories,''
   {{\tt hep-th/9903153}}.

\bibitem{ars97} A.~Armoni, J.~Sonnenschein,
  ``Screening and Confinement in Large $N_f$ $QCD_2$ and in $N=1$ $SYM_2$,"
   {{\tt hep-th/9703114}}.

%=================================  BBB

\bibitem{bfss97} T.~Banks, W.~Fischler, S.~Shenker, L.~Susskind,
  "M theory as a matrix model: a conjecture,"
  {\em Phys. Rev.} {\bf D55} (1997) 5112,{{\tt hep--th/9610043}}.

\bibitem{bpv94} C.~Bender, S.~Pinsky, B.~van de Sande,
  ``Spontaneous symmetry breaking of $\phi^4_{1+1}$ in light front field
   theory,''
  {\em Phys. Rev.} {\bf D48 } (1993) 816.

\bibitem{bdk93} G.~Bhanot, K.~Demeterfi, I.R.~Klebanov,
"$(1+1)$--Dimensional Large  N QCD Coupled to Adjoint fermions,"
{\em Phys. Rev.} {\bf D48} (1993) 4980, {{\tt hep--th/9307111}}.

\bibitem{bis97} D.~Bigatti, L.~Susskind,
  "Review of Matrix Theory," {{\tt hep-th/9712072}}.

\bibitem{BF}
D.~Birmingham, M.~Blau, M.~Rakowski, and G.~Thompson, ``Topological field
   theory,'' {\em Phys. Rept.} {\bf 209} (1991) 129--340.

\bibitem{kub94} J.~Boorstein, D.~Kutasov,
  "Symmetries and mass splittings in QCD$_2$ coupled to adjoint fermions,"
   {\em Nucl.Phys.} {\bf B421} (1994) 263, {{\tt hep-th/9401044}}.

\bibitem{BISY}
A.~Brandhuber, N.~Itzhaki, J.~Sonnenschein, and S.~Yankielowicz, ``Wilson
   loops, confinement, and phase transitions in large N gauge theories from
   supergravity,'' {\em JHEP} {\bf 06} (1998) 001,
   {{\tt hep-th/9803263}}.

\bibitem{BPP}
S.~J. Brodsky, H.-C. Pauli, and S.~S. Pinsky, ``Quantum chromodynamics and
   other field theories on the light cone,'' {\em Phys. Rept.} {\bf 301} (1998)
   299, {{\tt hep-ph/9705477}}.

\bibitem{zm2}
M.~Burkardt, F.~Antonuccio, and S.~Tsujimaru, ``Decoupling of zero modes and
   covariance in the light front formulation of supersymmetric theories,'' {\em
   Phys. Rev.} {\bf D58} (1998) 125005,
   {{\tt hep-th/9807035}}.


%=================================  CCC

\bibitem{ccg}
C.~G. Callan, N.~Coote, and D.~J. Gross, ``Two-Dimensional Yang-Mills Theory: A
   Model of Quark Confinement,'' {\em Phys. Rev.} {\bf D13} (1976) 1649.

\bibitem{coo98}
  C.~Csaki, H.~Ooguri, Y.~Oz, J.Terning,
  ``Glueball mass spectrum from supergravity,''
   {\em JHEP} {\bf 9901} (1999) 017, {{\tt hep-th/9806021}}.

%=================================  DDD

\bibitem{dak93} S.~Dalley and I.R.~Klebanov,
  "String Spectrum of 1+1-Dimensional Large $N$ QCD with Adjoint Matter",
   {\em Phys. Rev.} {\bf D47} (1993) 2517, {{\tt hep-th/9209049}}.

\bibitem{KresIgor}
K.~Demeterfi and I.~R. Klebanov, ``Matrix models and string theory,''. Lectures
   given at Spring School on String Theory, Gauge Theory and Quantum Gravity,
   Trieste, Italy, 19-27 Apr 1993.

\bibitem{DVV}
R.~Dijkgraaf, E.~Verlinde, and H.~Verlinde, ``Matrix string theory,'' {\em
   Nucl. Phys.} {\bf B500} (1997) 43,
   {{\tt hep-th/9703030}}.

\bibitem{Dirac}
P.~A.~M. Dirac, ``Forms of relativistic dynamics,'' {\em Rev. Mod. Phys.} {\bf
   21} (1949) 392.

%=================================  EEE

%=================================  FFF

\bibitem{fer65} S.~Ferrara,
  ``Supersymmetric gauge theories in two dimensions,''
   {\em Lett. Nuovo. Cimen} {\bf 13} (1975) 629.

\bibitem{ferrara}
S.~Ferrara, C.~Fronsdal, and A.~Zaffaroni, ``On N=8 supergravity on AdS(5) and
   N=4 superconformal Yang- Mills theory,'' {\em Nucl. Phys.} {\bf B532} (1998)
   153, {{\tt hep-th/9802203}}.


%=================================  GGG

\bibitem{givkut} A.~Giveon, D.~Kutasov,
  "Brane dynamics and gauge theory,"
  {{\tt hep-th/9802067}}.

\bibitem{GSW} M.B.~Green, J.H.~Schwarz, E.~Witten, {\em
  Superstring Theory}, Vol.1, CUP (1987).

\bibitem{gri78} V.N.~Gribov,
  ``Quantization of non--abelian gauge theories,''
  {\em Nucl. Phys.}{\bf B139} (1978) 1.

\bibitem{gkm96} D.J.~Gross, I.R.~Klebanov, A.V.~Matytsin, A.V.~Smilga,
  "Screening vs confinement in $1+1$ dimensions,"
  {\em Nucl.Phys} {\bf B461} (1996) 109, {{\tt hep-th/9511104}}.

\bibitem{ghk97} D.J.~Gross, A.~Hashimoto and I.R.~Klebanov,
  "The Spectrum of a Large N Gauge Theory Near Transition From Confinement
   to Screening," {{\tt hep-th/9710240}}.

\bibitem{krasnitz2}
S.~S. Gubser, A.~Hashimoto, I.~R. Klebanov, and M.~Krasnitz, ``Scalar
   absorption and the breaking of the world volume conformal invariance,'' {\em
   Nucl. Phys.} {\bf B526} (1998) 393,
   {{\tt hep-th/9803023}}.

\bibitem{GKP}
S.~S. Gubser, I.~R. Klebanov, and A.~M. Polyakov, ``Gauge theory correlators
   from noncritical string theory,'' {\em Phys. Lett.} {\bf B428} (1998) 105,
   {{\tt hep-th/9802109}}.


%========================    HHH

\bibitem{akisunny}
A.~Hashimoto and N.~Itzhaki, ``A Comment on the Zamolodchikov c function and
   the black string entropy,''
   {{\tt hep-th/9903067}}.

\bibitem{hak95}
A.~Hashimoto and I.R.~Klebanov,
  "Non-perturbative solution of matrix models modified by trace--squared
  terms,''
  {\em Nucl.Phys} {\bf B434}, (1995) 264, {{\tt hep-th/9409064}}.

\bibitem{klebhash} A.~Hashimoto, I.R.~Klebanov,
``Matrix Model Approach to $d>2$ Non-critical Superstrings'',
{\em Mod.Phys.Lett.} {\bf A10} (1995) 2639.

\bibitem{zm1}
S.~Hellerman and J.~Polchinski, ``Compactification in the lightlike limit,''
   {\em Phys. Rev.} {\bf D59} (1999) 125002,
   {{\tt hep-th/9711037}}.

\bibitem{thooft}
G.~'t~Hooft, ``A Two-Dimensional Model for Mesons,'' {\em Nucl. Phys.} {\bf
   B75} (1974) 461.

\bibitem{Horn2}
K.~Hornbostel, ``The Application of Light Cone Quantization to Quantum
   Chromodynamics in (1+1)-Dimensions,''. PhD Thesis, SLAC-0333.

\bibitem{Horn1}
K.~Hornbostel, S.~J. Brodsky, and H.~C. Pauli, ``Light Cone Quantized QCD in
   (1+1)-Dimensions,'' {\em Phys. Rev.} {\bf D41} (1990) 3814.

\bibitem{garyjoe}
G.~T. Horowitz and J.~Polchinski, ``A Correspondence principle for black holes
   and strings,'' {\em Phys. Rev.} {\bf D55} (1997) 6189--6197,
   {{\tt hep-th/9612146}}.


%=================================  III

\bibitem{IMSY}
N.~Itzhaki, J.~M. Maldacena, J.~Sonnenschein, and S.~Yankielowicz,
   ``Supergravity and the large N limit of theories with sixteen supercharges,''
   {\em Phys. Rev.} {\bf D58} (1998) 046004,
   {{\tt hep-th/9802042}}.


%=================================  JJJ

%=================================  KKK

\bibitem{kall} A.C.~Kalloniatis,
  ``On zero zodes and the vacuum problem -- a study of scalar
     adjoint matter in two-dimensional Yang-Mills theory via light-cone
     quantisation,''
   {\em Phys.Rev} {\bf D54} (1996) 2876.

\bibitem{kpp94}  A.C.~Kalloniatis, H.C.~Pauli, S.S.~Pinsky,
  ``Dynamical Zero Modes and Pure Glue ${\bf{\rm{QCD}}_{1+1}}$ in
     Light-Cone Field Theory,''
   {\em Phys. Rev.} {\bf D50} (1994) 6633.

\bibitem{kpp94a} H.C.~Pauli, A.C.~Kalloniatis, S.S.~Pinsky,
  ``Towards solving QCD - the transverse zero modes in light-cone
     quantization,''
  {\em Phys. Rev.} {\bf D52} (1995) 1176.

\bibitem{KRvN}
H.~J. Kim, L.~J. Romans, and P.~van Nieuwenhuizen, ``Mass spectrum of chiral
   ten-dimensional $N=2$ supergravity on $S^5$,'' {\em Phys. Rev.} {\bf D32}
   (1985) 389--399.

\bibitem{klt99} I.~Klebanov and A.~Tseytlin,
``A Non-Supersymmetric Large N CFT from Type 0 String Theory",
{{\tt hep-th/9901101}}

\bibitem{krasnitz1}
M.~Krasnitz and I.~R. Klebanov, ``Testing effective string models of black
   holes with fixed scalars,'' {\em Phys. Rev.} {\bf D56} (1997) 2173--2179,
   {{\tt hep-th/9703216}}.

\bibitem{kut93} D.~Kutasov,
"Two Dimensional QCD Coupled to Adjoint Matter and String Theory,"
{\em Phys. Rev.} {\bf D48} (1993) 4980,  {{\tt hep--th/9306013}}.

%=================================  LLL

\bibitem{lnt94}  F.~Lenz, H.W.L.~Naus, M.~Theis,
  ``QCD in the axial gauge representation,''
  {\em Ann. Phys.} (N.Y.) {\bf 233} (1994) 317.

\bibitem{lst95}  F.~Lenz, M.~Shifman, M.~Thies,
  ``Quantum mechanics of the vacuum state in two-dimensional QCD with
     adjoint fermions,''
  {\em Phys. Rev.} {\bf D51} (1995) 7060

\bibitem{Li95}
M.~Li, ``Large N solution of the 2-d supersymmetric Yang-Mills theory,'' {\em
   Nucl. Phys.} {\bf B446} (1995) 16--34,
   {{\tt hep-th/9503033}}.


%=================================  MMM

\bibitem{adscft}
J.~Maldacena, ``The Large N limit of superconformal field theories and
   supergravity,'' {\em Adv. Theor. Math. Phys.} {\bf 2} (1998) 231,
   {{\tt hep-th/9711200}}.

\bibitem{juanwilson}
J.~Maldacena, ``Wilson loops in large N field theories,'' {\em Phys. Rev.
   Lett.} {\bf 80} (1998) 4859,
   {{\tt hep-th/9803002}}.

\bibitem{MaYam}
T.~Maskawa and K.~Yamawaki, ``The problem of $p^+ = 0$ mode in the null plane
   field theory and Dirac's method of quantization,'' {\em Prog. Theor. Phys.}
   {\bf 56} (1976) 270.

\bibitem{sakai}
Y.~Matsumura, N.~Sakai, and T.~Sakai, ``Mass spectra of supersymmetric
   Yang-Mills theories in (1+1)-dimensions,'' {\em Phys. Rev.} {\bf D52} (1995)
   2446--2461, {{\tt hep-th/9504150}}.

\bibitem{mrp97} G.~McCartor, D.G.~Robertson, S.~Pinsky,
  ``Vacuum structure of two-dimensional gauge theories on the light front,''
  {\em Phys.Rev} {\bf D56} (1997) 1035, {{\tt hep-th/9612083}}.

\bibitem{kalla} A.S.~Mueller, A.C.~Kalloniatis, H.C.~Pauli,
  ``,Effect of zero modes on the bound-state spectrum in light-cone
     quantisation''
  {\em Phys.Lett} {\bf B435} (1998) 189.

%=================================  NNN

%=================================  OOO

\bibitem{Oda97}
H.~Oda, N.~Sakai, and T.~Sakai, ``Vacuum structures of supersymmetric
   Yang-Mills theories in (1+1)-dimensions,'' {\em Phys. Rev.} {\bf D55} (1997)
   1079--1090, {{\tt hep-th/9606157}}.

%=================================  PPP

\bibitem{BP85}
H.~C. Pauli and S.~J. Brodsky, ``Discretized light cone quantization: solution
   to a field theory in one space one time dimensions,'' {\em Phys. Rev.} {\bf
   D32} (1985) 1993, 2001.

\bibitem{pin97}
S.~Pinsky,
  "The Analog of the t'Hooft Pion with Adjoint Fermions,"
   Invited talk at New Nonperturbative Methods and Quantization of the Light
   Cone, Les Houches, France, 24 Feb - 7 Mar 1997. {{\tt hep-th/9705242}}.

\bibitem{pin97a} S. Pinsky,
  ``(1+1)-dimensional Yang-Mills theory coupled to adjoint fermions on the
     light front,"
   {\em Phys.Rev} {\bf D56} (1997) 5040, {{\tt hep-th/9612073}}.

\bibitem{pk96} S.~Pinsky, A.~Kalloniatis,
  ``Light-front QCD$_{1+1}$ coupled to adjoint scalar matter,''
  {\em Phys.Lett} {\bf B 365} (1996) 225.

\bibitem{mrp97a}  S.~Pinsky, D.~Robertson,
  ``Light-front QCD$_{1+1}$ coupled to chiral adjoint fermions,''
  {\em Phys.Lett } {\bf B 379} (1996) 169

%=================================  QQQ

%=================================  RRR

\bibitem{RY}
S.-J.~Rey and J.~Yee, ``Macroscopic strings as heavy quarks in large N gauge
   theory and anti-de Sitter supergravity,''
   {{\tt hep-th/9803001}}.

%=================================  SSS

\bibitem{seiberg} N.~Seiberg,
  "Electric-magnetic duality in supersymmetric non-Abelian gauge theories,"
  {\em Nucl.Phys.} {\bf B435} (1995) 129

\bibitem{seibergwitten} N.~Seiberg, E.~Witten,
  "Monopoles, duality and chiral symmetry breaking in N=2 supersymmetric
     QCD,"
  {\em Nucl.Phys.} {\bf B431} (1994) 484.

\bibitem{str96} M.J.~Strassler,
``Manifolds of fixed Points and Duality in Supersymmetric Gauge theories,"
   {\em Prog.Theor.Phys.Suppl.} {\bf 123} (1996) 373,  {{\tt hep-th/960202}}.

\bibitem{sus97} L.~Susskind, "Another Conjecture About Matrix Theory",
{{\tt hep-th/9704080}}.

%=================================  TTT
\bibitem{tay98} W.~Taylor,
  ``Lectures on D-Branes, Gauge Theory and M(atrices),''
  {{\tt  hep-th/9801182}}.

%=================================  VVV

\bibitem{van96} B. van de Sande,
  "Convergence of discretized light cone quantization in the
   small mass limit," {\em Phys.Rev.} {\bf D54} (1996) 6347,
   {{\tt hep-ph/9605409}}.

%=================================  WWW

\bibitem{WessBag} J.~Wess, J.~Bagger,
  {\em Supersymmetry and supergravity}, Princeton University Press(1992).

\bibitem{wit79}  E.~Witten,
  ``Theta vacua in two--dimensional quantum chromodynamics,''
{\em Nuovo Cim.} {\bf A51} (1979) 325.

\bibitem{witt96} E.~Witten,
  "Bound states of strings and p--branes,"
  {\em Nucl.Phys.} {\bf B460}, (1996) 335, {{\tt hep-th/9510135}}.

\bibitem{Wit}
E.~Witten, ``Anti-de Sitter space and holography,'' {\em Adv. Theor. Math.
   Phys.} {\bf 2} (1998) 253, {{\tt hep-th/9802150}}.

\bibitem{Witten99}
E.~Witten, ``Supersymmetric index of three-dimensional gauge theory,''
   {{\tt hep-th/9903005}}.


%================================= YYY

\bibitem{zm5}
K.~Yamawaki, ``Zero mode problem on the light front,''
   {{\tt hep-th/9802037}}.
\end{thebibliography}
\end{document}